\documentclass[fleqn,usenatbib,useAMS]{mnras}

\usepackage{graphicx}	
\usepackage{amsmath}	
\usepackage{multicol}        
\usepackage{bm}		
\usepackage{pdflscape}	
\usepackage{grffile}
\usepackage{hyperref}

\hypersetup{
    colorlinks=true,
    allcolors=blue
    }

\usepackage[T1]{fontenc}
\usepackage{ae,aecompl}
\usepackage{newtxtext,newtxmath}

\def\spose#1{\hbox to 0pt{#1\hss}}
\def\simlt{\mathrel{\spose{\lower 3pt\hbox{$\mathchar"218$}}
     \raise 2.0pt\hbox{$\mathchar"13C$}}}
\def\simgt{\mathrel{\spose{\lower 3pt\hbox{$\mathchar"218$}}
     \raise 2.0pt\hbox{$\mathchar"13E$}}}

\def\gtsima{$\; \buildrel > \over \sim \;$}
\def\ltsima{$\; \buildrel < \over \sim \;$}
\def\gsim{\lower.5ex\hbox{\gtsima}}
\def\lsim{\lower.5ex\hbox{\ltsima}}
\def\simgt{\lower.5ex\hbox{\gtsima}}
\def\simlt{\lower.5ex\hbox{\ltsima}}
\def\simpr{\lower.5ex\hbox{\prosima}}

\newcommand\Eqn[1]     {Eq.\,(\ref{#1})}

\newcommand\Fig[1]     {Fig.~{\ref{#1}}}

\newcommand{\be}{\begin{equation}}
\newcommand{\ee}{\end{equation}}
\newcommand{\ba}{\begin{eqnarray}}
\newcommand{\ea}{\end{eqnarray}}

\def\pp1{{\prime}}
\def\pp2{{\prime\prime}}

\def\2D{{\rm 2D}}

\def\Vmu{{V_{\mu}}}

\def\Vu{{V_{\mu}}}

\def\bx{{\bf x}}
\def\br{{\bf r}}

\def\bk{{\bf k}}
\def\bq{{\bf q}}

\def\1Loop{{\rm 1Loop}}

\def\rhob{\bar{\rho}}

\def\Gpc{\, h^{-1}{\rm Gpc}}

\def\kMpc{\, h \, {\rm Mpc}^{-1}}

\def\dx{{\rm d}{\bf x}}
\def\dk{{\rm d}{\bf k}}

\def\deltaK{\delta^{\rm K}}

\def\nbar{\bar{n}}

\def\dirac{\delta^{\rm D}}
\def\kronecker{\delta^{\rm K}}

\def\fun#1#2{\lower3.6pt\vbox{\baselineskip0pt\lineskip.9pt
        \ialign{$\mathsurround=0pt#1\hfill##\hfil$\crcr#2\crcr\sim\crcr}}}

\newcommand{\mmathbfss}[1]{\textbf{\textsf{#1}}}
\def\Cov{\mmathbfss{C}}

\def\Imat{\mmathbfss{I}}
\def\Dmat{\mmathbfss{D}}
\def\rmat{\mmathbfss{r}}

\def\Ymat{\mmathbfss Y}
\def\Amat{\mmathbfss A}
\def\Covm{\mmathbfss{C}_{\rm m}}
\def\Covt{\mmathbfss{C}_{\rm t}}
\def\Dmatt{\mmathbfss{D}_{\rm t}}
\def\rmatt{\mmathbfss{r}_{\rm t}}
\def\Rmat{\mmathbfss{R}}

\def\br{\mmathbfss{r}}

\newcommand{\bki}{\bk_{1}}
\newcommand{\bkj}{\bk_{2}}

\begin{document}

\title[Cosmology dependence of the covariance matrix]{An efficient model of cosmology dependence in the covariance matrix of the matter power spectrum}

\author[Steele \& Smith]{Theodore Steele$^{1}$\thanks{Contact e-mail: \href{mailto:T.W.Steele@sussex.ac.uk}{T.W.Steele@sussex.ac.uk}} and Robert E. Smith$^{1}$\thanks{Contact e-mail: \href{mailto:R.E.Smith@sussex.ac.uk}{R.E.Smith@sussex.ac.uk}}\\$^{1}$Astronomy Centre, Department of Physics and Astronomy, University of Sussex, BN1 9RH, UK}
\date{\today}

\maketitle

\begin{abstract}
  Covariance matrices are
  essential for cosmological likelihood analyses.  They can be estimated from large suites of
  cosmological simulations, but this is computationally intensive, so various approximations are often adopted, such as treating it as parameter independent to avoid having to compute matrices for every variation in physical parameters. In this paper, we develop a new
  approach to model the parameter dependence of the covariance matrix
  of the matter power spectrum. We do this by Taylor expanding the
  `true covariance matrix' around a fiducial point in parameter space
  projected onto a theoretically known precision matrix. In the limit
  that the known model approaches the true matrix then the Taylor
  expansion truncates at the first term; for more realistic models,
  additional terms are required.   We use the publicly available Quijote
  suite of simulations, giving us both a fiducial model and variations in the six parameters $\{\Omega_{\rm m}, \Omega_{\rm
  b}, n_{\rm s},\sigma_{8}, h,w_{0}\}$. For each variation, we generate model
  covariance matrices with and without super-sample covariance. Using different approximations, we develop three
  versions of our method and validate them by recreating all of our input cosmologies, both with and without super-sample covariance corrections, showing that the model provides good recreations of
  the original results, with the different approximations being optimal
  in different regimes. Our method enables rapid calculation of
  parameter dependent covariance matrices, and consequently precision matrices, for
  deployment in likelihood modelling codes.  
\end{abstract}

\begin{keywords}
cosmology: theory -- large-scale structure 
\end{keywords}



\section{Introduction}
\label{sec:introduction}

Modern cosmology is primarily focused on using cosmological probes to constrain fundamental physical parameters such as those of dark energy, dark
matter, and the origins and evolution of structure in the Universe.
This has led to the development of a vast array of large-scale
multi-wavelength imaging and spectroscopic surveys such as the Dark
Energy Spectroscopic Instrument (DESI) mission \citep{Wang:2021moa},
ESA's Euclid mission \citep{EuclidPaper1_2025}, the 4MOST cosmology
redshift survey (CRS) \citep{Richard:2019dwt}, the Vera Rubin
Observatory's legacy survey of space and time (LSST)
\citep{2019ApJ...873..111I}, and SPHEREx \citep{Dore:2014cca}, along
with the upcoming Nancy Grace Roman telescope's WFIRST survey
\citep{interloperbias}. Taken together, these will provide us with an
unprecedented volume of cosmological data.

One of the main cosmological probes studied in such surveys is the statistical distribution of matter, studied through the overdensity field that defines fluctuations of matter and tracer densities against a mean background. This is usually used to construct power spectra, the Fourier transform of the two-point correlator of overdensity fields, as well as higher order spectra correlating larger numbers of such fields \citep{GilMarinetal2015,NovellMasotetal2025}.  

However, the accurate interpretation of power spectrum
measurements requires us to perform Bayesian inference. This involves
specifying a likelihood function and a set of priors \citep[for a
  detailed discussion in the context of cosmic shear surveys,
  see][]{Eifleretal2009,Kilbingeretal2013}. 

The treatment of likelihood is a topic of much debate. In many studies it is common
practice to model it as a multi-variate Gaussian
\citep{Eifleretal2009,NovellMasotetal2025}. This requires the
specification of a model mean and a covariance matrix. 

If the covariance matrix is to be estimated from the
data, then noise in the covariance modifies the likelihood, deviating it from a multivariate t-distribution \citep{SellentinHeavens2016}. 

If the covariance matrix is to be treated as a model
dependent quantity, then it is to be obtained
from a theoretical model. In that case one must understand and
characterise its variations with cosmology just as one
models the power spectrum itself. 

If one has an analytic model for the
covariance then the precision matrix can, in principle, be obtained
through standard matrix inversion techniques and the likelihood goes
back to being Gaussian.

Another element of the debate circles on the fact that it is the
real-space fractional over-density field that is, at least in its
initial primordial state, Gaussian distributed. Consequently, the power spectrum for a given mode should be exponentially
distributed, such that the simplest power spectrum estimator should be
$\chi^2$-distributed \citep{Takahashietal2009}. 

In the limit of a
large number of modes, the central limit theorem leads to the
$\chi^2$-distribution being well approximated by that of a
Gaussian. 

However, as shown in \citet{Carron2013},
treating the power spectrum likelihood as a multivariate Gaussian
with cosmology dependent mean and covariance leads to violations
of the Cramer-Rao bounds on the Fisher information\footnote{For a
discussion of these bounds, see \citet{Heavens2009}.} for a Gaussian
density field. The conclusion drawn from this is that one should
not vary the covariance matrix, as this would lead to artificially
improved constraints.  

However, there are reasons to question this conclusion. A number of studies have shown that the covariance matrix is meaningfully model dependent
\citep{Eifleretal2009}; so, adopting the wrong covariance
may lead to inaccuracies in parameter estimates. Furthermore, nonlinear
evolution of large-scale structure generates
off-diagonal elements in the covariance matrix
through the onset of mode-coupling, which slows or saturates the
information contained in the two-point statistics
\citep{Byunetal2017}.  In addition, the overdensity distribution at late
times is non-Gaussian with a hierarchy of higher-order connected
correlators, all of which contain non-trivial information
\citep{TaylorWatts2001,Sefusattietal2006}.  So, results from Gaussian
based models become increasingly irrelevant, especially when studying late-time
observables.

Ultimately, the covariance matrix, and
consequently the precision matrix, do vary with respect to the chosen
cosmological model and neglecting this variation means that one may
bias the best-fit cosmological parameters and inaccurately estimate
confidence contours, with the magnitude of such errors depending on the survey size and the dimensionality of the parameter space under study
\citep{Eifleretal2009,labatieetal2012}.

In the Gaussian case, the covariance matrix of the power spectrum is
given by a diagonal matrix of variance terms entirely derived from the
power spectrum itself \citep{Feldmanetal1994}. As structure grows under
gravitational instability, large-scale modes couple to
small-scales modes and power cascades from large-scales to small.  

The
most common analytic approach to model this growth of structure
is Eulerian perturbation theory, commonly known as Standard Perturbation
Theory (SPT) \citep{Bernardeauetal2002}.  In this model, the nonlinear density field is estimated from a power series expansion around the linear density field as $\delta_{\mathrm{nl}}=\sum_{i=1}^{\infty}\delta_{i}$ and correlators are constructed from products of these perturbative fields, with a Feynman diagram representation used to portray each contribution to the fully non-linear correlators.  Couplings between studied modes and those of arbitrarily small physical scales are represented in these diagrams by loops, such that higher order contributions are referred to as loop terms.



As nonlinearities grow, the
covariance between modes in the power spectrum grows
\citep{MeiksinWhite1999,Scoccimarroetal1999,NeyrinckSzapudi2008} and off-diagonal terms in the
covariance matrix arise, which are sourced by the
connected four point function, known in Fourier space as the trispectrum.

A further important contribution to the covariance arises from Fourier modes with wavelengths longer than the scales being modelled. These modes can couple to and affect smaller scale modes, thus affecting the results of a given model; this is a phenomenon
known as super-sample covariance (SSC)
\citep{TakadaHu2013,Lietal2014}.  Corrections, therefore, should
be added to the predicted covariances to account for this effect.
Shot-noise terms will also arise due to the underlying density field being continuous while observable tracers, such as halos and galaxies, are discrete
\citep{MeiksinWhite1999,Smith2009,WadekarScoccimarro2020}
as well as biasing terms which account for disparities between the matter field, which treats matter as purely gravitational, and observable tracers such as baryonic matter
\citep{WadekarScoccimarro2020}. For the purposes of this work, biasing is neglected.

Empirical determination of the covariance matrix for a given set of
cosmological parameters can be achieved through the generation of
large ensembles of $N$-body simulations
\citep{Takahashietal2009,Blotetal2015,Villaescusa-Navarroetal2020}.
This is computationally intensive and becomes unfeasible for mapping
multi-dimensional parameter spaces, as each point in parameter space
will need a new run of the simulations suite. This naturally raises the question of how one can model parameter dependence without recourse to simulations for every permutation. Some authors have explored the possibility of
building emulators across the prior space \citep{WhitePadmanabhan2015}; this is a viable approach, but still requires a large amount of
computational resources, especially as one expands the cosmological
model space. 

Another approach that has been tried is that of \citet{Blotetal2021},
who computed for the first time the relative variations in the
covariance matrices from a suite of $N$-body simulations, where large
ensembles of simulations were run at various points in parameter space
around a fiducial point. In this work, the authors showed that a
Taylor expansion could provide a starting basis for model building.
 
In this paper, we build on this latter approach and develop an
efficient method for estimating the parameter dependence of the
covariance matrix which requires explicit calculations only at a
few single points in the parameter space.  This method is based upon a
Taylor expansion of the covariance matrix projected onto the precision
matrix of a theoretical model.  By defining new expressions from the
ensuing expanded matrix which quantify sensitivity to parameter
variations, we are able to derive an expression for the covariance
matrix as a function of input parameters which may be analytically
calculated from a pre-existing set of covariance matrices, negating
the need for additional computations of the full matrix with new
simulation data.  This permits parameter spaces to be explored with
trivial computational time and no need for additional simulations
beyond those that were used to establish the model.


To demonstrate the method, we employ the Quijote simulation results
from the work of \citet{Villaescusa-Navarroetal2020}, to generate both
non-linear and SPT covariance matrices that map the parameter space
with dimensions
$\{\Omega_{\rm m}, \Omega_{\rm
  b}, n_{\rm s},\sigma_{8}, h,w_{0}\}$ (see App.~\ref{app:sims} for more details).

The rest of this paper is organised as follows: in
\S\ref{sec:background}, we summarise in more detail the theoretical concepts
that we will make use of in this work. In \S\ref{sec:model}, we present
our approach for modelling cosmological dependence of the covariance
matrix from a set of ensembles of simulations around some fiducial
point in parameter space. In \S\ref{sec:simulations}, we present an
overview of some basic results from the Quijote simulations. In
\S\ref{sec:reconstruction}, we show how one can reconstruct the
covariance matrix and validate the model. Finally, in
\S\ref{sec:discussion}, we discuss our findings and conclude.


\section{Theoretical background}\label{sec:background}

In this section we will review the standard form for the theoretical
covariance matrix as modelled with SPT (itself summarised in
App.~\ref{app:SPT}) in both the Gaussian and non-Gaussian cases, before
incorporating shot noise terms and SSC corrections.  We will then band
average the matrix, permitting studies in discrete bins, as is
standard in both theoretical and observational analyses of LSS.


\subsection{Spatial statistics}

In this work we will be primarily focused on over-density
perturbations in the distribution of matter. These can be defined at a
given coordinate time $t$ and comoving position $\bx$ as:
\be \delta(t,\bx) = \left[\rho(t,\bx)-\rhob(t)\right]/\rhob(t)\ . \ee
where $\rho$ and $\rhob$ denote the local density of matter and the
background density of the homogeneous and isotropic background, respectively.
Assuming space is flat and neglecting dependence on time, we can use statistical homogeneity and isotropy to
Fourier transform the field, giving us the time independent Fourier density modes:
\be \delta(\bk) = \int \dx\, \delta(\bx) e^{i\bk\cdot\bx} \ .\ee
We now wish to compute statistical averages of products of the Fourier
modes. Owing to the fact that that, by construction,
$\left<\delta\right>=0$, the lowest order non-trivial statistic of
interest is that of the product of two Fourier modes. This leads us to
define the power spectrum $P(k)$ of the field as:
\be \left<\delta(\bk_1)\delta(\bk_2)\right> \equiv (2\pi)^3 \dirac(\bk_1+\bk_2)P(k_1) \ .\ee
The Dirac delta function, $\dirac(\bk_1+\bk_2)$ results from the
application of statistical homogeneity to the correlation function
$\left<\delta(\bx)\delta(\bx+\br)\right>$, which means that it is
dependent only on $\mathbf{r}$ and not $\mathbf{x}$. The fact that $P(k)$ depends only on the
length of the $k$-vector encodes the statistical isotropy of the
spatial statistic \citep[for details see][]{Peacock1999}.  

One can
further generalise this idea to higher order correlators:
\begin{align}
  \left<\delta(\bk_1)\delta(\bk_2)\delta(\bk_3)\right>  \equiv& (2\pi)^3
  \delta^{\mathrm{D}}\left(\sum_{i=1}^{3}\bk_i\right)B(\bk_1,\bk_2,\bk_3) \ , \\
  \left<\delta(\bk_1)\delta(\bk_2)\delta(\bk_3)\delta(\bk_4)\right>  \equiv& (2\pi)^3
  \delta^{\mathrm{D}}\left(\sum_{i=1}^{4}\bk_i\right) T(\bk_1,\bk_2,\bk_3,\bk_4) \ , \label{eq:trispectrum}
\end{align}
where in the above $B$ and $T$ denote the bispectrum and trispectrum,
respectively.


\subsection{Contributions to the covariance of the power spectrum}

In order to accurately extract cosmological information from the full
shape of the power spectrum, one must account for
errors on the estimates.

\vspace{0.2cm}\noindent $\bullet$ {\bf Gaussian field:} In the case of purely
Gaussian fluctuations, the covariance between two
different power spectrum estimates is non-vanishing, and is simply given by
\citep{Feldmanetal1994}:
\be
C^{\rm G}(\bk_1,\bk_2)=P(k_1)P(k_2)\left(\deltaK_{\bk_{1},\bk_{2}}+\deltaK_{\bki,-\bkj}\right)\ .
\label{eq:gCmatrix}
\ee
This is also known as the `background', or `disconnected'
contribution.

\vspace{0.2cm}\noindent $\bullet$ {\bf Non-Gaussian field:}
Incorporating non-Gaussian contributions necessitates including the
connected trispectrum in configurations corresponding to coupled pairs of
power spectra \cite{Mohammedetal2017}, providing off-diagonal
elements as well as modifying the diagonals:
\begin{align}
    C(\bki,\bkj) & =\frac{1}{\Vu}T^{\rm
      c}(\bki,\bkj,-\bki,-\bkj)+C^{\rm G}(\bk_1,\bk_2)\ ,
    \label{eq:CNG}
\end{align}
where $T^{\rm c}$ identifies the connected contribution to the
trispectrum and $\Vu$ is the sample volume.

\vspace{0.2cm}\noindent $\bullet$ {\bf Discrete sampling of Gaussian
  field:} When dealing with discrete tracers such as galaxies and
halos, additional discreteness terms become relevant
\citep{Peebles1980,MeiksinWhite1999,Smith2009}. Shot noise terms
primarily come in three forms: a constant addition that accounts for
Poissonian variance from random point sampling, a bispectrum term that
accounts for non-Gaussianity introduced by gravitational collapse, and
an additional power spectrum term that relates to the scale dependence
of discrete tracer formation that arises from the scale dependent
distribution of matter in the continuous matter field. Upon
incorporating the discreteness terms, we find that the background
covariance gains the additional terms:
\begin{equation}
\begin{split}
  C^{\mathrm{G}}_{\rm d}(\mathbf{k}_{1},\bk_{2}) & = \left[P(k_1)+\frac{1}{\nbar}\right]\left[P(k_2)+\frac{1}{\nbar}\right]
  \left(\delta^K_{\bk_1+\bk_2,0}+\delta^K_{\bk_1-\bk_2,0}\right)\\
  &+\frac{1}{\nbar^2\Vmu} \left[P(\bk_1)+P(\bk_2)+P(-\bk_1)+P(-\bk_2)\right]  \\
  & +
  \frac{1}{\nbar^2\Vmu} \left[P(\bk_1+\bk_2)+P(\bk_1-\bk_2)\right]
  +\frac{1}{\nbar^3\Vmu}\ ,
  \label{eq:CdGauss}
  \end{split}
\end{equation}
where $\bar{n}$ is the number density of the tracers.

\vspace{0.2cm}\noindent $\bullet$ {\bf Discrete sampling of
  non-Gaussian field:} On including the additional non-Gaussian
contributions from the connected trispectrum and bispectrum, this
further becomes:
\begin{equation}
\begin{split}
  C_{\rm d}(\bk_1,\bk_2)  & = C^{\mathrm{G}}_{\rm d}(\mathbf{k}_{1},\bk_{2}) +
  \frac{T^{\rm c}(\bk_1,\bk_2,-\bk_1,-\bk_2)}{\Vmu} \\
    & + \frac{1}{\nbar\Vmu}\left[
    B(\bk_1,\bk_2,-\bk_1-\bk_2)+B(\bk_1,-\bk_2,\bk_2-\bk_1)
    \right.\\&\left. +B(-\bk_1,\bk_2,\bk_1-\bk_2)+B(-\bk_1,-\bk_2,\bk_1+\bk_2)\right]  \ .
  \label{eq:CdNG}
  \end{split}
\end{equation}

\vspace{0.2cm}\noindent $\bullet$ {\bf Super-sample covariance}:
Structures on scales larger than those measured in a given survey or
simulation will have gravitational effects on those within the
measured volume.  These effects can become significant and are termed
super-sample covariance (SSC). These were discussed in
\citet{Sirko2005} and their contributions to the covariance matrix
were originally derived by \citet{TakadaHu2013}.  In the simple cases that we are interested in, this takes the form:
\begin{equation}
\mathrm{SSC}(\bki,\bkj)= P(k_{1})P(k_{2})\int_{0}^{k_{\mathrm{min}}}\frac{dk}{2\pi^{2}} k^{2} P_{11}(k)~.
\label{eq:SSC}
\end{equation}
where in the above integral $k_{\rm min}\approx 2\pi/L$ is computed
over all scales physically larger than the fundamental mode of the
survey volume. These scales are coupled at each point in momentum
space to the power spectra at the scales of interest. This
contribution is added directly to the covariance matrix to correct for
coupling between the modes being sampled and those beyond the sample
limits.


\subsection{Band Averaging}

The covariance and correlation matrices discussed so far have applied
to individual pairs of wavenumbers.  In practice, it is not practical
to correlate every mode individually and they are instead band
averaged, with those averages then being correlated to generate a band
averaged set of correlators. The covariance matrix of the band
averaged power spectrum is given by the band average of the mode-pair
covariance matrix itself:
\begin{equation} 
  \overline{C}(k_{i},k_{j}) \equiv \int_{V_{s,i}}
  \frac{\dk_1} {V_{s,i}}\int_{V_{s,j}}\frac{\dk_2}{V_{s,j}} C(\bk_1,\bk_2)~ \ .
\label{eq:band}
\end{equation}
where $k_i$ represents the centre of the $i$-th $k$-space shell and
where its volume is given by:
\be
  V_{s,i}  \equiv \int_{V_{s,i}} \!\!\dk = 4\pi \int_{k_i-\Delta /2}^{k_i+\Delta /2} \!\!k^2dk 
  = 4\pi k^2 \Delta\left[1+\frac{1}{12}\left(\frac{\Delta}{k}\right)^2\right]\ , \label{eq:shell}
\ee
where $\Delta$ is the width of the $k$-space shell.

On inserting \Eqn{eq:gCmatrix} into \Eqn{eq:band}, we obtain the band
averaged covariance matrix of a continuous Gaussian matter field:
\begin{align} 
\overline{C}^{\rm G}(k_i,k_j) & = \frac{2}{N_{i}}
\left(\overline{P}(k_i)+\frac{1}{\nbar}\right)^2\kronecker_{i,j}  \ , \label{eq:GaussShell}
\end{align}
where $N_i$ is the total number of modes in the shell and where we
have defined the band-averaged power spectrum as:
\be \overline{P}(k_i)  \equiv \int_{V_{s,i}} \frac{\dk_1}{V_{s,i}} P(\bk_1) \ .\ee 

For the corresponding Non-Gaussian case, as given by \Eqn{eq:CNG}, the
above expression is modified to:
\begin{align} 
  \overline{C}(k_{i},k_{j}) & = \frac{1}{\Vu}\overline{T}^{\rm c}(k_i,k_j) + \overline{C}^{\rm G}(k_{i},k_{j})\ , \label{eq:covTri}
\end{align}
where we have defined the band-averaged connected trispectrum as
\be \overline{T}^{\rm c}(k_i,k_j) \equiv \int_{V_{s,i}}
\frac{\dk_1}{V_{s,i}}\int_{V_{s,j}} \frac{\dk_2}{V_{s,j}} T^{\rm
  c}(\bk_1,\bk_2,-\bk_1,-\bk_2) . \ee

In the case of a Gaussian field with discrete tracers, this becomes
\begin{equation}
\begin{split}
\overline{C}^{\rm G}_{\rm d}(k_{i},k_{j}) & = \frac{2}{N_{k_i}}
\left(\overline{P}(k_i)+\frac{1}{\nbar}\right)^2\delta^{K}_{k_i,k_j}  \\
& + \frac{2}{\nbar^2\Vu}\left[ \overline{P}(k_{i},k_{j})+ \overline{P}(k_i)+
  \overline{P}(k_j) \right] +\frac{1}{\nbar^3 V_{\mu}} \ .
\label{eq:gBACovM}
\end{split}
\end{equation}
where in the above expression we defined
\begin{align}
  \overline{P}(k_i,k_j) & \equiv \int_{V_{s,i}} \frac{\dk_1}{V_{s,i}}\int_{V_{s,j}} 
  \frac{\dk_2}{V_{s,j}} P(\bk_1+\bk_2) \ .
\end{align}
The band averaged non-Gaussian covariance matrix can likewise be
obtained by inserting Eq.~\eqref{eq:CdNG} into Eq.~\eqref{eq:band}
to obtain
\begin{align} 
  \overline{C}_{\rm d}(k_{i},k_{j}) & = \overline{C}^{\rm G}_{\rm d}(k_{i},k_{j}) +
  \frac{\overline{T^{\rm c}}(k_i,k_j)}{\Vmu} 
   + \frac{4}{\nbar\Vmu}\overline{B}(k_i,k_j)  \ ,
  \label{eq:BACovM}
\end{align}
where we introduce the bin-averaged bispectrum as:
\begin{align}
  \overline{B}(k_i,k_j) & \equiv \int_{V_{s,i}}
  \frac{\dk_1}{V_{s,i}}\int_{V_{s,j}} \frac{\dk_2}{V_{s,j}}
  B(\bk_1,\bk_2,-\bk_1-\bk_2) \ .
\end{align}
We note that in the limit that $N=\nbar\Vmu\rightarrow\infty$ the
discreteness contributions to the covariance are suppressed.

To incorporate super-sample covariance, we would likewise band average
over the power spectra in Eq.~\eqref{eq:SSC} to obtain
\begin{equation}
  \overline{\mathrm{SSC}}(k_{i},k_{j})= \overline{P}(k_{i})
  \overline{P}(k_{j})\int_{0}^{k_{\mathrm{min}}}\frac{dk}{2\pi^{2}} k^{2} P(k)~,
\label{eq:BASSC}
\end{equation}
which can be added directly to Eqs.~\eqref{eq:gBACovM} and
\eqref{eq:BACovM} to incorporate super-sample covariance effects.
Given the dominance of the linear power spectrum on super-sample
scales, we calculate Eq.~\eqref{eq:BASSC} with only the linear power
spectrum in all of our subsequent analysis.


\subsection{Correlation matrix and reduced covariance matrix}

Finally, in what follows we will also make use of the correlation
matrix and introduce a quantity that we have defined as the reduced
covariance matrix. The correlation matrix can be defined as
\be
    r(\bki,\bkj)\equiv\frac{C(\bki,\bkj)}{\sqrt{C(\bki,\bki)C(\bkj,\bkj)}}
\ee
and is constrained to have values $r\in[-1,1]$. We also define a
useful quantity that we have dubbed the `reduced covariance matrix':
\begin{equation}
    C^{\mathrm{R}}(\bki,\bkj)\equiv\frac{C(\bki,\bkj)}{\sqrt{C^{\mathrm{G}}(\bki,\bki)C^{\mathrm{G}}(\bkj,\bkj)}}\ .
    \label{eq:CovReduced}
\end{equation}
This latter matrix is useful measure for model building purposes, as
it highlights the deviations away from the Gaussian model more clearly
and rescales the Non-Gaussian (off-diagonal) terms.


\section{Modelling the parameter dependencies of the covariance matrix}\label{sec:model}

\subsection{Perturbative expansion around an approximate model}

We would like to model the dependence of the covariance matrix on the available set of parameters. 

Let us begin by assuming that we have a
model for the covariance matrix which is accurate to roughly 10\% and
for which we can analytically compute both the precision matrix and the matrix's derivatives with respect to the parameters being varied. Without loss of generality, we can then say
\begin{align} 
  \Covt(\bm\theta)
  & = \Covm(\bm\theta)\Covm^{-1}(\bm\theta)\Covt(\bm\theta) = \Covm(\bm\theta)\Ymat(\bm\theta)
  \label{eq:model1}\ ,
\end{align}
where $\Covt$ is the true covariance matrix that we wish to obtain, $\Covm$ is the known model covariance matrix, and we defined
\be \Ymat(\bm\theta) \equiv \Covm^{-1}(\bm\theta)\Covt(\bm\theta) \ . \ee
If the model was perfect then the $\Ymat(\bm\theta)$ would simply be
an identity matrix. However, no analytic model can actually achieve such accuracy, resulting in non-trivial forms for the matrix. 

In what follows, we will aim to perturbatively
expand $\Ymat$ around some fiducial point in the cosmological
parameter space for a given model.  We will show that in
order to do this we will need \textit{a priori} knowledge of the true
non-linear covariance matrices at the fiducial point and in a compact
space of parameter variations around that point. 

This approach
builds on the method that was described in \citet{Blotetal2021}, which
focuses on a perturbative expansion of the covariance matrix. We note that \Eqn{eq:model1} is, by construction, guaranteed to be symmetric.


\subsection{Approximating the $\Ymat$ matrix}

Let us now perform the perturbative expansion of \Eqn{eq:model1}.  To
begin, we note that each matrix element of $\Ymat(\bm\theta)$ can be
considered as a function of cosmological parameters $\bm\theta$
and so can be Taylor expanded around some fiducial point in
parameter space $\bm\theta_0$:
\begin{equation}
\begin{split} 
  \Ymat(\bm\theta)
  & = \Ymat(\bm\theta_0)
  +\sum_{\alpha} \left.\frac{\partial\Ymat(\bm\theta)}{\partial \theta_{\alpha}}\right|_{\bm\theta_0} \Delta\theta_\alpha + \dots 
  \end{split}
\end{equation}
As was shown in \citet{Blotetal2021} the second order contributions to
the Taylor expansion of the covariance matrix are relatively weak for
most parameters. We expect that this should be an even better
approximation here, since the projection of the approximate model's
precision matrix on to the true covariance matrix should reduce parameter dependence. We therefore truncate the sequence at linear order in
$\Delta\theta_{\alpha}$. Upon doing this, we define our approximate model as:
\begin{equation}
\begin{split}
  \widetilde{\Ymat}(\bm\theta)
  & \equiv
  \Ymat(\bm\theta_0)\left[\Imat+\Ymat^{-1}(\bm\theta_0)\sum_{\alpha}
    \left.\frac{\partial \Ymat(\bm\theta)}{\partial
      \theta_{\alpha}}\right|_{\bm\theta_0} \Delta\theta_\alpha
    \right] \\
    & =
  \Ymat(\bm\theta_0)\left[\Imat+\sum_{\alpha} {\bm\Rmat}_{\alpha}(\bm\theta_0)
    \Delta\theta_\alpha \right] \ ,
  \label{eq:Ycovmodel1}
  \end{split}
\end{equation}
where in the last line we have defined a set of constant matrices that
control the strength of the response to a variation in the covariance
to changes in a chosen cosmological parameter $\alpha$, which we call the response matrices:
\be {\bm\Rmat}_{\alpha}(\bm\theta_0) \equiv \Ymat^{-1}(\bm\theta_0) \left.\frac{\partial
      \Ymat(\bm\theta)}{\partial \theta_{\alpha}}\right|_{\bm\theta_0}\ .  \label{eq:response0}\ee
Thus, we define our approximation of the true covariance matrix as
\begin{align} 
  \widetilde{\Covt}(\bm\theta)
  & \equiv \Covm(\bm\theta)\widetilde\Ymat(\bm\theta)
  \label{eq:model1b}\ ,
\end{align}
For our model to be a
valid covariance matrix it must satisfy two important properties:
it should be symmetric and positive
semi-definite. This latter statement means that for any vector
${\bx}$, the matrix must obey the relation
$\sum_{ij}x_iC_{ij}x_j\ge0$. In other words the eigenvalues of the
matrix must all be real and non-negative. In truncating the Taylor
expansion as above, we have likely broken both of these properties,
except at the fiducial point $\bm\theta_0$, for which they are
guaranteed to be satisfied.

Regarding the symmetry, while $\Covm(\bm\theta)\Ymat(\bm\theta)$ must
be symmetric, $\Covm(\bm\theta)\widetilde\Ymat(\bm\theta)$ is not
guaranteed to be so. Therefore, in order to ensure that our model
covariance matrix is always symmetric, we work with the following
quantity:
\begin{align} 
  \Cov_{\parallel}(\bm\theta)
  & = \frac{1}{2}\left[\Covm\widetilde{\Ymat} + \widetilde{\Ymat}^{T}\Covm\right] \ .
  \label{eq:Cparallel}
\end{align}
Interestingly, the construction \Eqn{eq:Cparallel}
also suggests a way in which we can gain an understanding of the error
on the perturbative expansion's contribution to the covariance
modelling. That is we can also construct the anti-symmetric part of
the covariance model as:
\begin{align} 
  \Cov_{\perp}(\bm\theta)
  & \equiv \frac{1}{2}\left[\Covm\widetilde{\Ymat} - \widetilde{\Ymat}^{T}\Covm\right]\ ,
\end{align}
which should vanish any well-behaved covariance matrix and so any
residual non-zero contributions should provide an indication of the
magnitude of the failure of the method.

Regarding the positive semi-definiteness requirement, one can compute
the eigenvalues of the resulting approximation and demand that they
obey the non-negative requirement, such as $\widetilde{\lambda}_i\ge
0$. Again, any departures from this requirement provide some degree
of failure of the method and the validity of the approximate model.


\subsection{Developing the response matrices $\Rmat_{\alpha}$}

The response matrices ${\bm\Rmat}_{\alpha}$, as defined in
\Eqn{eq:response0}, depend on the derivatives of the matrix $\Ymat$
with respect to the parameters under study, which is given by
\be \frac{\partial \Ymat(\bm\theta)}{\partial \theta_{\alpha}}
= \frac{\partial \Covm^{-1}(\bm\theta)}{\partial\theta_\alpha}\Covt(\bm\theta) 
+\Covm^{-1}(\bm\theta)\frac{\partial\Covt(\bm\theta)}{\partial\theta_{\alpha}} \ .
\ee
This suggests that we need to compute the
derivatives of the inverse of the model covariance matrix, which may contribute noise to the calculation. This can be
avoided if we recall that for any
invertible square matrix $A$, the derivative of the inverse obeys the
relation 
\be \frac{\partial\Amat^{-1}(x)}{\partial x}  = - \Amat^{-1}(x)\frac{\partial\Amat(x)}{\partial x}\Amat^{-1}(x)  \ .\ee
Upon applying this well known result, we see that
\begin{align} \frac{\partial \Ymat(\bm\theta)}{\partial \theta_{\alpha}}
& = - \Covm^{-1}\frac{\partial\Covm}{\partial\theta_{\alpha}}\Covm^{-1} \Covt(\bm\theta) 
+\Covm^{-1}(\bm\theta)\frac{\partial\Covt(\bm\theta)}{\partial\theta_{\alpha}} \ .
\end{align}
Evaluating this at the fiducial point $\bm\theta_0$ and multiplying
by $\Ymat^{-1}(\bm\theta_0)$, we see the constant response matrix to
be
\begin{equation}
\begin{split}
  \Rmat_{\alpha} & = \Ymat^{-1}\frac{\partial \Ymat}{\partial \theta_{\alpha}}  \\
& = (\Covm^{-1}\Covt)^{-1}\left[-\Covm^{-1}\frac{\partial\Covm}{\partial\theta_{\alpha}}\Covm^{-1} \Cov_{\rm t} 
+\Covm^{-1}\frac{\partial\Covt}{\partial\theta_{\alpha}}\right]  \\  
& = - \Covt^{-1}\Covm\Covm^{-1}\frac{\partial\Covm}{\partial\theta_{\alpha}}\Covm^{-1} \Cov_{\rm t} 
+\Covt^{-1}\Covm\Covm^{-1}\frac{\partial\Covt}{\partial\theta_{\alpha}}  \\
& = \Covt^{-1}\frac{\partial\Covt}{\partial\theta_{\alpha}}
- \Covt^{-1}\frac{\partial\Covm}{\partial\theta_{\alpha}}\Covm^{-1} \Cov_{\rm t}  \\
& = \Covt^{-1}\left[\frac{\partial\Covt}{\partial\theta_{\alpha}}
- \frac{\partial\Covm}{\partial\theta_{\alpha}}\Covm^{-1} \Cov_{\rm t} \right] \ , \label{eq:resp}
\end{split}
\end{equation}
where the right-hand-side of the above expression is to be evaluated
at $\bm\theta_0$.  Thus, to model the matrix $\Rmat_{\alpha}$ at the
fiducial point, we need to be able to compute $\Covt$, its inverse, and
its derivative with respect to the parameters under study at the
fiducial point ${\bm\theta}_0$. It is also
worth noting that if $\Covm\rightarrow\Covt$ then from the above
expression we see that $\Ymat^{-1} \partial \Ymat/\partial
\theta_{\alpha}\rightarrow {\bf 0}$, as it should.

When implementing the parameter derivatives of the covariance matrices, we defined the finite difference estimator
\begin{equation}
\frac{dC}{d\theta_{\alpha}}\bigg\vert_{\mathrm{FD}}=\frac{C(\theta_{0}+\Delta\theta_{\alpha})-C(\theta_{0}-\Delta\theta_{\alpha})}{2\Delta\theta_{\alpha}}
\end{equation}
and also interpolated our covariance matrices across the parameter variations before using the inbuilt derivative function in Julia, the results from which we here label $dC/d\theta_{\alpha}\vert_{\mathrm{Julia}}$.  We found that both methods gave us the same results.  For the rest of the paper, we use $dC/d\theta_{\alpha}\vert_{\mathrm{Julia}}$ for our calculations of the derivatives.

The full and reduced response matrices in both their complete and approximate forms for our various cosmologies are shown in App.~\ref{app:response}.


\subsection{Decomposition of the derivatives of the covariance matrix}

We wish to estimate the derivatives of the true covariance matrix with
respect to the parameters under study. To that end, one may follow the
direct approach developed in \citet{Blotetal2021}.  However, here we
will exploit the well known property that the covariance matrix can be
decomposed as
\be
\Cov = \Dmat \rmat \Dmat \label{eq:Cdecomp1}~,\ee 
where $\Dmat$ is the diagonal matrix of
standard deviations of the $N$ variables and $\rmat$ is the
correlation matrix, which is the covariance matrix normalised to its own diagonal. Thus, the
inverse of the covariance matrix, known as the precision matrix,
can be decomposed as
\be
\Cov^{-1}
= (\Dmat \rmat \Dmat)^{-1}
= \Dmat^{-1} \rmat^{-1} \Dmat^{-1} ~.\label{eq:CinvDecomp1}
\ee

Taking the derivative of Eq.~\eqref{eq:Cdecomp1} with respect to parameter $\alpha$ gives us
\begin{equation}
\begin{split}
  \frac{\partial\Covt}{\partial\theta_\alpha}
  & =
  \frac{\partial\Dmatt}{\partial\theta_\alpha} \rmatt \Dmatt +
  \Dmatt\frac{\partial\rmatt}{\partial\theta_\alpha}  \Dmatt +
  \Dmatt \rmat \frac{\partial\Dmatt}{\partial\theta_\alpha}  \\
  & =
  \frac{\partial\Dmatt}{\partial\theta_\alpha} \rmatt \Dmatt +
  \Dmatt\frac{\partial\rmatt}{\partial\theta_\alpha}  \Dmatt +
  \left(\frac{\partial\Dmatt}{\partial\theta_\alpha} \rmatt \Dmatt\right)^{T}\ .
  \label{eq:diffcov}
  \end{split}
\end{equation}
So, by estimating
the derivative of the diagonal matrix of standard deviations $\Dmatt$,
the derivative of the correlation matrix $\rmatt$, and evaluating $\Dmatt$
and $\rmatt$ at the fiducial point, we can estimate the derivative of the covariance matrix with respect to a given parameter.


\subsection{Estimating the response matrix $\Rmat$}

Inserting Eq.~\eqref{eq:diffcov} into Eq.~\eqref{eq:resp}, we see that 

\begin{align} \widehat{\Rmat}_{\alpha} 
  & = \widehat{\Covt^{-1}}
  \left[
    \widehat{\frac{\partial\Dmatt}{\partial \theta_{\alpha}}} \widehat{\rmatt}\widehat{ \Dmatt}
   +\widehat{\Dmatt}\widehat{\frac{\partial\rmatt}{\partial\theta_\alpha} }\widehat{ \Dmatt}
   +\left(\widehat{\frac{\partial\Dmatt}{\partial \theta_{\alpha}}} \widehat{\rmatt}\widehat{ \Dmatt}\right)^T  
      - \frac{\partial\Covm}{\partial\theta_{\alpha}}\Covm^{-1} \widehat{\Cov_{\rm t}} \right] \ ,\label{eq:fullR}
\end{align}
where all terms are to be evaluated at the fiducial point
$\bm\theta_0$. 

In what follows we will also consider the case where the
correlation matrix is roughly constant and only slowly varies with
cosmology, such that $\partial \rmat/\partial \theta_\alpha \approx
0$. In this case, we can approximate the response matrix as
\begin{align} \widehat{\Rmat}^{\rm approx}_{\alpha}
  & \approx \widehat{\Covt^{-1}}\left[
    \widehat{\frac{\partial\Dmatt}{\partial \alpha}} \widehat{\rmatt}\widehat{ \Dmatt}
    +\left(\widehat{\frac{\partial\Dmatt}{\partial \alpha}} \widehat{\rmatt}\widehat{ \Dmatt}\right)^T
    - \frac{\partial\Covm}{\partial\theta_{\alpha}}\Covm^{-1} \widehat{\Cov_{\rm t}} \right]\label{eq:aR}
 \ .
\end{align}
The motivation for this is to remove what we would expect to
be the largest source of noise from the measured covariance matrix
while retaining most of the parameter dependence.

Furthermore, we note that the inverse covariance matrix, even for a
measured matrix with good signal to noise, may be extremely noisy.  As
such, we define the reduced response matrix as
\begin{align} \widehat{\Rmat}_{\alpha,\mathrm{red}} 
  & \equiv 
  \widehat{\frac{\partial\Dmatt}{\partial \theta_{\alpha}}} \widehat{\rmatt}\widehat{ \Dmatt}
   +\widehat{\Dmatt}\widehat{\frac{\partial\rmatt}{\partial\theta_\alpha} }\widehat{ \Dmatt}
   +\left(\widehat{\frac{\partial\Dmatt}{\partial \theta_{\alpha}}} \widehat{\rmatt}\widehat{ \Dmatt}\right)^T
      - \frac{\partial\Covm}{\partial\theta_{\alpha}}\Covm^{-1} \widehat{\Cov_{\rm t}} \ ,
      \label{eq:redR}
\end{align}
and its constant correlation function approximation as
\begin{align} \widehat{\Rmat}^{\rm approx}_{\alpha,\mathrm{red}}
  & \equiv
    \widehat{\frac{\partial\Dmatt}{\partial \alpha}} \widehat{\rmatt}\widehat{ \Dmatt}
    +\left(\widehat{\frac{\partial\Dmatt}{\partial \alpha}} \widehat{\rmatt}\widehat{ \Dmatt}\right)^T
    - \frac{\partial\Covm}{\partial\theta_{\alpha}}\Covm^{-1} \widehat{\Cov_{\rm t}}
 \ ,
 \label{eq:redaR}
\end{align}
in order to study the response matrix in the context of a
comparatively noiseless environment.


\begin{figure*}
\centering{
  \includegraphics[width=8cm,clip=]{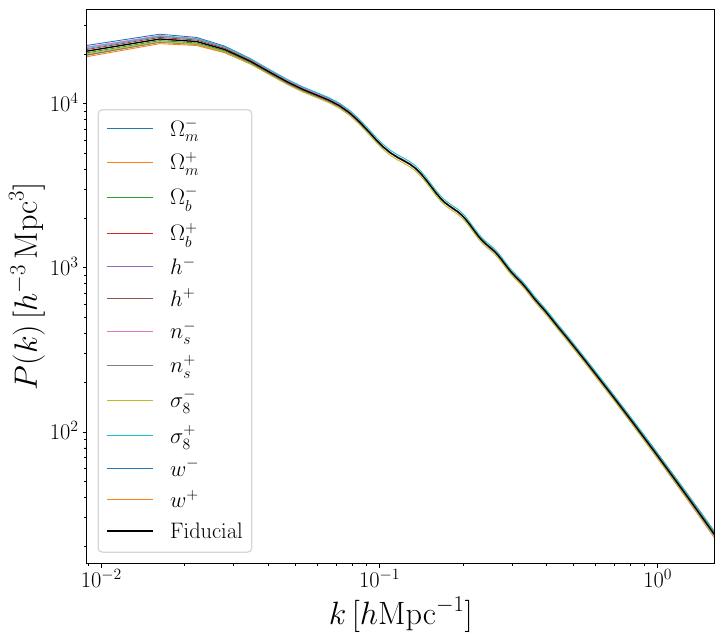}\hspace{0.3cm}
  \includegraphics[width=8cm,clip=]{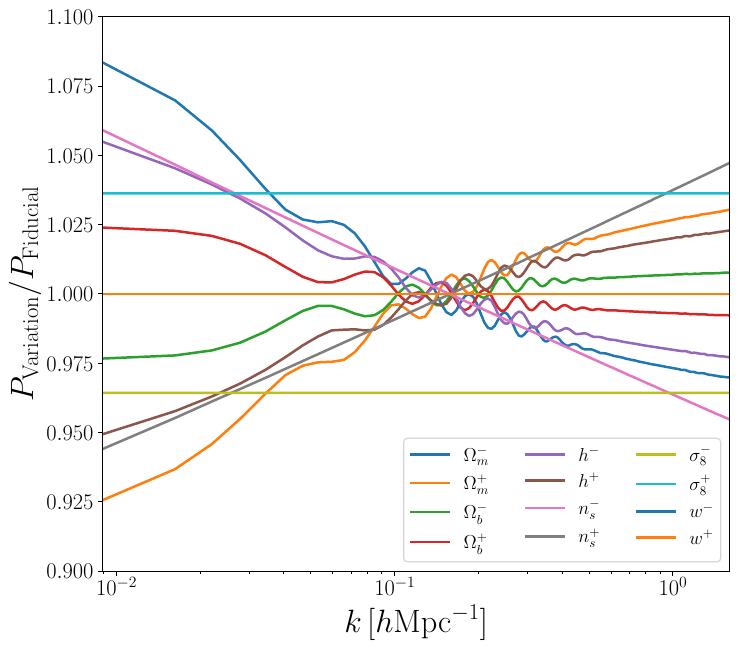}}
\caption{\small{Linear theory power spectra for the Quijote
    Simulations. Left panel: the power spectra for the fiducial model
    and the 12 variations in cosmology at $z=0$. Right panel: the
   ratio of the power spectra from each variation with the fiducial
    spectrum. }
  \label{fig:PkLinear} }
\end{figure*}


\section{The Quijote simulations}\label{sec:simulations}

\subsection{Overview}

In this work we make use of the publicly available
results from the Quijote suite of $N$-body simulations
\citep{Villaescusa-Navarroetal2020}\footnote{\tt
https://Quijote-simulations.readthedocs.io/en/latest/}. These
cosmological simulations have been designed to explore the statistical
properties of various large-scale structure spatial statistics through
averages over a vast ensemble of realisations with $\Lambda$CDM cosmology.

Each simulation follows the evolution
under gravity of $N=512^3$ matter particles in a cubic, periodic
volume, where the length of each side is $L=1\Gpc$, giving a comoving
volume of $V=1\Gpc^3$. For this work we make use of the gravity only
suite of runs. Each of these was simulated using the Tree-PM code {\tt
  Gadget-III}, which is an optimised version of {\tt Gadget-II}
\citep{Springel2005}. 

The
initial over-density fluctuations in each simulation are set to follow
a Gaussian random field, with mean zero and variance set by the power
spectrum. The initial linear power spectrum was computed using the
Einstein-Boltzmann solver {\tt CAMB} \citep{Lewisetal2000}, where the
linear theory matter power spectrum was evolved down to $z=0$. This
was then rescaled back to the initial start redshift of $z=127$ for
which,
\be P_{\rm L}(k,z) = \left(\frac{D(z)}{D(z=0)}\right)^2 P_{\rm L}(k,z=0) \ ,\ee
where $D(z)$ is the linear theory growth factor extrapolated to
redshift $z$. With the Gaussian random field computed, the particles
are evolved to their initial particle positions are evolved further
into the light nonlinear regime the {\tt 2LPT} code
\citep{Scoccimarro1998}, to suppress transients. For further details
about each simulation refer the reader to
\citep{Villaescusa-Navarroetal2020}.


\subsection{Ensembles and parameters}

The parameters used in the
fiducial model are those of standard $\Lambda$CDM
cosmology with massless neutrinos.  Each of the other 12 cosmologies
takes one of the parameters of the fiducial model and varies it by a small step. We make
use of the variations in the parameters $\{\Omega_{\rm m}, \Omega_{\rm
  b}, n_{\rm s},\sigma_{8}, h,w_{0}\}$.  The values taken by these parameters are given in App.~\ref{app:sims}.  For the fiducial model there are 15,000
realisations and for the each variation there are 500 realisation.

The parameters that were used to define the cosmological
model are those that are most closely derived from late-time
observations.  However, for some of them, it is less transparent how
they connect back to the underlying physics.  Other equivalent
parameterisations of the cosmological model connect back more directly
to primordial physics, but are considered less intuitive. In
particular, a more physical parameterisation is based upon the
physical densities of cold dark matter, $\omega_{\rm cdm}\equiv(\Omega_{\rm
  m}-\Omega_{\rm b})h^2$, and baryons, $\omega_{\rm b}\equiv\Omega_{\rm
  b}h^2$, and the amplitude of the primordial power spectrum of scalars,
$A_{\rm s}$. These can all be derived from the original set of
parameters. 

The left-hand panel of
Fig.~\ref{fig:PkLinear} shows the linear theory spectra for the
fiducial model and the variations. The right panel shows the ratio of
the variations with respect to the fiducial model.



\begin{figure}
  \centering{
    \includegraphics[width=8cm,clip=]{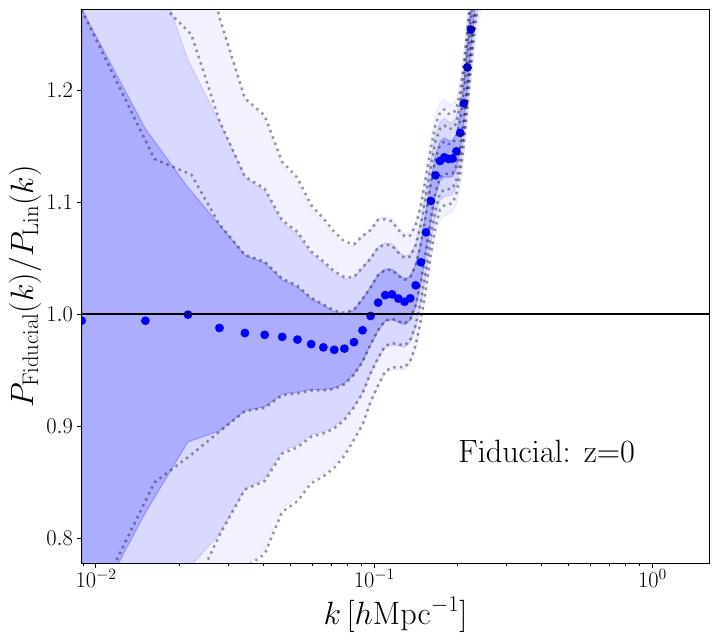}}
  \caption{\small{Ratio of the power spectra at $z=0$ from the fiducial
      runs of the Quijote simulations and the linear
      theory predictions from CAMB. The points show the ensemble
      average ratio. The three shaded blue regions show the
      1-$\sigma$, 2-$\sigma$ and 3-$\sigma$ confidence regions. The
      dotted line shows the simple Gaussian predictions for the
      errors.  }\label{fig:Pk}}
\end{figure}


\begin{figure*}
  \centering{
    \includegraphics[width=16cm,clip=]{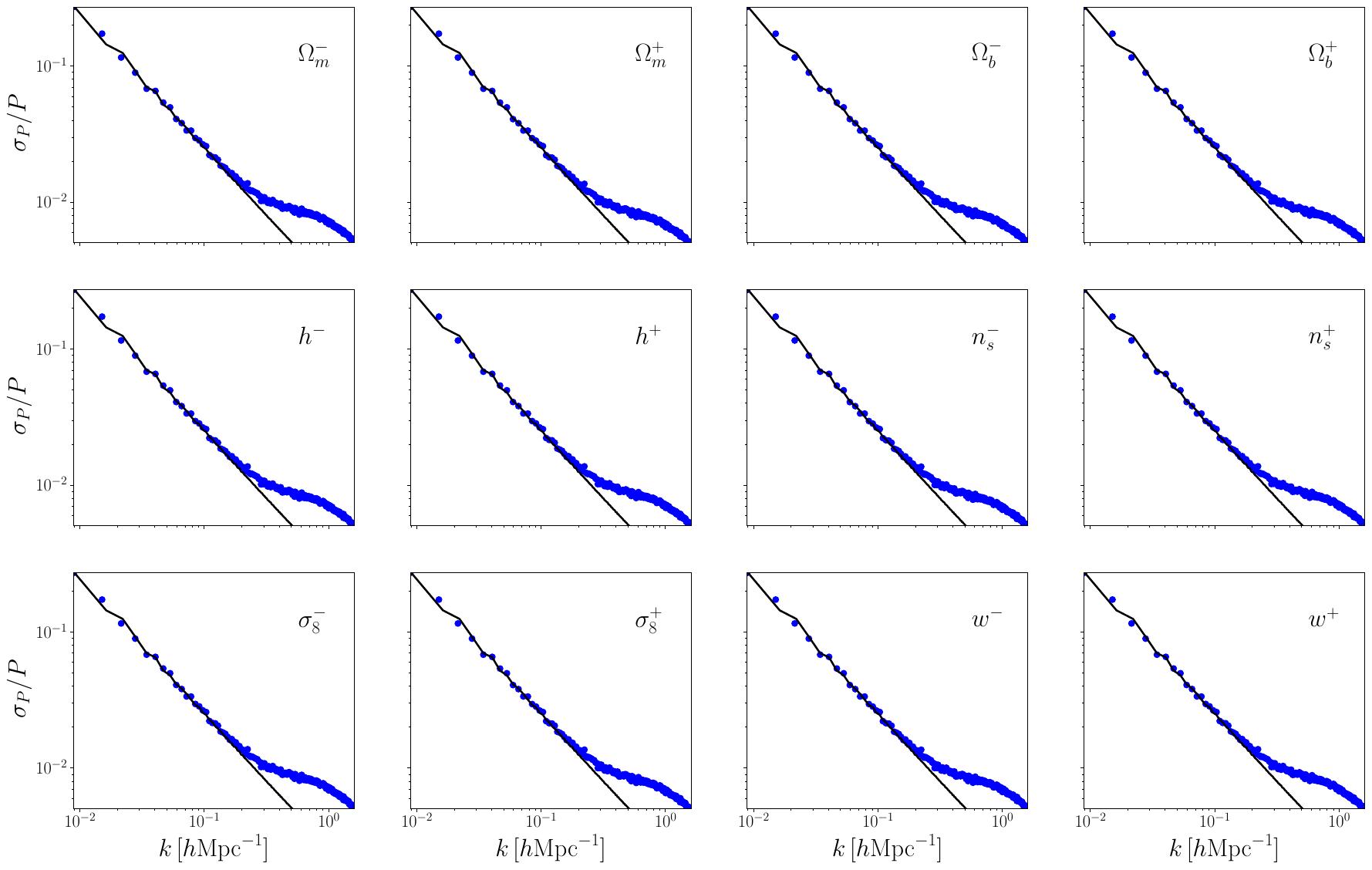}}
  \caption{\small{Ratio of the square root of the diagonal component
      of the covariance matrix to the power spectra
      ($\sigma_P/P\equiv\sqrt{C_{ii}}/P_i$) for the 12 variations in
      cosmology from the Quijote simulations as a function of scale.
      The points show the ensemble average ratio. The black solid line
      shows the predictions from the Gaussian
      formula given by \Eqn{eq:GaussShell}.}\label{fig:Pk3}}
\end{figure*}


\subsection{Ensemble average power spectra}

In this work we make use of the publicly available power spectra,
computed with the {\tt Pylians} code \citep{Pylians}. All of the
algorithmic details can be found in that work. We make one note
concerning the code and that is that the returned spectra are not at
the geometrical bin centres of the $k$-space shells, but at the mean
of the $k$-modes within those shell. The $k$-space shells are
spaced by the fundamental mode of the simulation cube \mbox{$k_{\rm
  fun}=2\pi/L\approx6.3\times10^{-3}\kMpc$}. When comparing linear
theory expectations with the expected signal from the Quijote Fourier
lattice expectation, we find that there is a small $\lesssim1.5\%$
impact on scales $k\lesssim3\times10^{-2}\kMpc$. The deviation is an
oscillating feature that impacts the largest scales of the spectra.
We remedy this by computing linear theory expectations using only the
exact Fourier modes that go into the band-averaged from {\tt Pylians}.

Figure~\ref{fig:Pk} shows the ensemble average of the matter power
spectra measured from the 15,000 runs of the fiducial cosmological
simulations, ratioed with respect to the linear theory. This shows
that the spectra on the largest scales match linear theory predictions
to high accuracy. The nonlinear evolution of the growth of structure
drives the collapse of structures on quasi-linear and small scales and
this explains the sharp upturn away from linear theory on scales
$k>0.1\kMpc$.  However, on the largest scales we also note the
suppression of power at the level of a few percent. This is known as
the pre-virialisation feature, where power is lost from large scales
due to nonlinear mode-coupling effects \citep{Smithetal2007}.


\begin{figure*}
  \centering{
    \includegraphics[width=8cm,clip=]{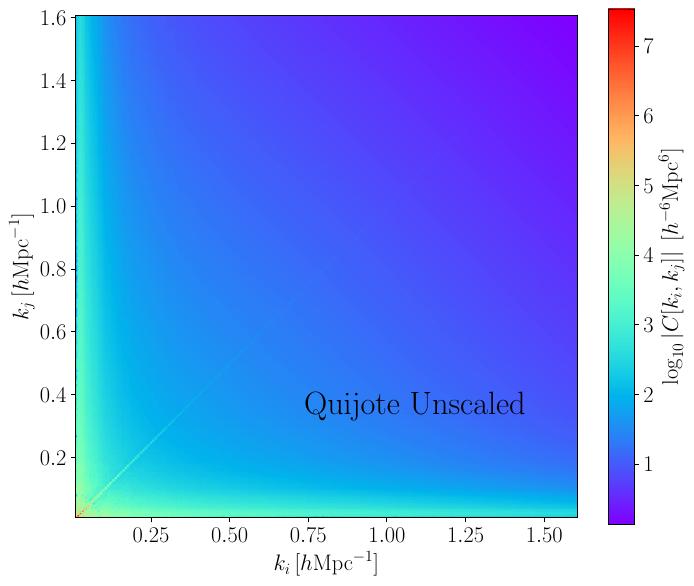}
    \includegraphics[width=8cm,clip=]{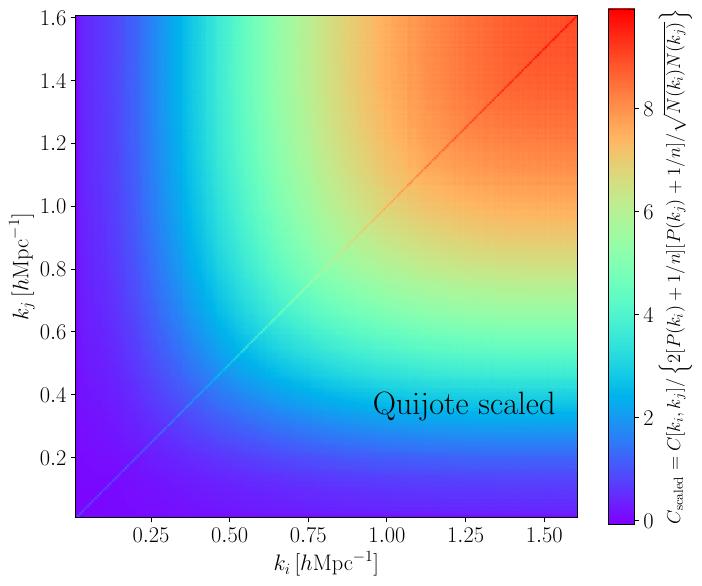}
    \caption{\small{{\bf Left panel}: Estimate of the true covariance
        matrix $\hat{\Covt}$ derived from the ensemble of 15,000
        fiducial Quijote cosmological simulations. {\bf Right panel}:
        shows the estimate of the reduced covariance matrix
        $\hat{\Cov}^{\rm R}$, which we obtain as described by
        \Eqn{eq:CovReduced}. Note that we use the mean band averaged 
        nonlinear power spectrum to construct the
        denominator. }\label{fig:cov}}}
\end{figure*}


\begin{figure*}
  \centering{
    \includegraphics[width=17cm,clip=]{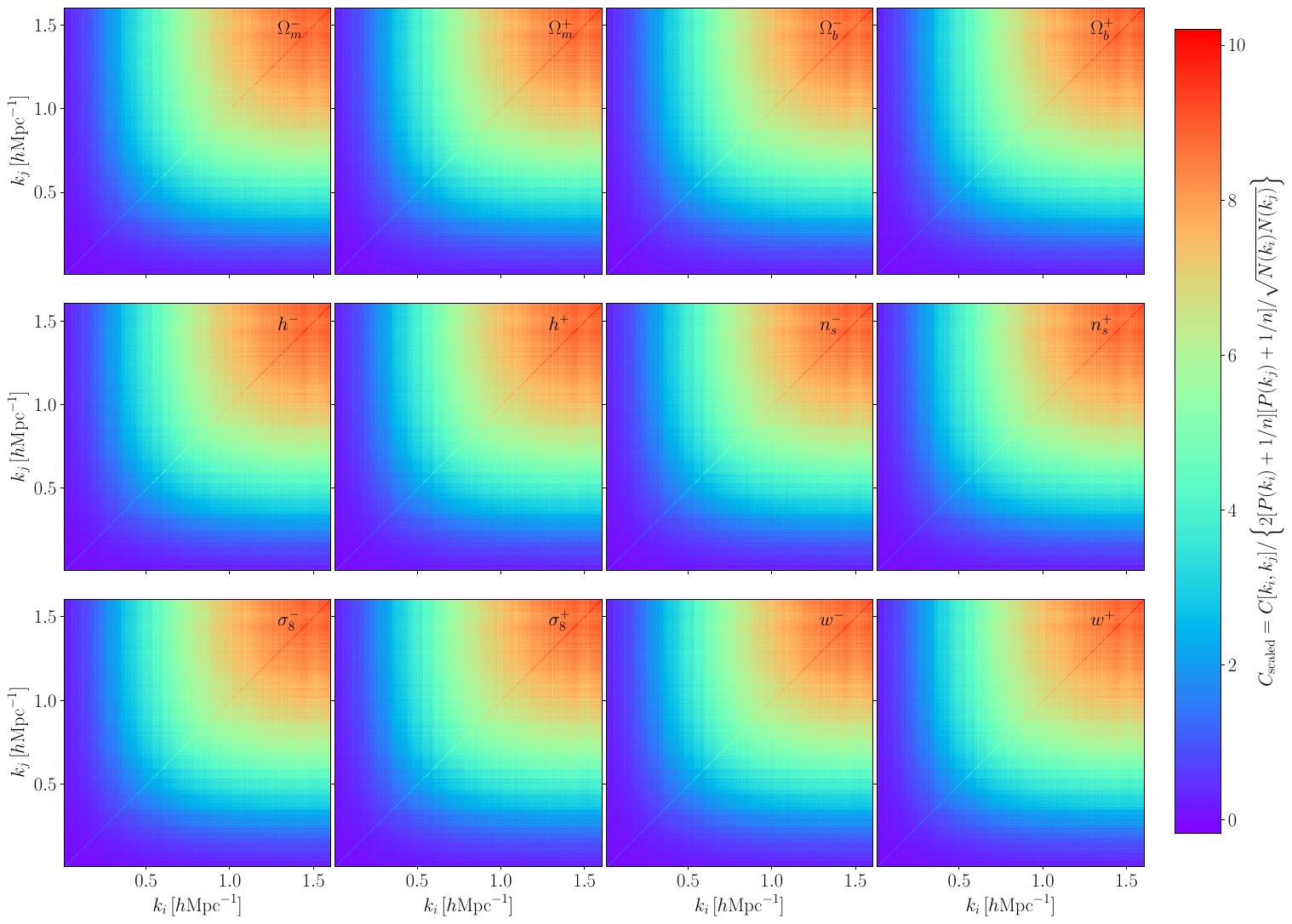} 
    \caption{\small{The reduced covariance matrices
        $\Cov^{\mathrm{R}}$ of the 12 variations in the cosmological
        model as a function of the wavenumbers $k_i$ and $k_j$.}\label{fig:CovScaledVar}}}
\end{figure*}



\subsection{Errors, covariance matrix and the reduced covariance matrix}

We estimate the covariance matrix of the matter power spectrum from
the ensemble using the standard unbiased covariance estimator:
\be \widehat{C}_{ij} = \frac{1}{N_{\rm ensemb}-1}\sum_{k=1}^{N_{\rm ensemb}}
\left[\hat{P}^{(k)}_{i}-\overline{P}_i^{(k)}\right]
\left[\hat{P}^{(k)}_{j}-\overline{P}_j^{(k)}\right] \label{eq:covest}
\ee
where $\hat{P}^{(k)}_{i}$ is the estimate of the power spectrum from
the $k$th cosmological simulation on the scale $k_i$, $N_{\rm ensemb}$
is the number of independent realisations, and where the standard mean
is given by
\be \overline{P}_i^{(k)} =
\frac{1}{N_{\rm ensemb}}\sum_{k=1}^{N_{\rm ensemb}}
\hat{P}^{(k)}_{i}\ .
\ee

In \Fig{fig:Pk} the blue shaded regions show the diagonal
errors on the power obtained by taking the square-root of the diagonal
components of $\widehat{C}_{ij}$. \Fig{fig:Pk3} shows this
information in more detail via the ratio of
$\sigma=\sqrt{\widehat{C}_{ij}}$ with respect to the mean power,
averaged over the ensemble, also known as the noise-to-signal ratio. We see
that this decreases in line with the Gaussian linear theory
predictions, but that at $k=0.2\kMpc$ it begins to saturate until
around $k\approx1\kMpc$, at which point it begins to slowly increase
again. The Gaussian theory predictions show that this ratio decreases
in proportion to the square-root of the number of independent Fourier
modes in the shell and that is when we use either the linear theory
spectrum or the measured non-linear theory spectrum to make predictions.

In the left panel of \Fig{fig:cov} we show the logarithm of the
absolute value of the covariance matrix $\log_{10}|\Cov|$, for the
fiducial simulations estimated from the ensemble using
\Eqn{eq:covest}. We see that this matrix varies by almost 7 orders of
magnitude, with the largest amplitudes being on the
largest scales. This is not the most informative way of exploring the
structure of the matrix.

In the right hand panel of \Fig{fig:cov} we show the reduced covariance
matrix $\Cov^{\mathrm{R}}$ as defined in \Eqn{eq:CovReduced}. This
shows that the band powers are uncorrelated on very large scales, but
that large-scale band powers couple strongly to small scale band
powers and that small scale modes are highly correlated. The red
colour shows that for $k_i\sim k_j\sim1\kMpc$ the variance is almost a
factor of 10 larger than the Gaussian predictions. In this case, this
is entirely driven by the non-Gaussian contributions from the
connected trispectrum.

In \Fig{fig:CovScaledVar} we present the reduced covariance matrices
of the 12 variations in cosmology presented in App.~\ref{app:sims}. We see
that these variations show a very similar structure to those of the
fiducial cosmological model, that is on large scales the band powers
are uncorrelated, but that on smaller scales the quasi-linear scales
and small-scale modes become highly correlated and that the Gaussian
approximation is a poor description of the covariances.


\begin{figure*}
\centering
    \begin{minipage}{0.48\textwidth}
      \includegraphics[width=\textwidth]{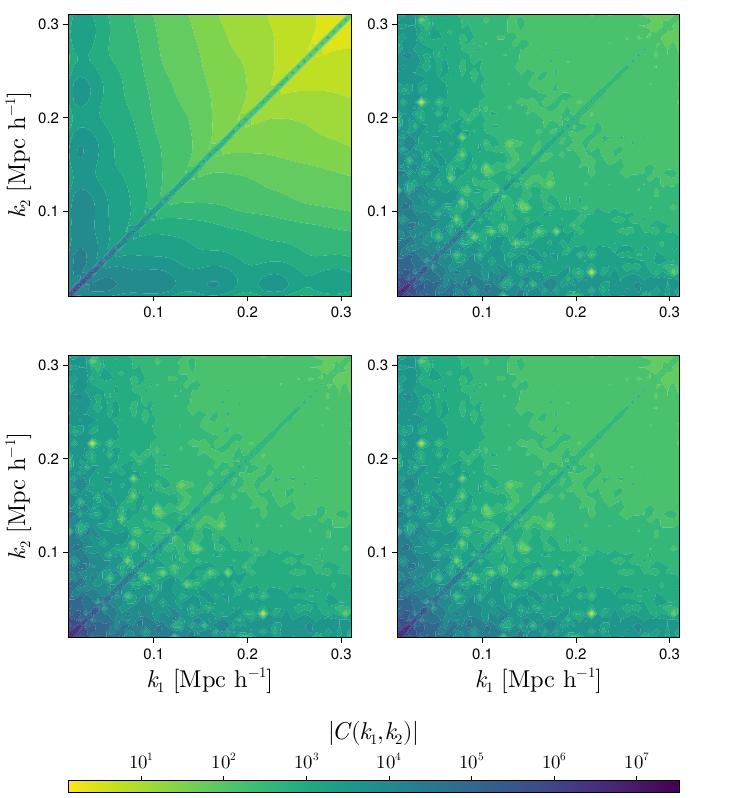}
    \end{minipage}
    \begin{minipage}{0.48\textwidth}
    \includegraphics[width=\textwidth]{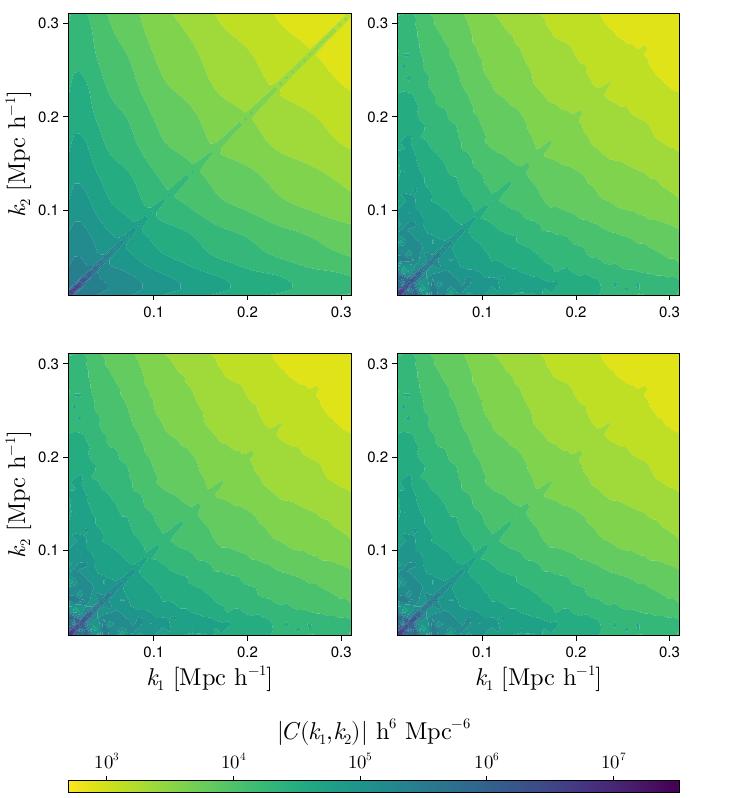}
    \end{minipage}
    \caption{The covariance matrices for the fiducial cosmology.  The
      left hand pair of columns omit SSC corrections, while the right
      hand pair of columns include them.  In each pair of columns, the
      top left panel shows the SPT model covariance matrix, the top
      right panel shows the non-linear covariance matrix from the
      simulations, the bottom left panel shows the reconstructed
      covariance matrix from Eq.~\eqref{eq:Cparallel} with the full
      response matrix as given in Eq.~\eqref{eq:fullR}, and the bottom
      right panel shows the reconstructed covariance matrix with the
      approximated response matrix as given in Eq.~\eqref{eq:aR}.}
    \label{fig:Cfid}
\end{figure*}


\begin{figure*}
  \centering{
    \includegraphics[width=0.3\textwidth]{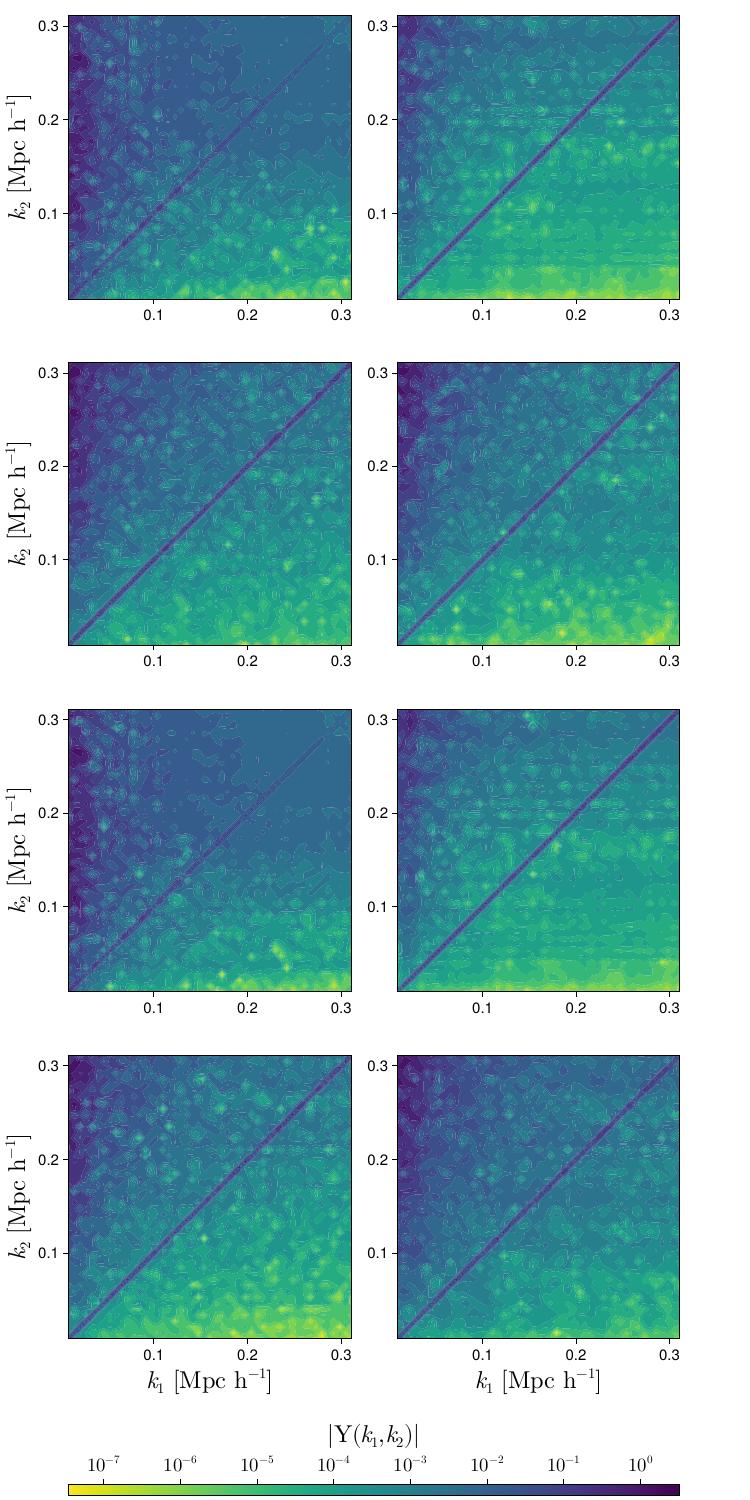}\hspace{0.4cm}
    \includegraphics[width=0.3\textwidth]{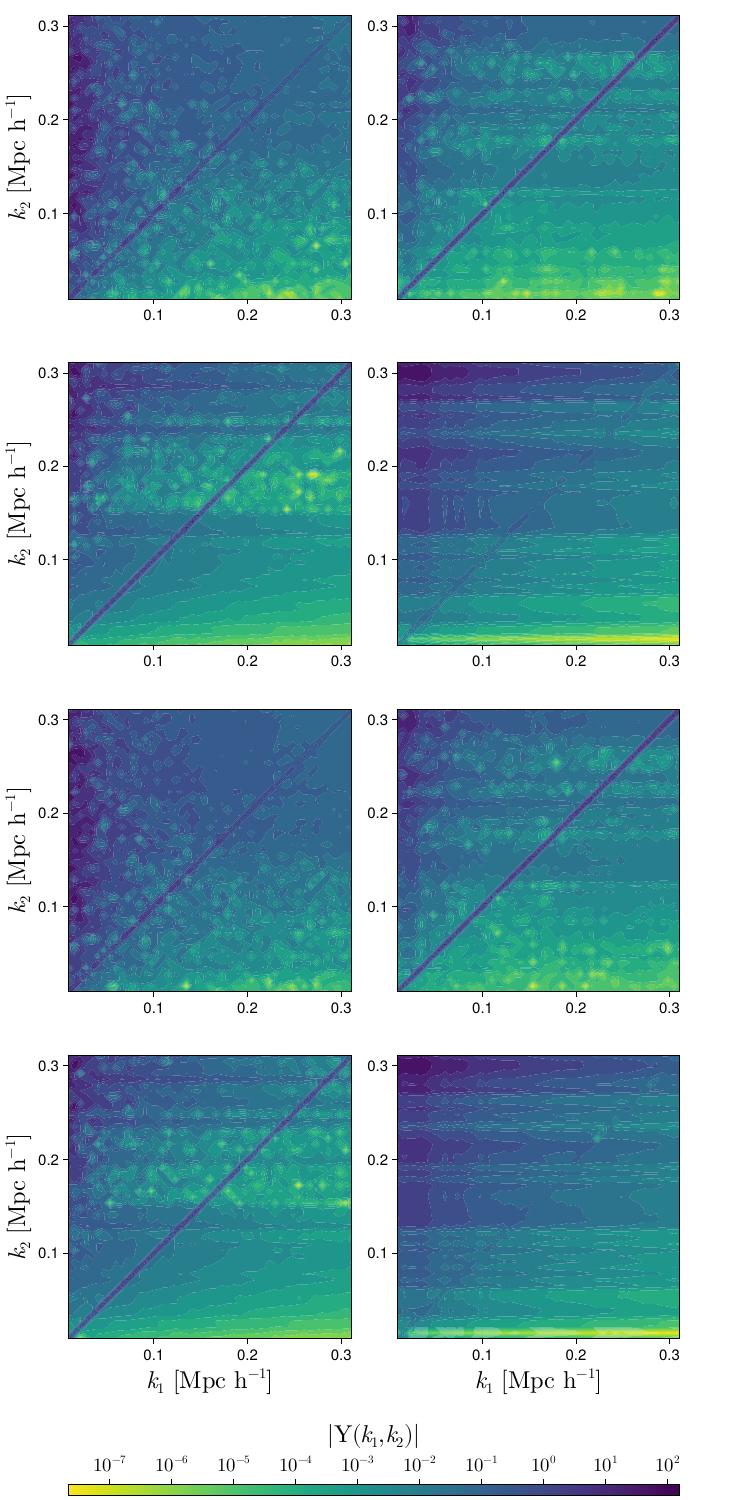}}
    \caption{The figures show our modelling of the matrix
      $\widetilde{\Ymat}$ for the variations in $\Omega_{\mathrm{m}}$.
      The top two rows show results for $\Omega_{\rm m}^+$ and the
      bottom two rows for $\Omega_{\rm m}^-$, respectively.  The left
      hand pair of columns show the results without SSC corrections
      and the right hand pair of columns include SSC corrections.  The
      top left panel shows the $\widetilde{\Ymat}$ for the SPT model,
      the top right panel shows $\widetilde{\Ymat}$ with the full
      response matrix $C_{\parallel}$, the bottom left panel shows
      $\widetilde{\Ymat}$ with the approximate response matrix
      $C_{\parallel,\mathrm{a}}$, and the bottom right panel shows
      $\widetilde{\Ymat}$ with $C_{\parallel,\mathrm{R}=0}$ in which
      the response matrix is set to zero.\label{fig:Y_Om}}
\end{figure*}


\begin{figure*}
  \centering{
    \includegraphics[width=0.3\textwidth]{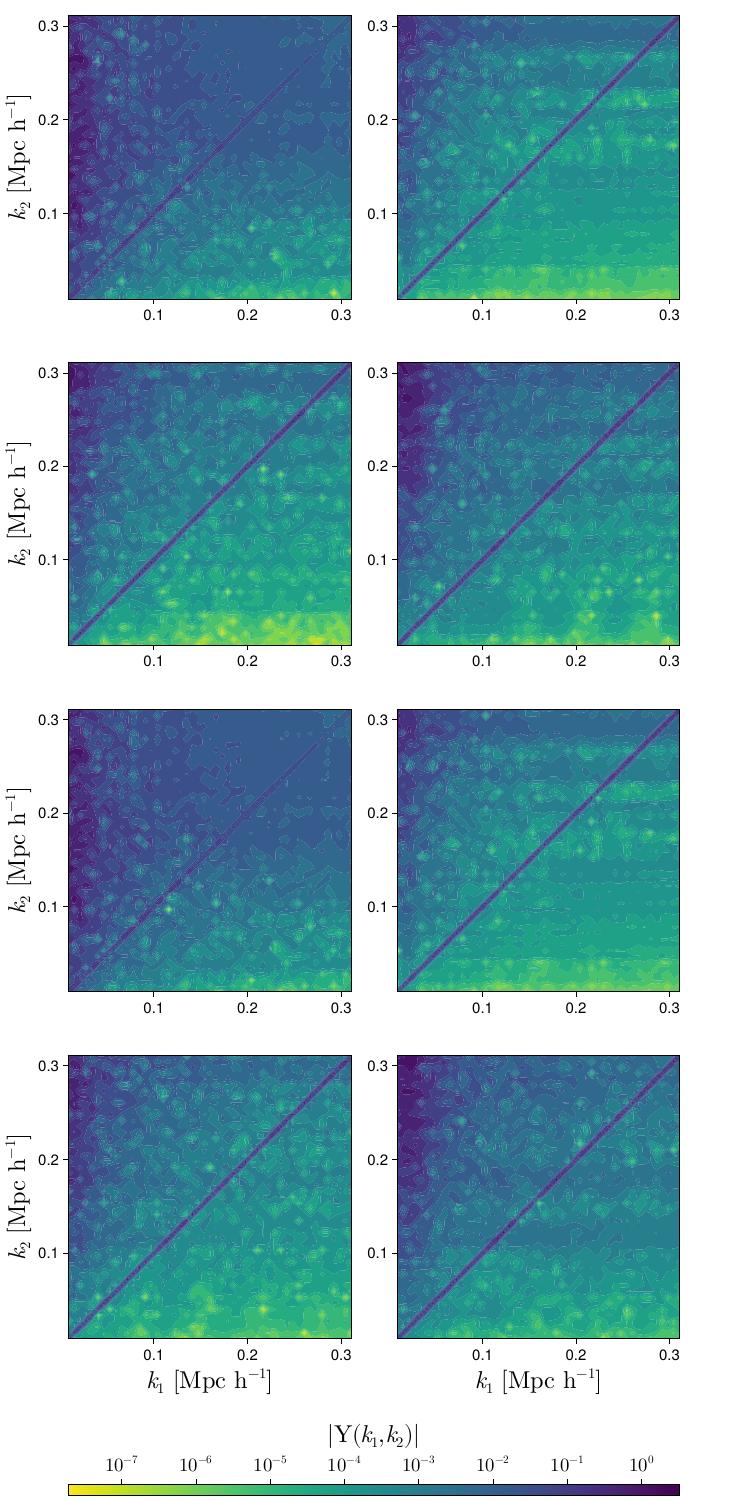}\hspace{0.4cm}
    \includegraphics[width=0.3\textwidth]{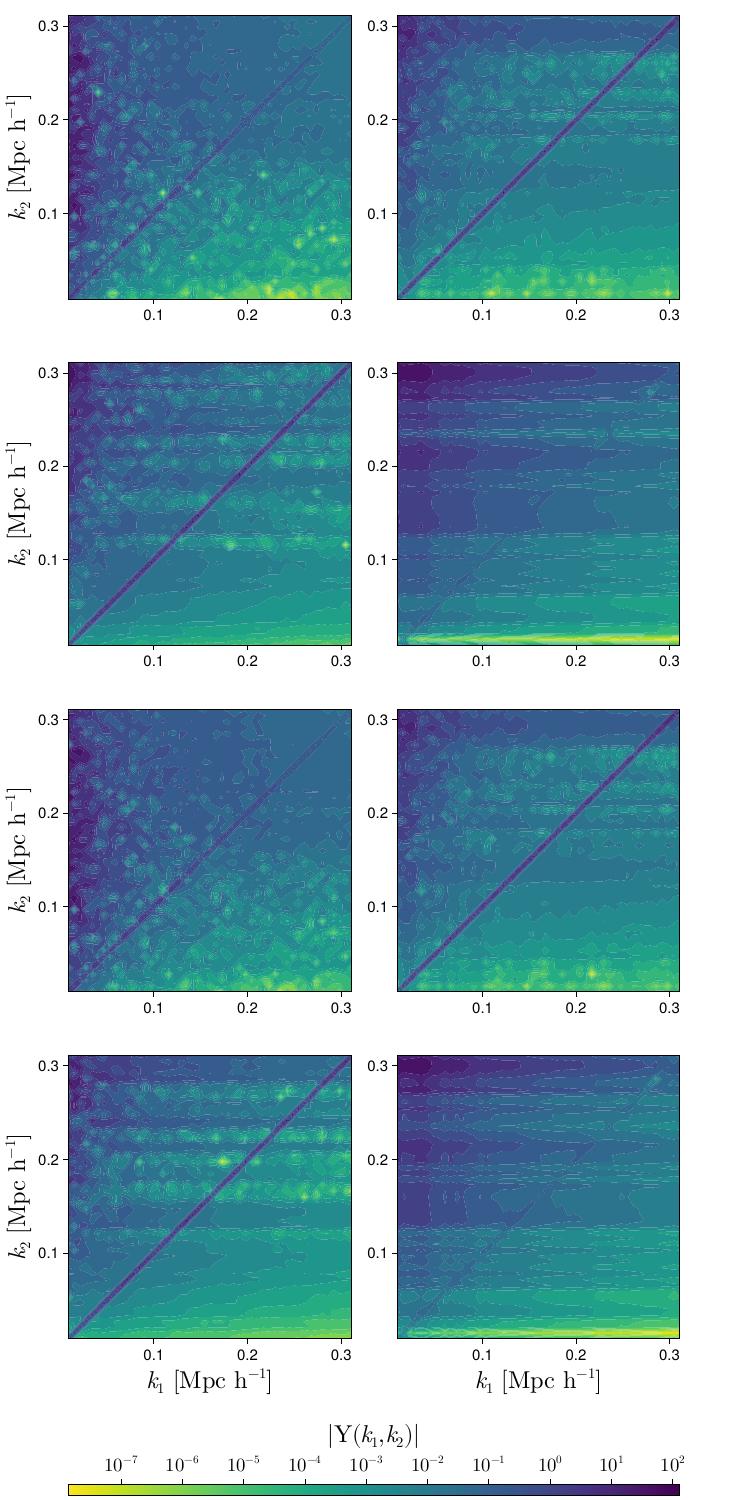}}
  \caption{The figure is the same as \Fig{fig:Y_Om}, but this time
    for $\Omega_{\mathrm{b}}$, where the top two rows show results
    for $\Omega_{\rm b}^+$ and the bottom two rows for $\Omega_{\rm
      b}^-$, respectively.\label{fig:Y_Ob}}
\end{figure*} 


\begin{figure*}
  \centering{
    \includegraphics[width=0.3\textwidth]{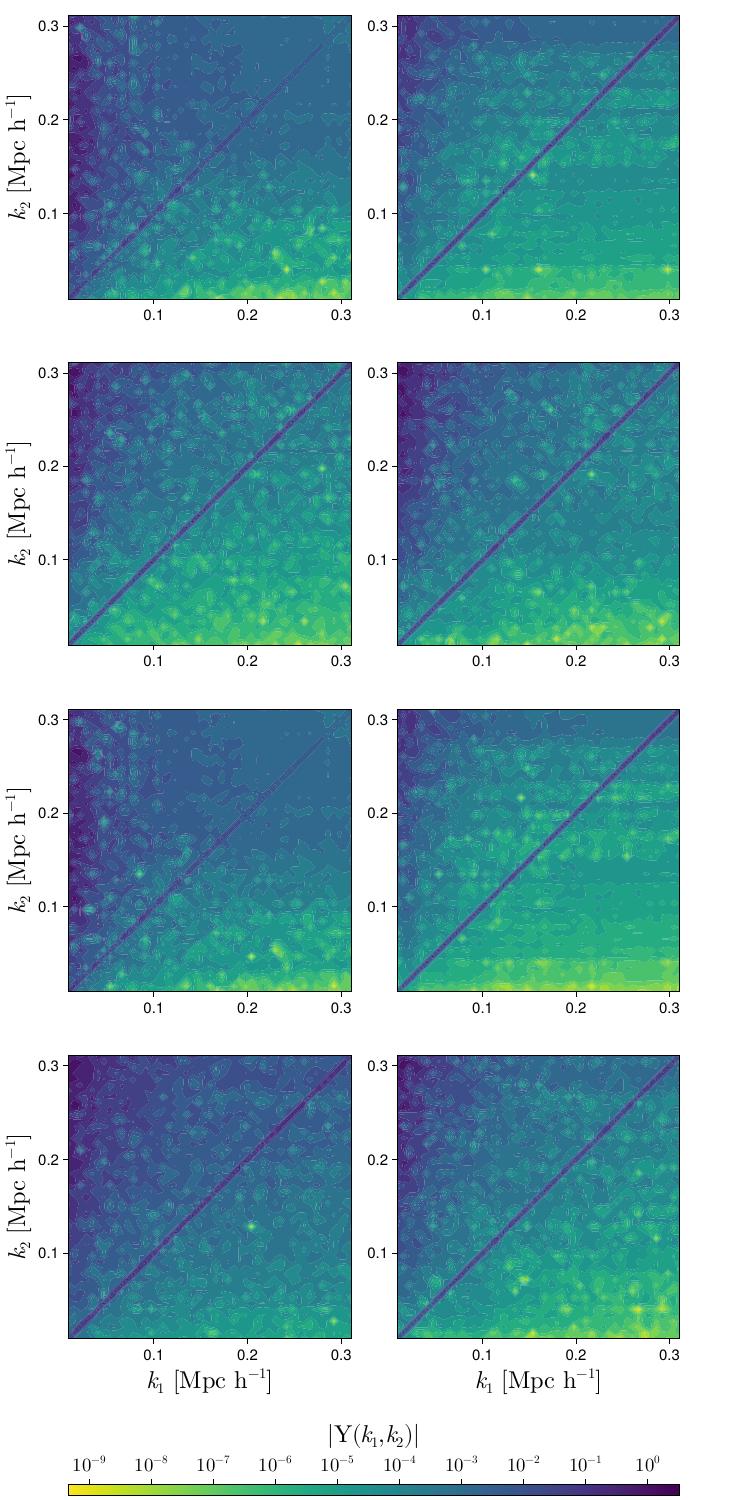}\hspace{0.4cm}
    \includegraphics[width=0.3\textwidth]{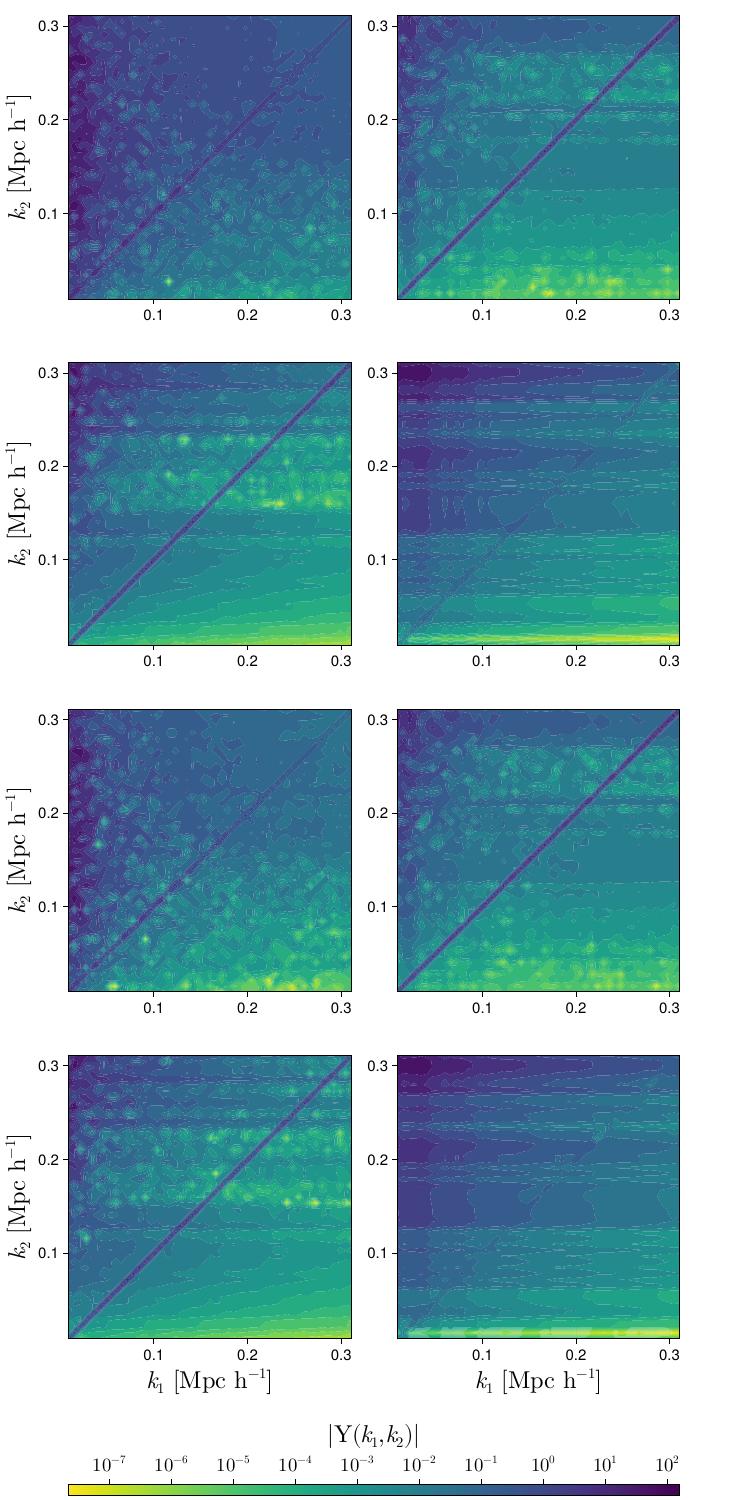}}
  \caption{The figure is the same as \Fig{fig:Y_Om} but this time for
    $n_{\mathrm{s}}$, where the top two rows show results for
    $n_{\mathrm{s}}^+$ and the bottom two rows for $n_{\mathrm{s}}^-$,
    respectively. \label{fig:Y_ns}}
\end{figure*}


\begin{figure*}
  \centering{
    \includegraphics[width=0.3\textwidth]{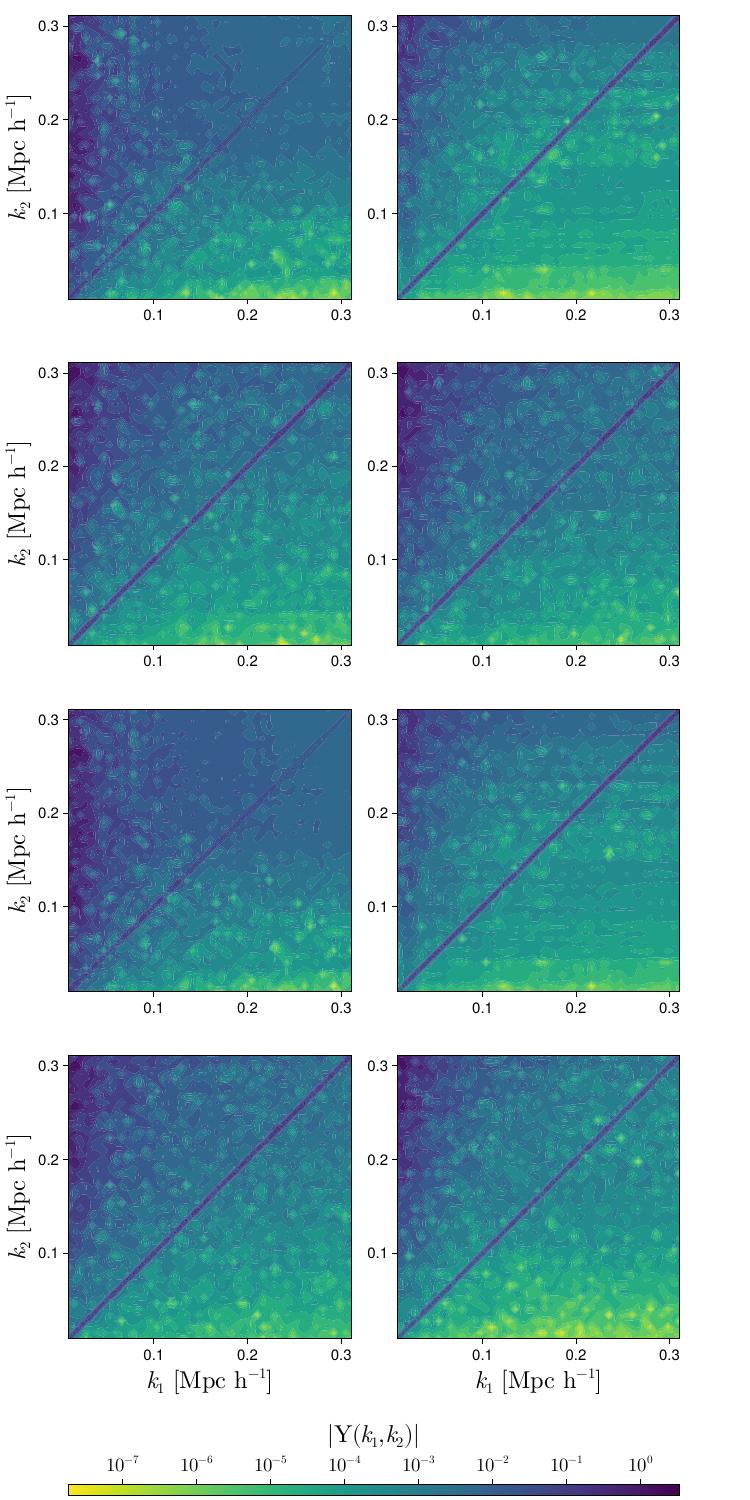}\hspace{0.4cm}
    \includegraphics[width=0.3\textwidth]{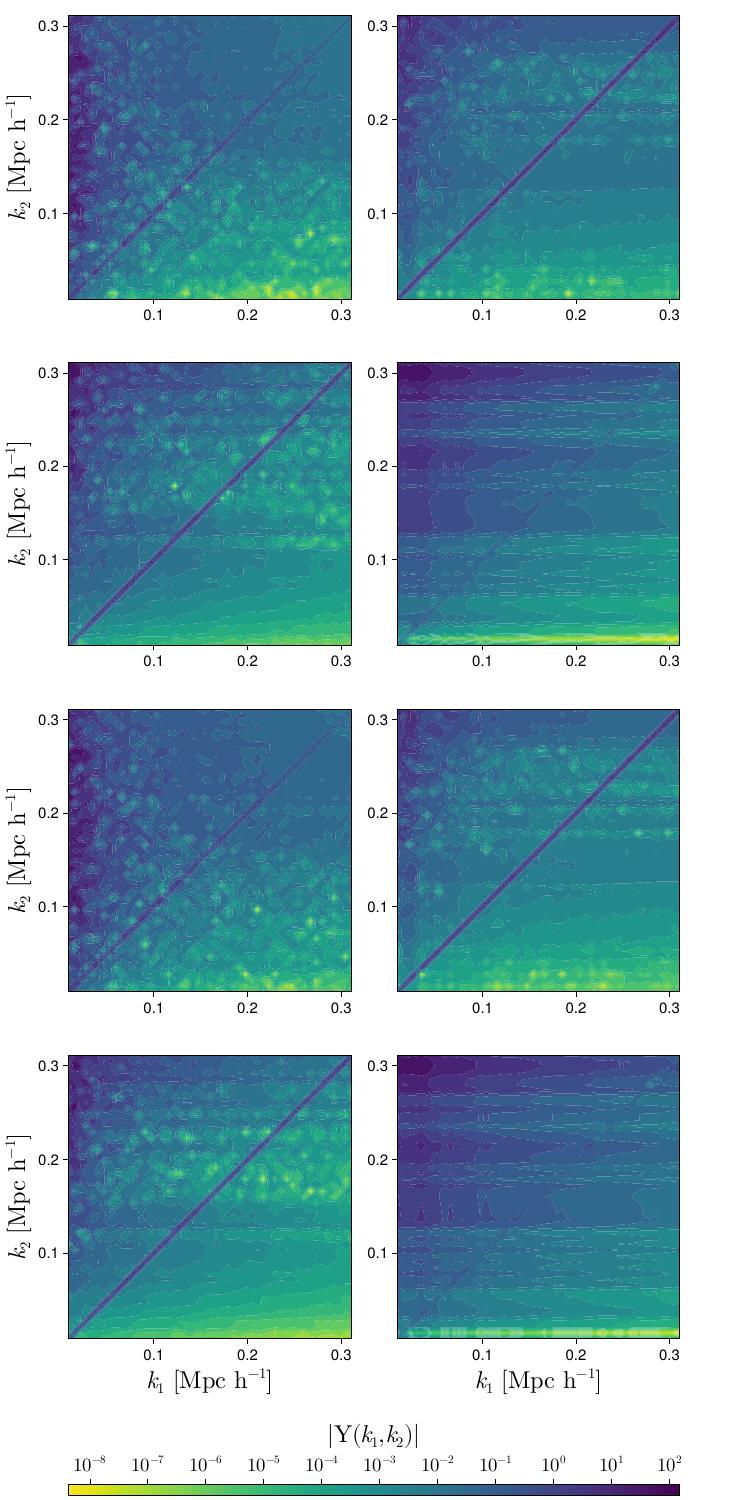}}
  \caption{The figure is the same as \Fig{fig:Y_Om} but this time for
    $\sigma_{8}$, where the top two rows show results for $\sigma_8^+$
    and the bottom two rows for $\sigma_8^-$,
    respectively. \label{fig:Y_s8}}
\end{figure*}


\begin{figure*}
  \centering{
    \includegraphics[width=0.3\textwidth]{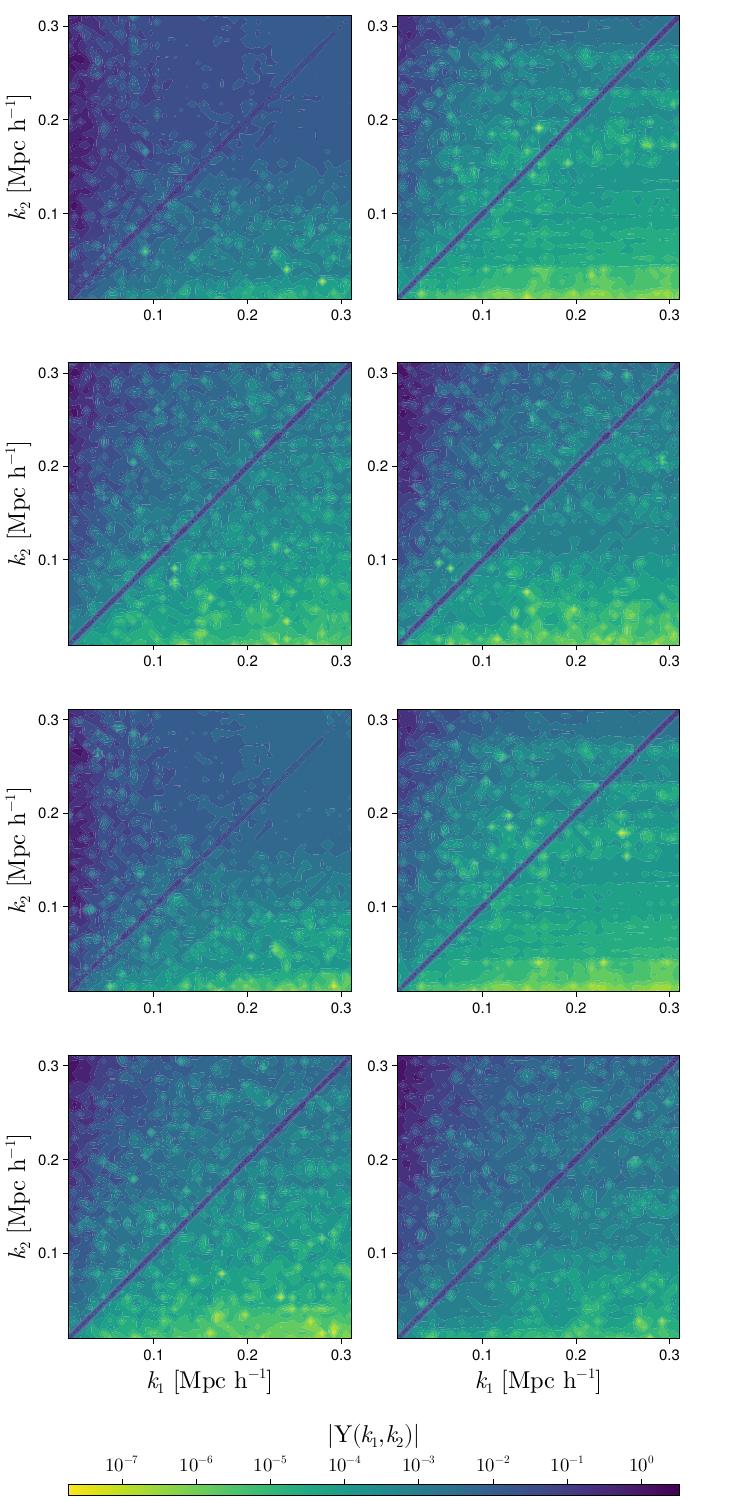}\hspace{0.4cm}
    \includegraphics[width=0.3\textwidth]{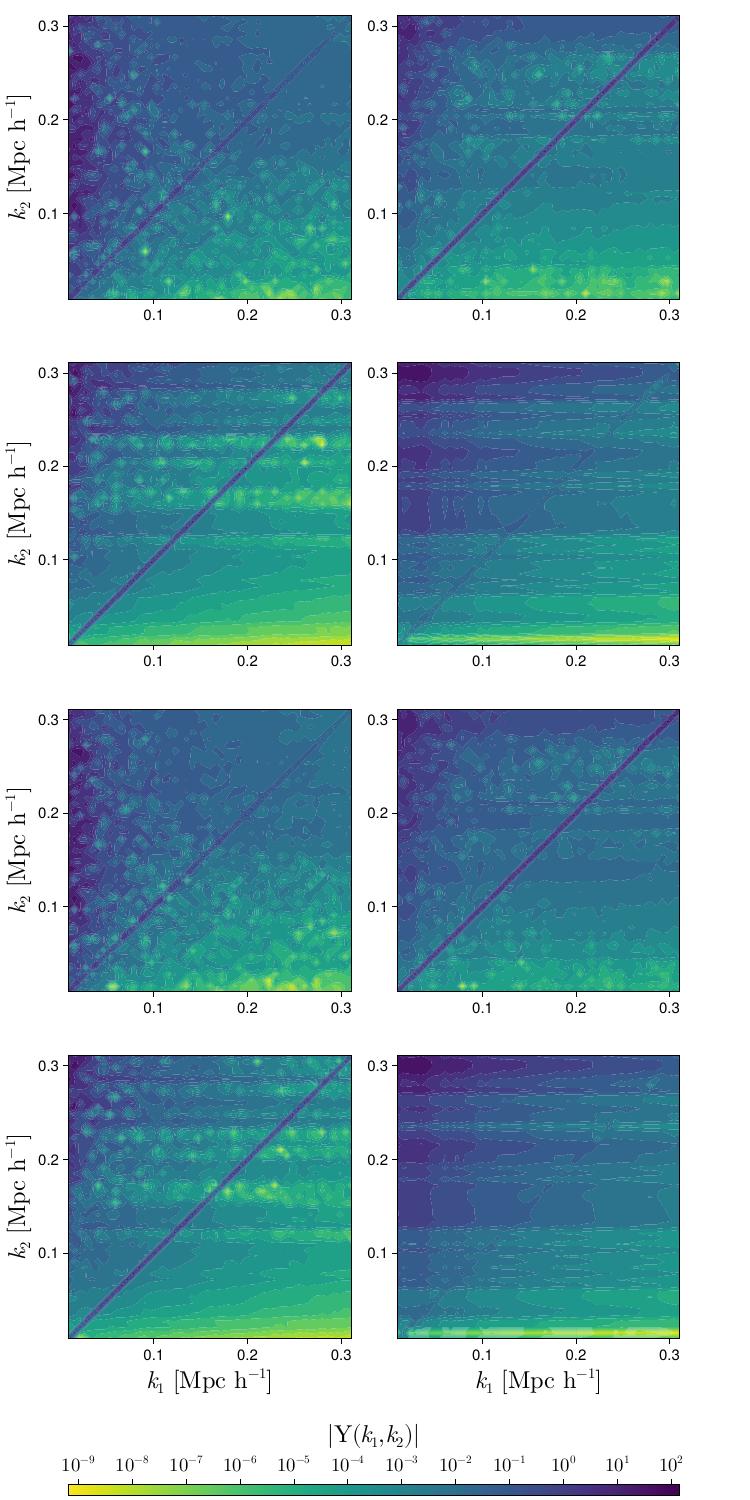}}
  \caption{The figure is the same as \Fig{fig:Y_Om} but this time for
    $h$, where the top two rows show results for $h^+$ and the bottom
    two rows for $h^-$, respectively. \label{fig:Y_h}}
\end{figure*}

\begin{figure*}
  \centering{
    \includegraphics[width=0.3\textwidth]{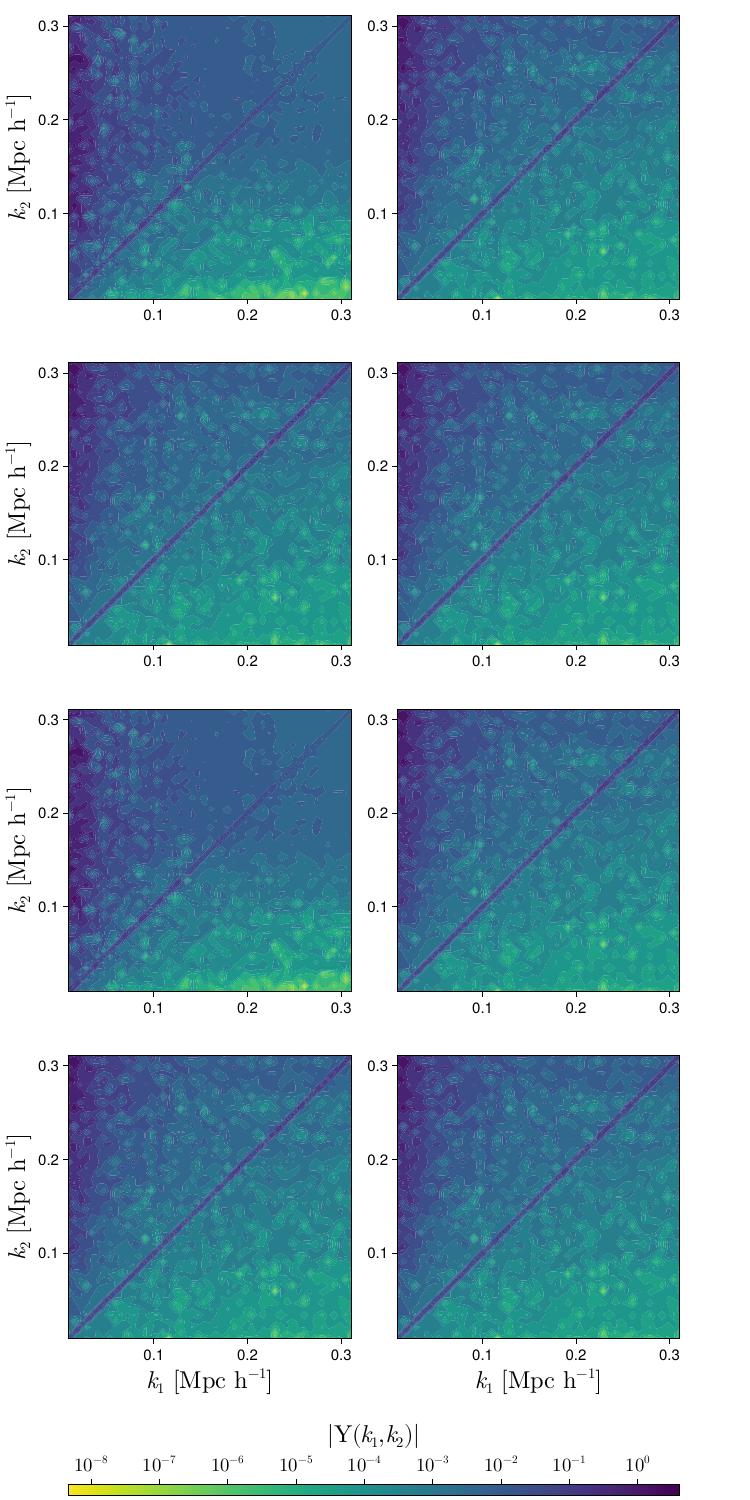}\hspace{0.4cm}
    \includegraphics[width=0.3\textwidth]{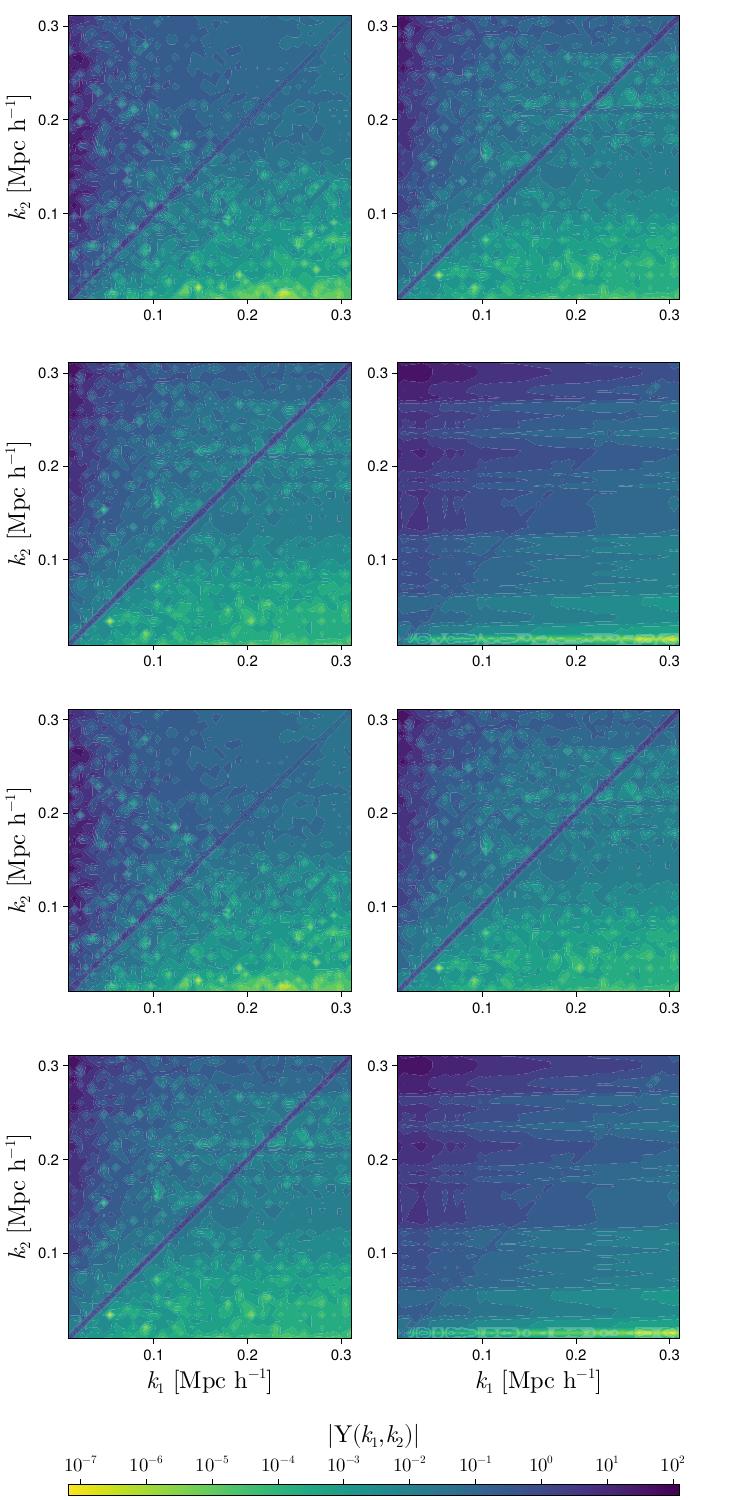}}
  \caption{The figure is the same as \Fig{fig:Y_Om} but this time for
    $w_{0}$, where the top two rows show results for $w_{0}^+$ and the bottom
    two rows for $w_{0}^-$, respectively. \label{fig:Y_h}}
\end{figure*}

\begin{figure*}
\centering
    \begin{minipage}{0.48\textwidth}\includegraphics[width=\textwidth]{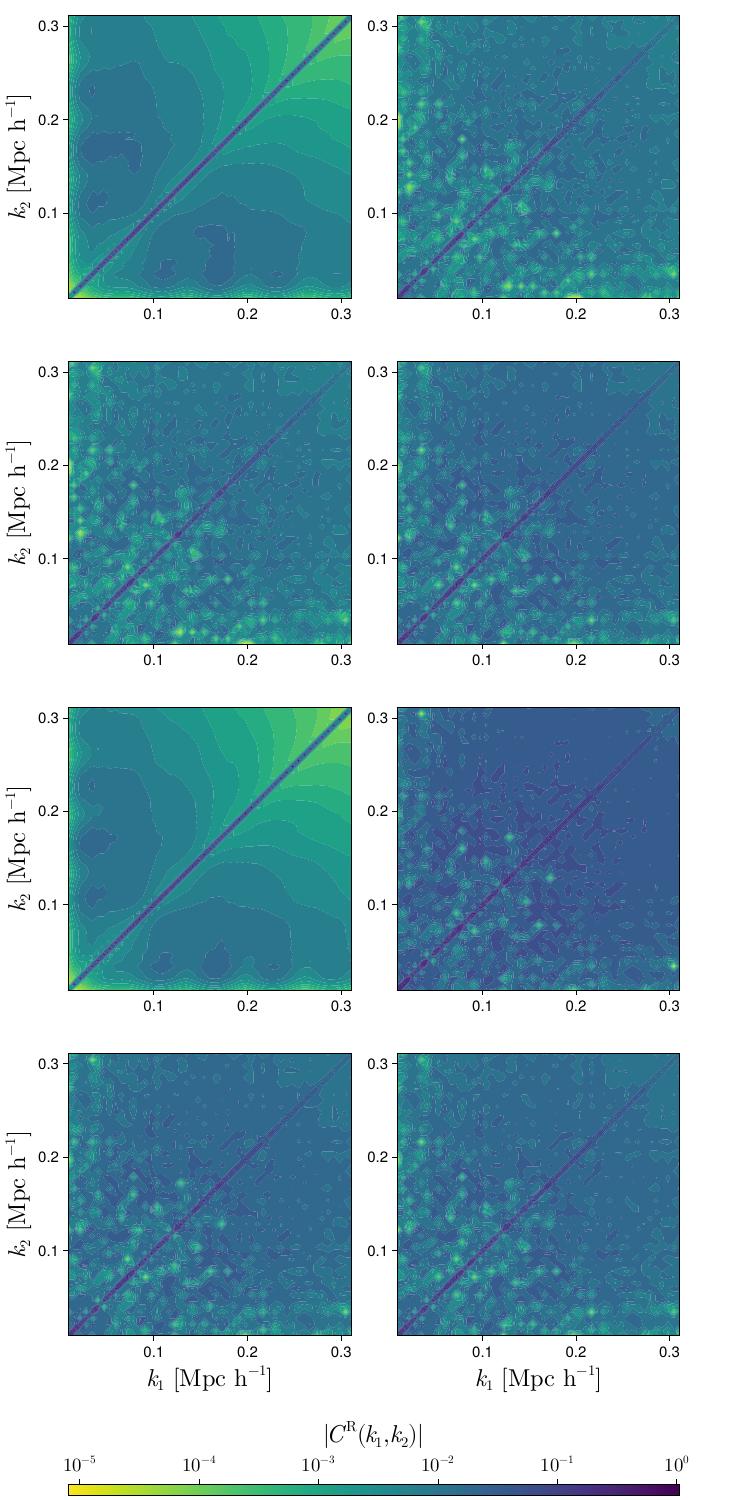}\end{minipage}
    \begin{minipage}{0.48\textwidth}\includegraphics[width=\textwidth]{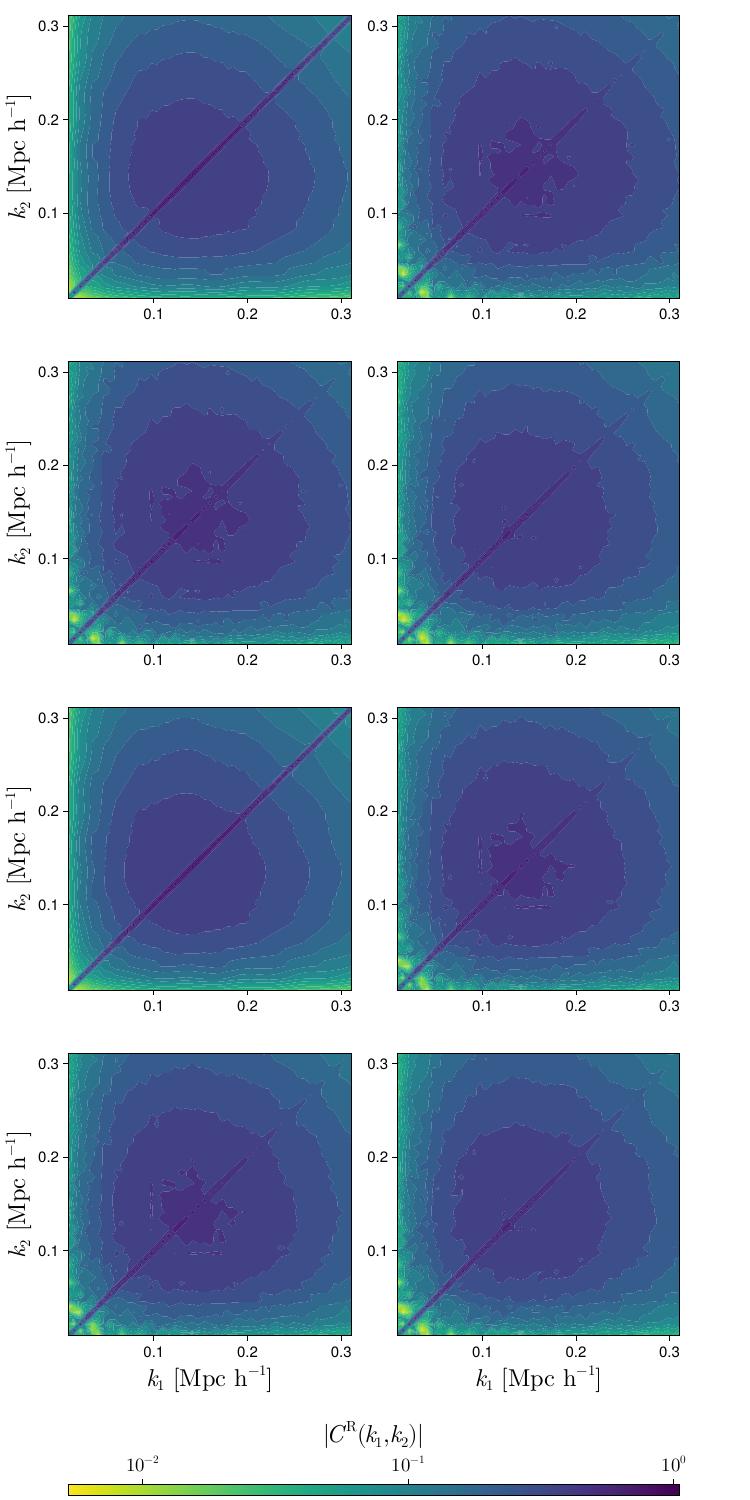}\end{minipage}
    \caption{The reduced covariance matrices for cosmologies with the increased (top two rows) and decreased (bottom two rows) values of the parameter $\Omega_{\mathrm{m}}$ and other parameters kept at their fiducial values.  The left hand pair of columns omit SSC corrections, while the right hand pair of columns include them.  In each pair of columns, the top left panel shows the SPT model covariance matrix, the top right panel shows the non-linear covariance matrix from the simulations, the bottom left panel shows the reconstructed covariance matrix from Eq.~\eqref{eq:Cparallel} with the full response matrix as given in Eq.~\eqref{eq:fullR}, and the bottom right panel shows the reconstructed covariance matrix with the approximated response matrix as given in Eq.~\eqref{eq:aR}.}
    \label{fig:ROm}
\end{figure*}

\begin{figure*}
\centering
    \begin{minipage}{0.48\textwidth}\includegraphics[width=\textwidth]{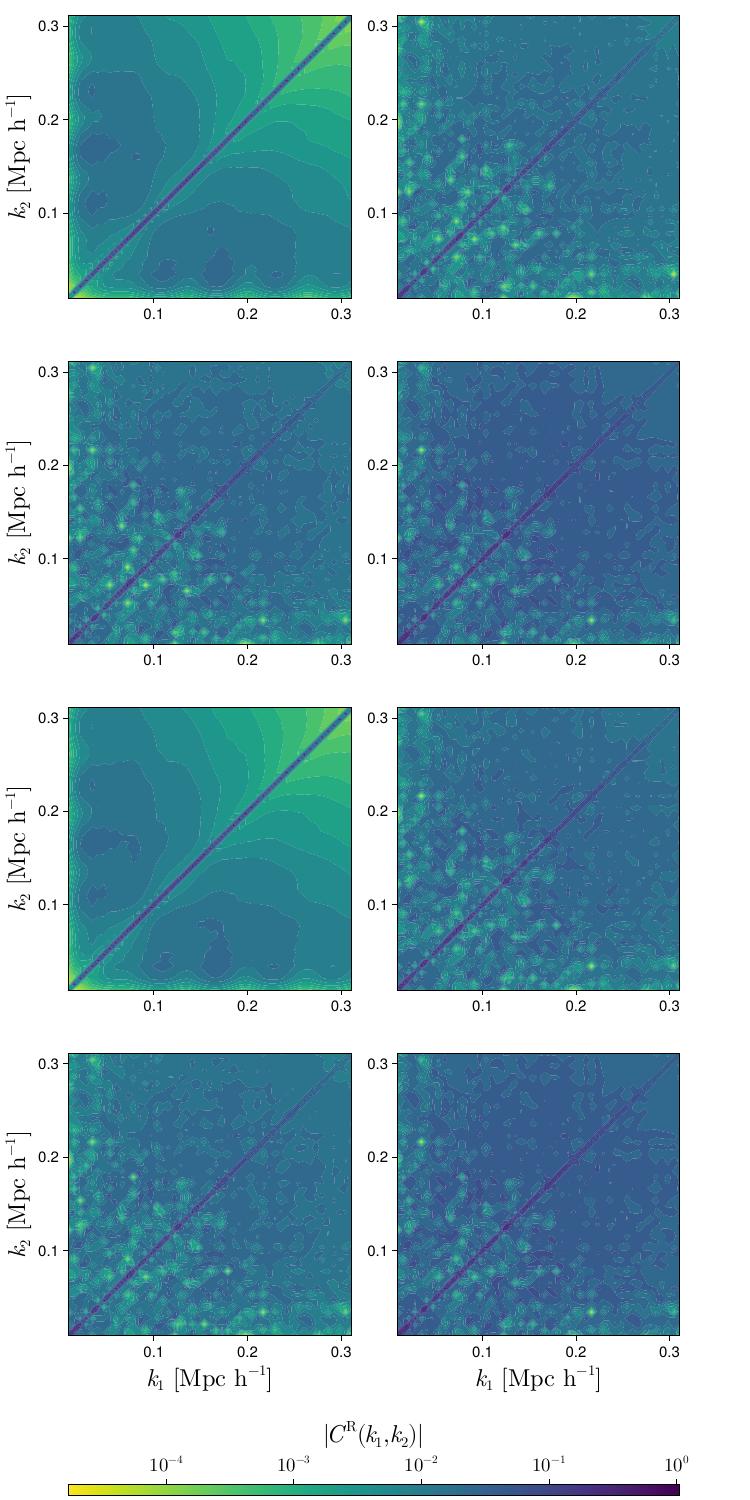}\end{minipage}
    \begin{minipage}{0.48\textwidth}\includegraphics[width=\textwidth]{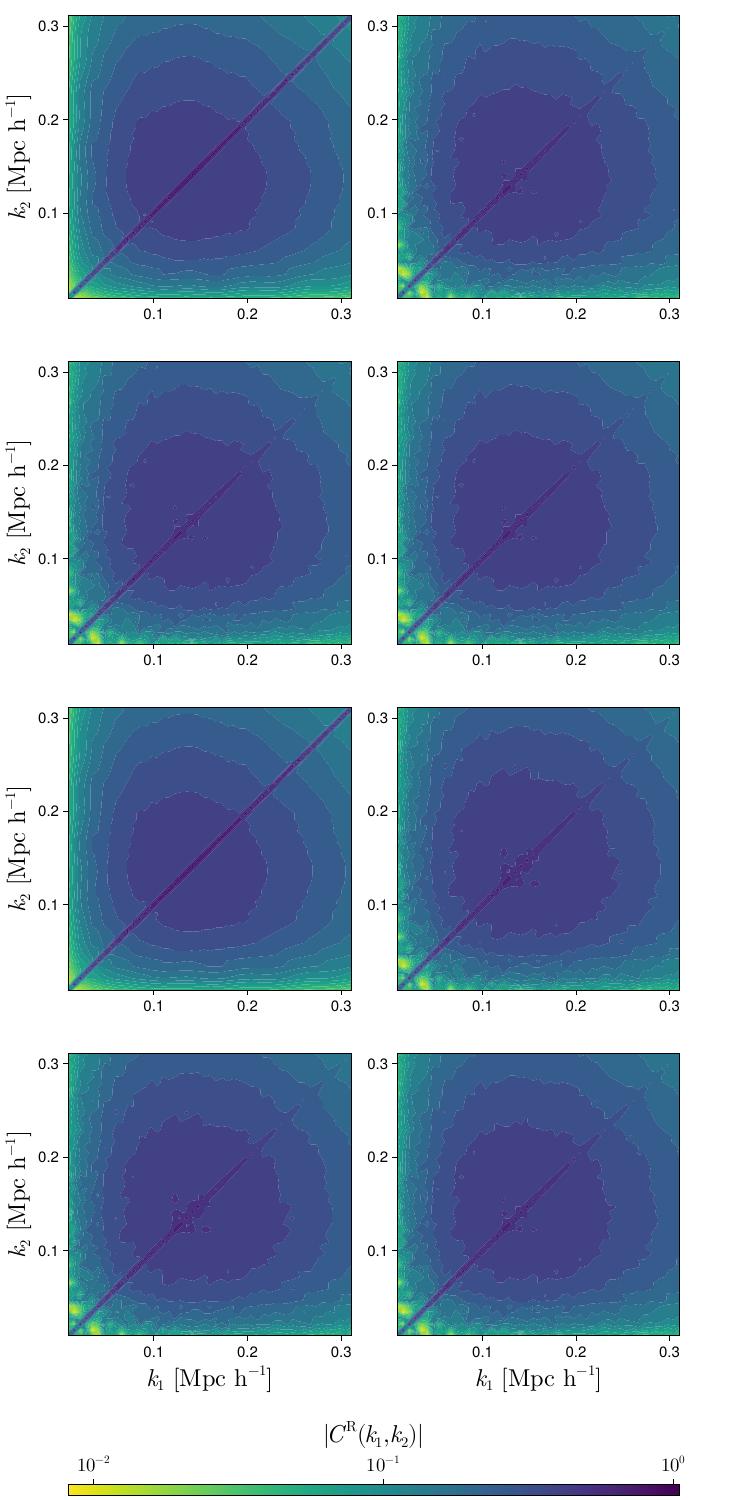}\end{minipage}
    \caption{The reduced covariance matrices for cosmologies with the increased (top two rows) and decreased (bottom two rows) values of the parameter $\Omega_{\mathrm{b}}$ and other parameters kept at their fiducial values.  The left hand pair of columns omit SSC corrections, while the right hand pair of columns include them.  In each pair of columns, the top left panel shows the SPT model covariance matrix, the top right panel shows the non-linear covariance matrix from the simulations, the bottom left panel shows the reconstructed covariance matrix from Eq.~\eqref{eq:Cparallel} with the full response matrix as given in Eq.~\eqref{eq:fullR}, and the bottom right panel shows the reconstructed covariance matrix with the approximated response matrix as given in Eq.~\eqref{eq:aR}.}
    \label{fig:ROb}
\end{figure*}

\begin{figure*}
\centering
    \begin{minipage}{0.48\textwidth}\includegraphics[width=\textwidth]{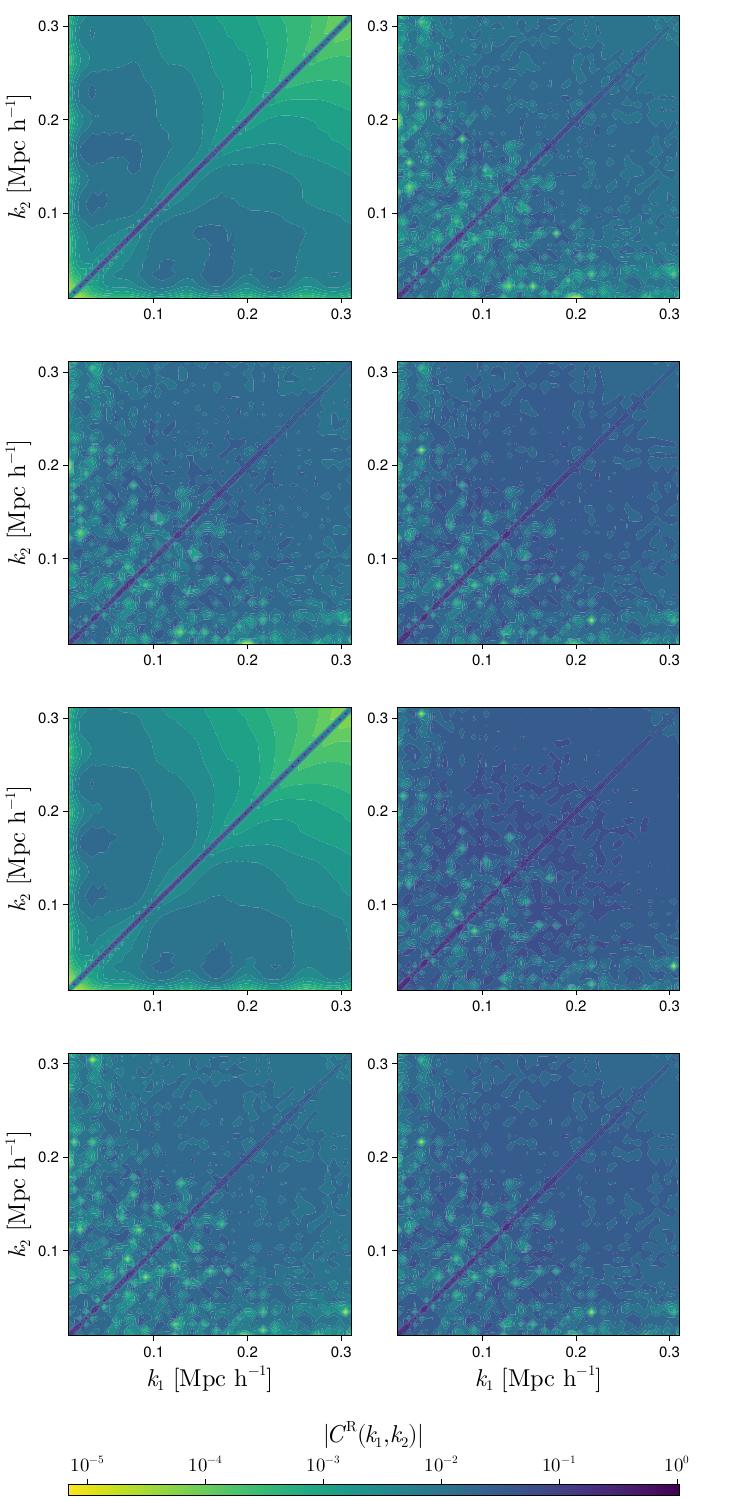}\end{minipage}
    \begin{minipage}{0.48\textwidth}\includegraphics[width=\textwidth]{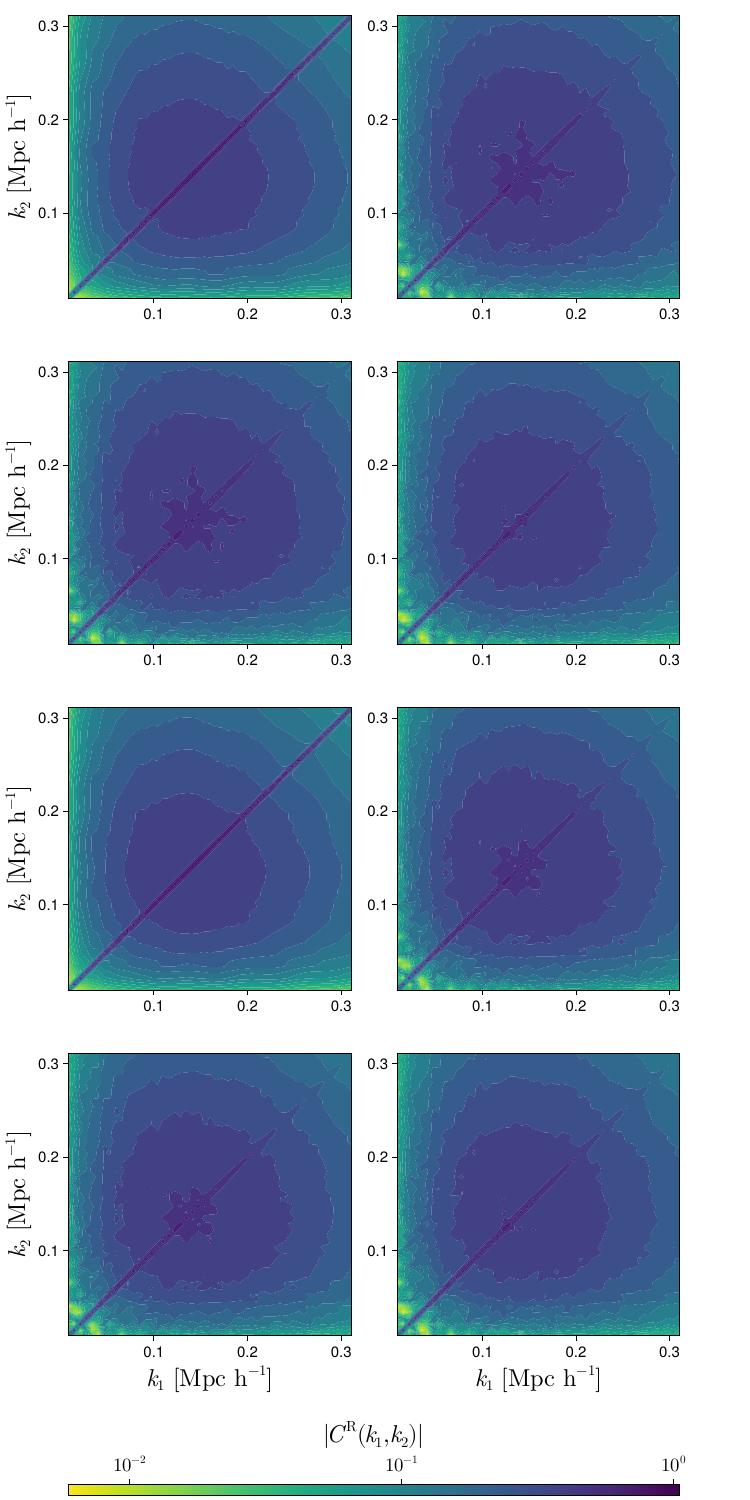}\end{minipage}
    \caption{The reduced covariance matrices for cosmologies with the increased (top two rows) and decreased (bottom two rows) values of the parameter $n_{\mathrm{s}}$ and other parameters kept at their fiducial values.  The left hand pair of columns omit SSC corrections, while the right hand pair of columns include them.  In each pair of columns, the top left panel shows the SPT model covariance matrix, the top right panel shows the non-linear covariance matrix from the simulations, the bottom left panel shows the reconstructed covariance matrix from Eq.~\eqref{eq:Cparallel} with the full response matrix as given in Eq.~\eqref{eq:fullR}, and the bottom right panel shows the reconstructed covariance matrix with the approximated response matrix as given in Eq.~\eqref{eq:aR}.}
    \label{fig:Rns}
\end{figure*}

\begin{figure*}
\centering
    \begin{minipage}{0.48\textwidth}\includegraphics[width=\textwidth]{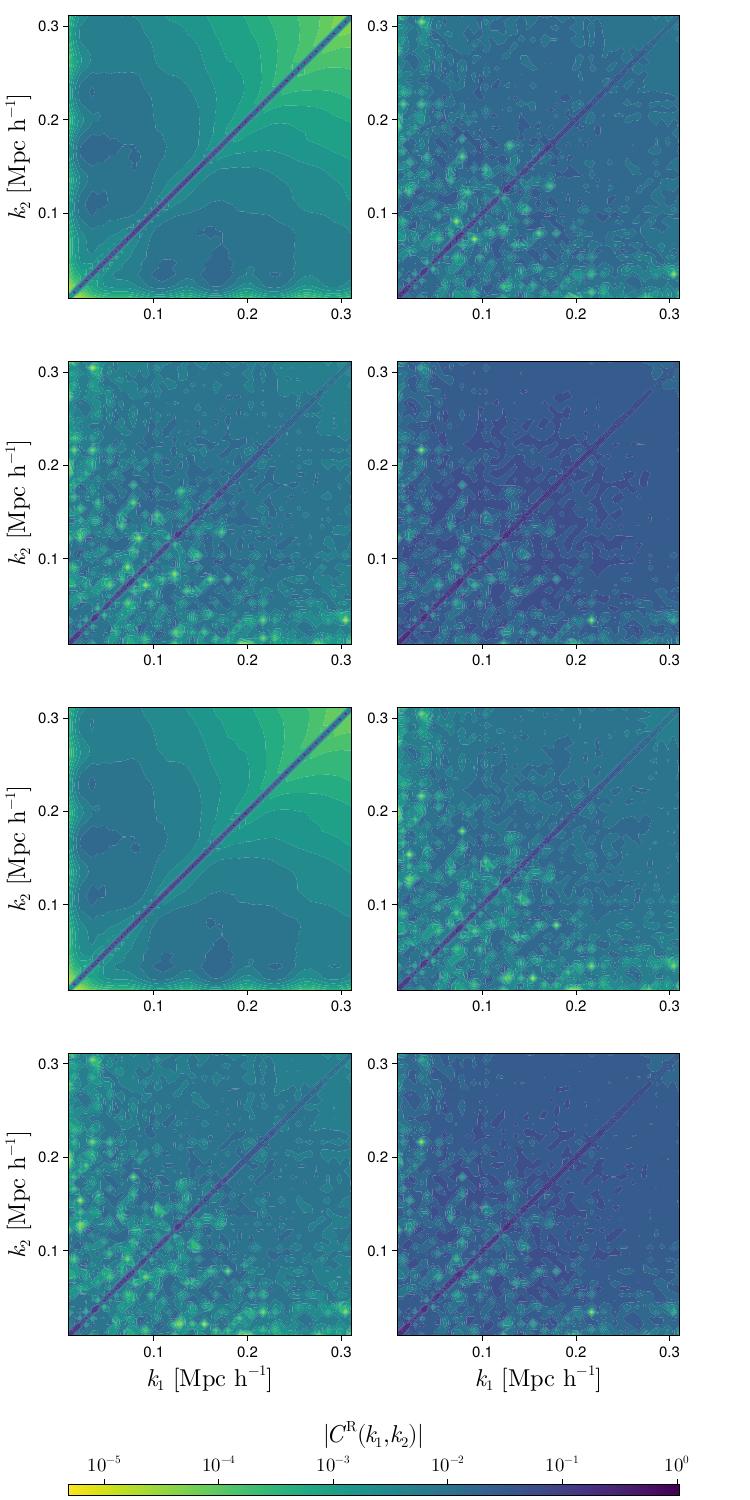}\end{minipage}
    \begin{minipage}{0.48\textwidth}\includegraphics[width=\textwidth]{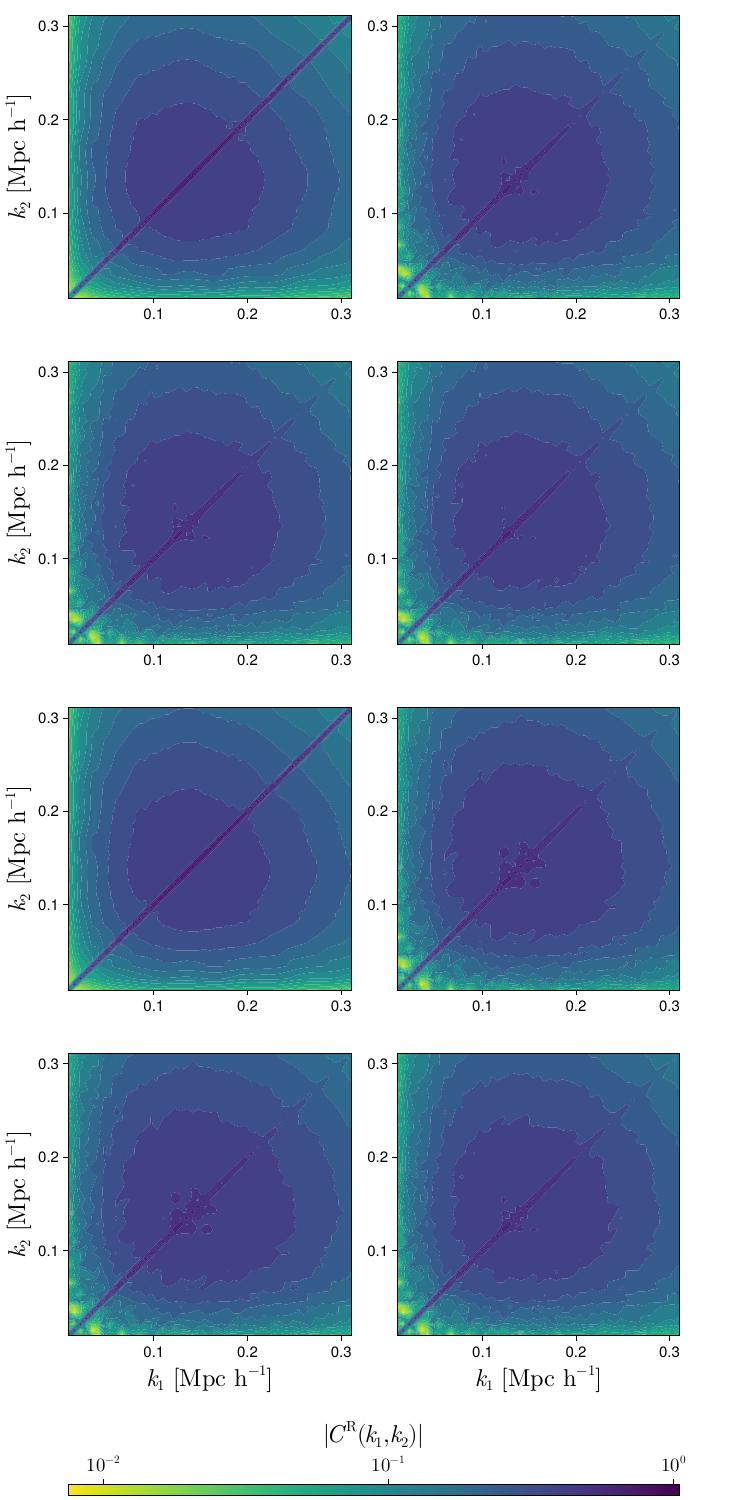}\end{minipage}
    \caption{The reduced covariance matrices for cosmologies with the increased (top two rows) and decreased (bottom two rows) values of the parameter $\sigma_{8}$ and other parameters kept at their fiducial values.  The left hand pair of columns omit SSC corrections, while the right hand pair of columns include them.  In each pair of columns, the top left panel shows the SPT model covariance matrix, the top right panel shows the non-linear covariance matrix from the simulations, the bottom left panel shows the reconstructed covariance matrix from Eq.~\eqref{eq:Cparallel} with the full response matrix as given in Eq.~\eqref{eq:fullR}, and the bottom right panel shows the reconstructed covariance matrix with the approximated response matrix as given in Eq.~\eqref{eq:aR}.}
    \label{fig:Rs8}
\end{figure*}

\begin{figure*}
\centering
    \begin{minipage}{0.48\textwidth}\includegraphics[width=\textwidth]{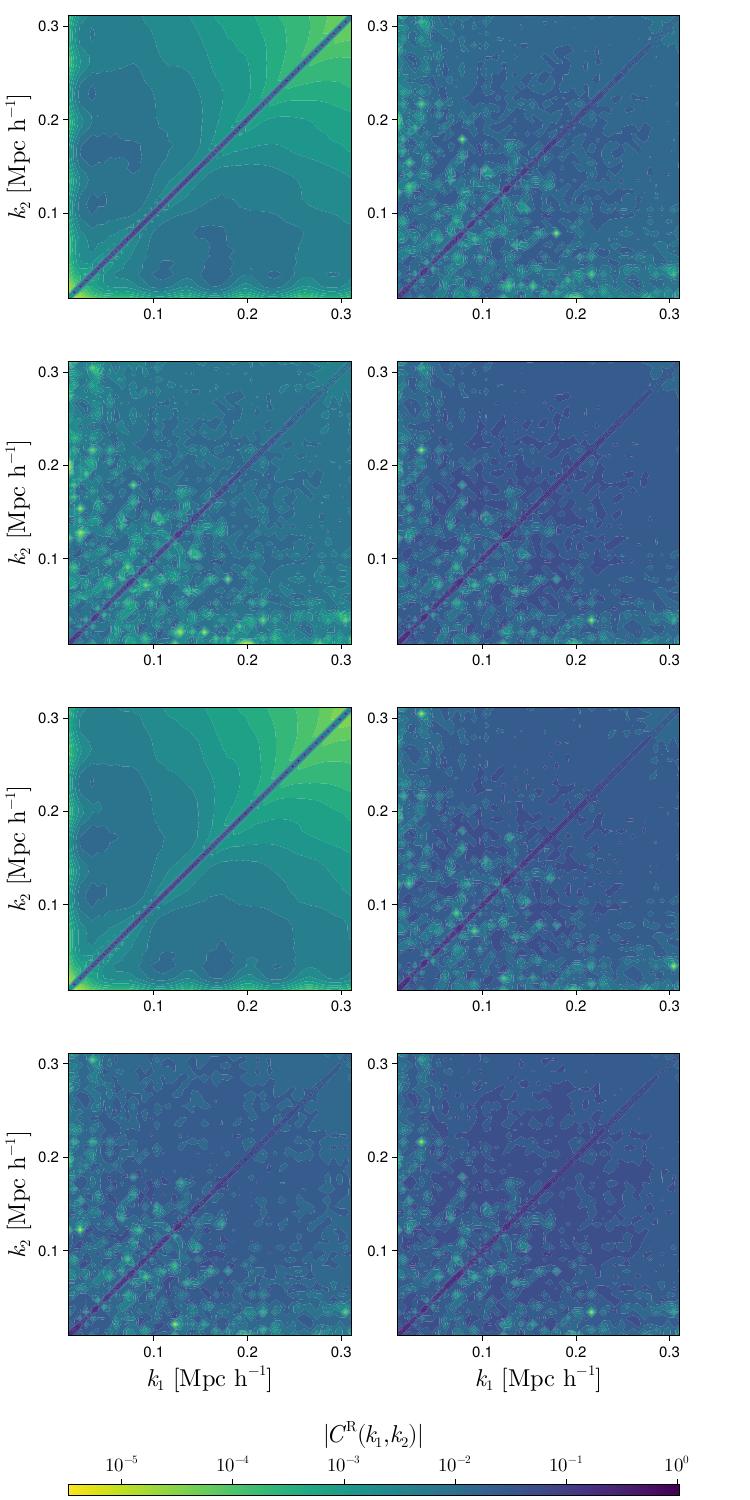}\end{minipage}
    \begin{minipage}{0.48\textwidth}\includegraphics[width=\textwidth]{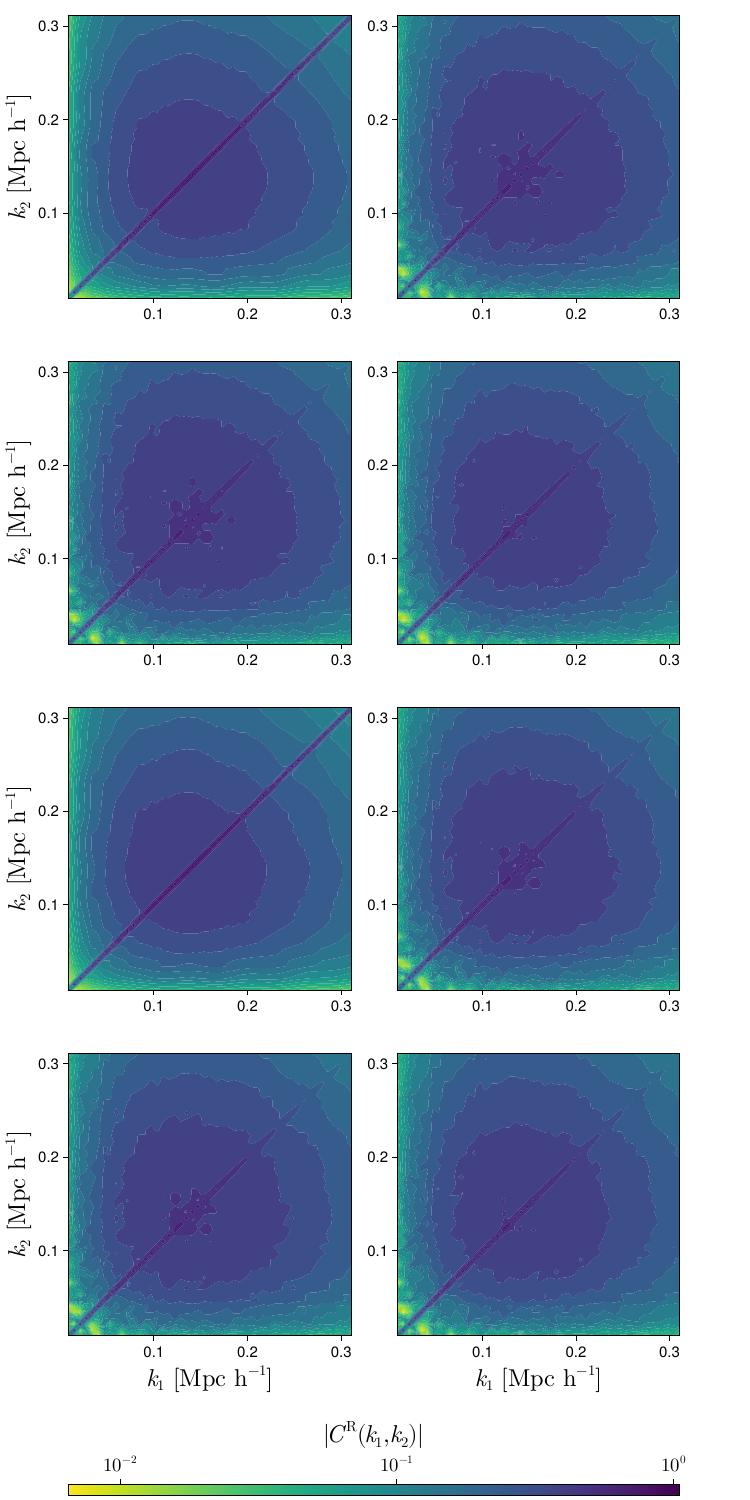}\end{minipage}
    \caption{The reduced covariance matrices for cosmologies with the increased (top two rows) and decreased (bottom two rows) values of the parameter $h$ and other parameters kept at their fiducial values.  The left hand pair of columns omit SSC corrections, while the right hand pair of columns include them.  In each pair of columns, the top left panel shows the SPT model covariance matrix, the top right panel shows the non-linear covariance matrix from the simulations, the bottom left panel shows the reconstructed covariance matrix from Eq.~\eqref{eq:Cparallel} with the full response matrix as given in Eq.~\eqref{eq:fullR}, and the bottom right panel shows the reconstructed covariance matrix with the approximated response matrix as given in Eq.~\eqref{eq:aR}.}
    \label{fig:Rh}
\end{figure*}

\begin{figure*}
\centering
    \begin{minipage}{0.48\textwidth}\includegraphics[width=\textwidth]{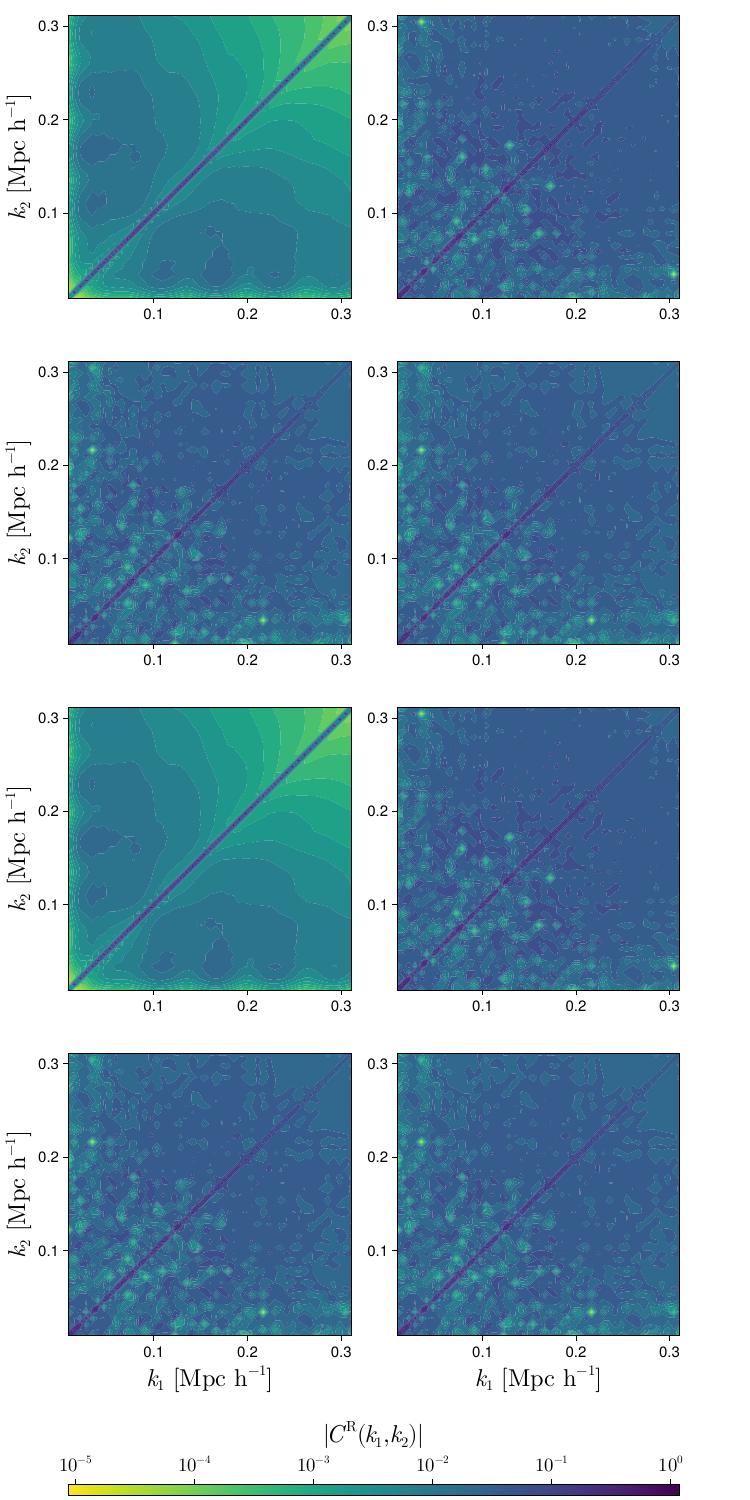}\end{minipage}
    \begin{minipage}{0.48\textwidth}\includegraphics[width=\textwidth]{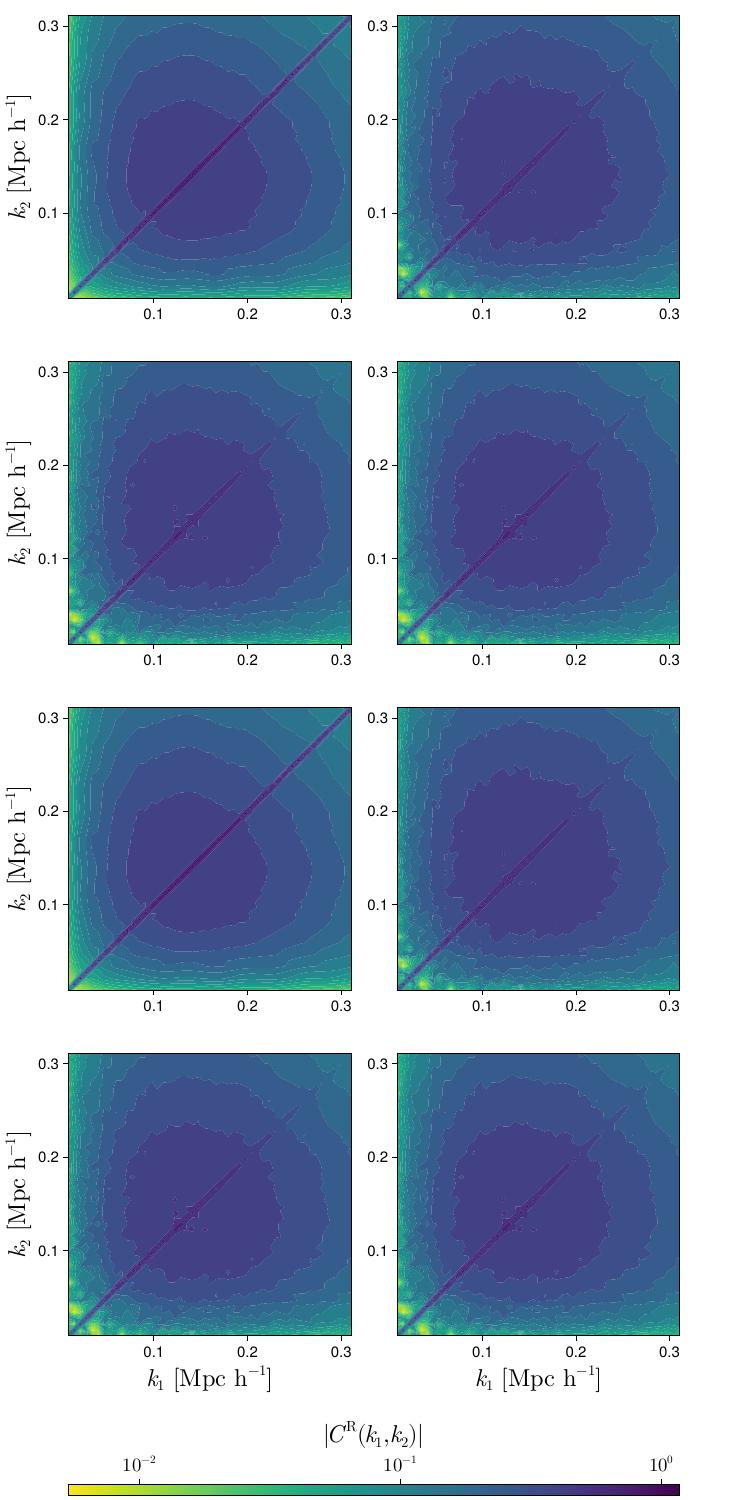}\end{minipage}
    \caption{The reduced covariance matrices for cosmologies with the increased (top two rows) and decreased (bottom two rows) values of the parameter $w_{0}$ and other parameters kept at their fiducial values.  The left hand pair of columns omit SSC corrections, while the right hand pair of columns include them.  In each pair of columns, the top left panel shows the SPT model covariance matrix, the top right panel shows the non-linear covariance matrix from the simulations, the bottom left panel shows the reconstructed covariance matrix from Eq.~\eqref{eq:Cparallel} with the full response matrix as given in Eq.~\eqref{eq:fullR}, and the bottom right panel shows the reconstructed covariance matrix with the approximated response matrix as given in Eq.~\eqref{eq:aR}.}
    \label{fig:Rw0}
\end{figure*}

\section{Results}\label{sec:results}

\subsection{Fiducial approximate model predictions: SPT and SSC}

To start, the top panels of \Fig{fig:Cfid} compare the SPT model
covariance matrices with the measured covariance matrix. The left-hand panels exclude the super-sample covariance corrections and the right-hand
panels include them. These corrections are estimated using Eq.~\eqref{eq:SSC} and are added to both the SPT
predictions and the simulation estimates; their purpose is to give an
indication of the level of super-sample effects in real surveys. 

As can be seen, our SPT model for the covariance matrix is a
reasonably good approximation for the measured fiducial power
spectrum on scales $k\lesssim0.3\kMpc$. 


\subsection{The matrix $\widetilde{\Ymat}$}

In Figs.~\ref{fig:Y_Om}-\ref{fig:Y_h} we show $\widetilde{\Ymat}$,
computed for the various cosmologies simulated in the Quijote
suite. This constitutes a measure of the
accuracy of each SPT model, as all deviations from identity matrices constitute disparities between the model and the measured nonlinear covariance.  

In
all cases, we notice that while there is a strong diagonal component
that is close to unity, there is off diagonal structure that is
largest where the covariance is between modes with maximally different wave numbers.

This suggests that the SPT model for the
covariance matrix evolution is effective on the scales of interest except when the modes being covaried approach opposite ends of the range under study. 

We
also note that the fixed correlation matrix approximation does not make a substantial difference to the structure of
$\widetilde{\Ymat}$ for the models that we have studied.  We also
note that the super-sample covariance makes a significant impact to
the structure of the matrix $\widetilde{\Ymat}$, suggesting that
for real surveys we do need to take care to add this contribution and
its cosmological dependence.

Note that in constructing the full and approximate response matrices
$\Rmat_{\alpha}$, we needed to compute the precision matrix of the
fiducial model. To do this we used both regular numerical matrix
inversion techniques and the Moore-Penrose pseudo-inverse
technique. We found that both gave the same results to a high degree
of precision, indicating that the noise contribution from the inverse
covariance matrix is relatively small. We therefore assume that the
inversion method introduces no further additional noise to our
analysis. 


\begin{figure*}
  \centering{
    \includegraphics[width=7.2cm]{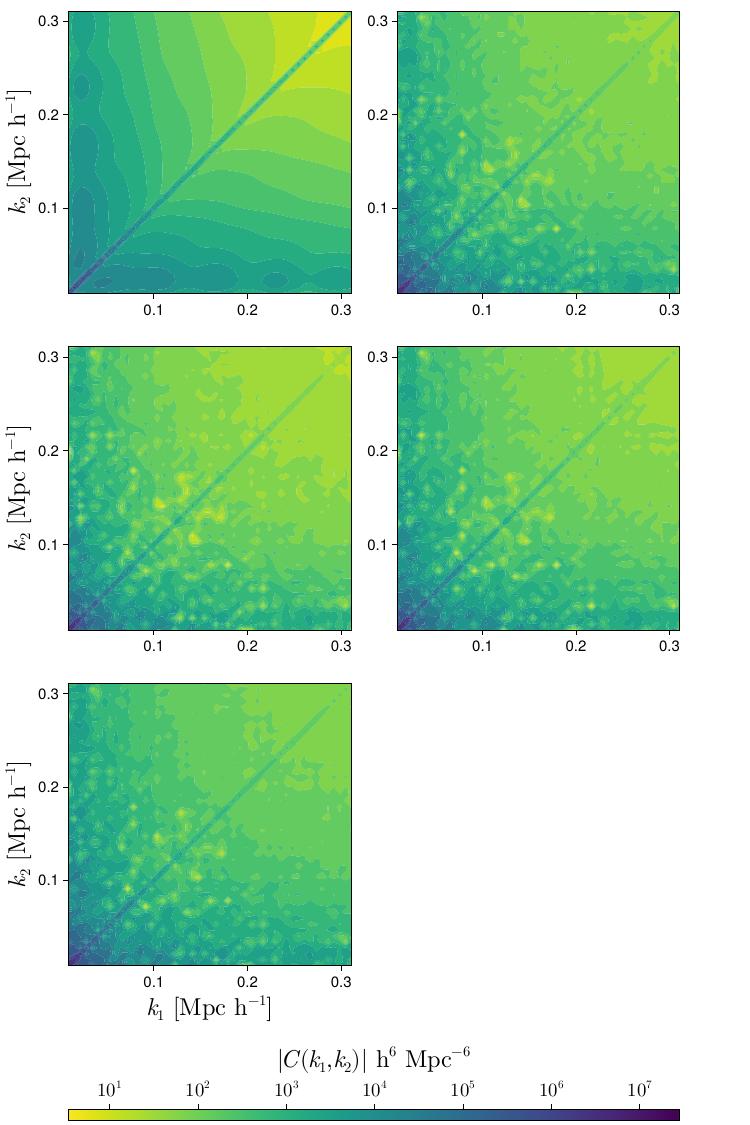}\hspace{0.5cm}
    \includegraphics[width=7.2cm]{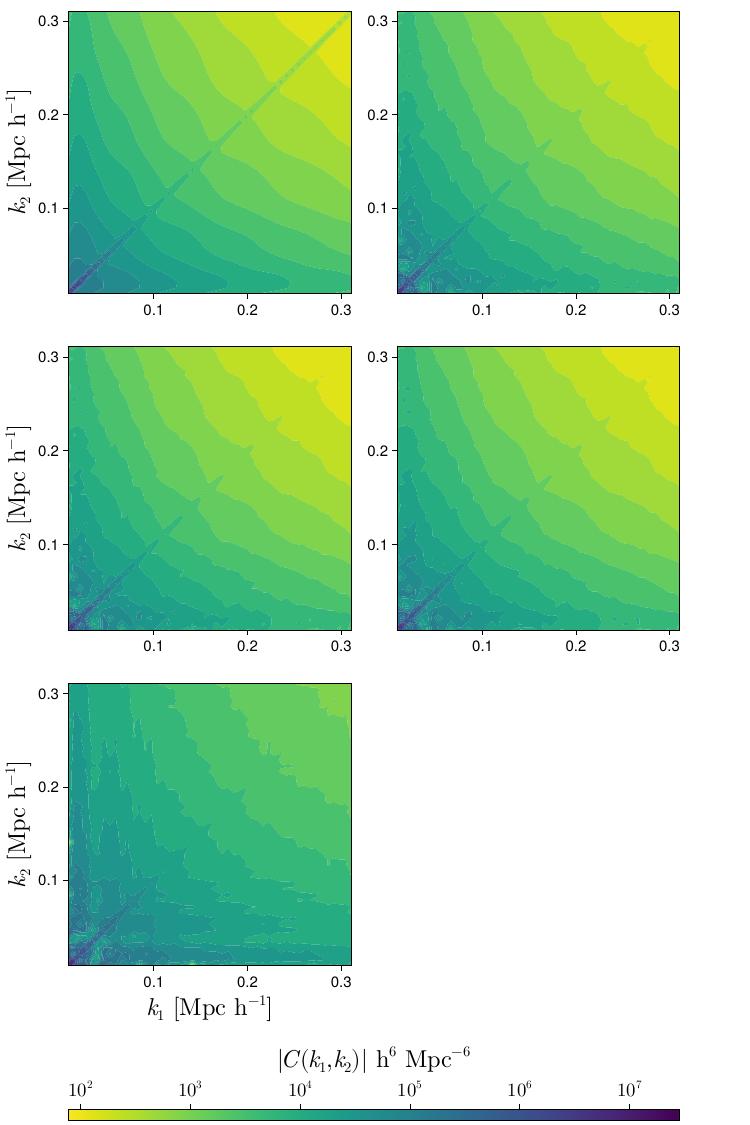}}
  \centering{
    \includegraphics[width=7.2cm]{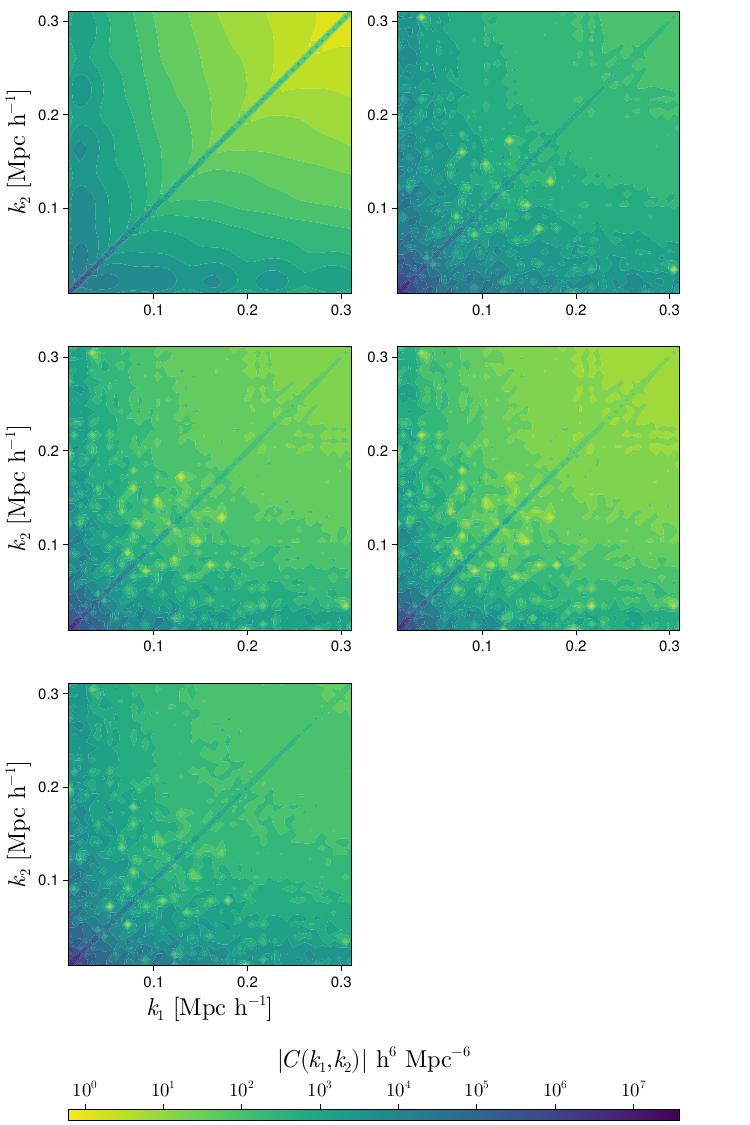}\hspace{0.5cm}
    \includegraphics[width=7.2cm]{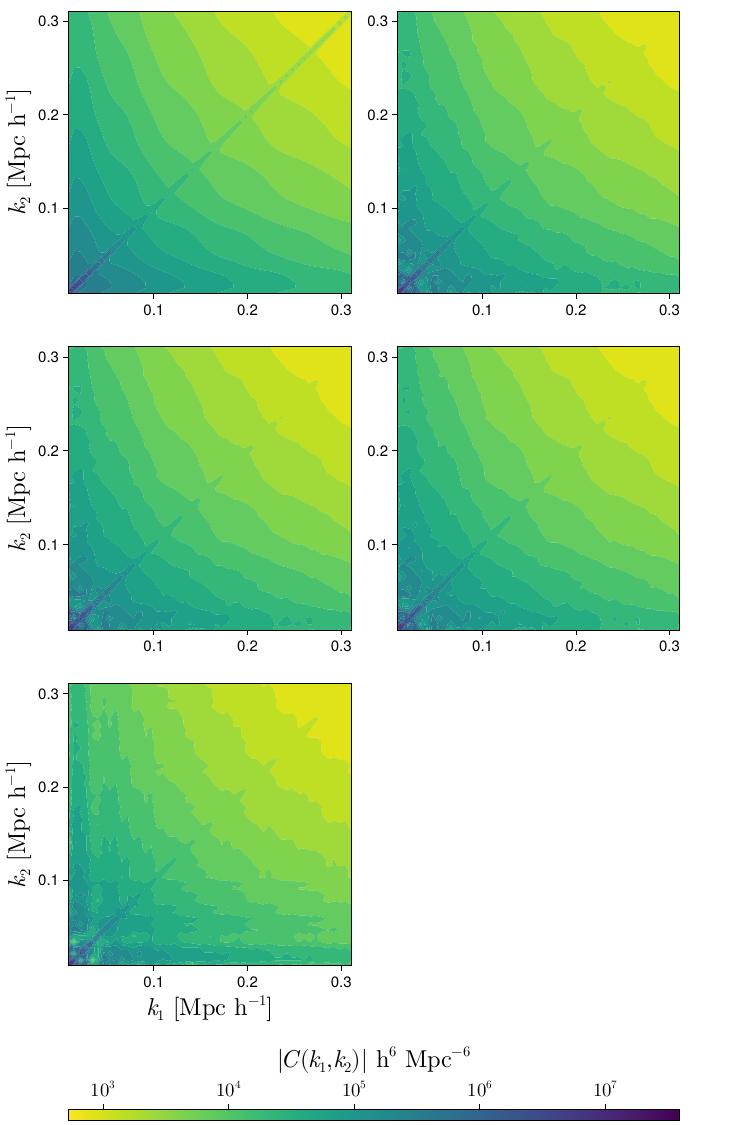}}
  \caption{The cosmological dependence of the covariance matrix on the
    matter density parameter $\Omega_{\mathrm{m}}$. The top and bottom
    sets of panels show results for the $\Omega_{\mathrm{m}}^+$ and
    $\Omega_{\mathrm{m}}^-$ variations in cosmology, respectively.
    Considering the top row of panel, the left hand pair of columns
    omit SSC corrections, while the right hand pair of columns include
    them.  In each set of five panels: the top left panel shows the
    SPT model covariance matrix; the top right panel shows the
    non-linear covariance matrix measured from the 500 simulations in
    the Quijote suite; the centre left panel shows $C_{\parallel}$,
    reconstructed with the full response matrix; the centre right
    panel shows $C_{\parallel,\mathrm{a}}$, where the correlation
    matrix is assumed to be independent of cosmology; the bottom panel
    shows $C_{\parallel,\mathrm{R}=0}$, where the response matrix has
    been set to zero.\label{fig:COm}}
\end{figure*}


\begin{figure*}
  \centering{
    \includegraphics[width=7.2cm]{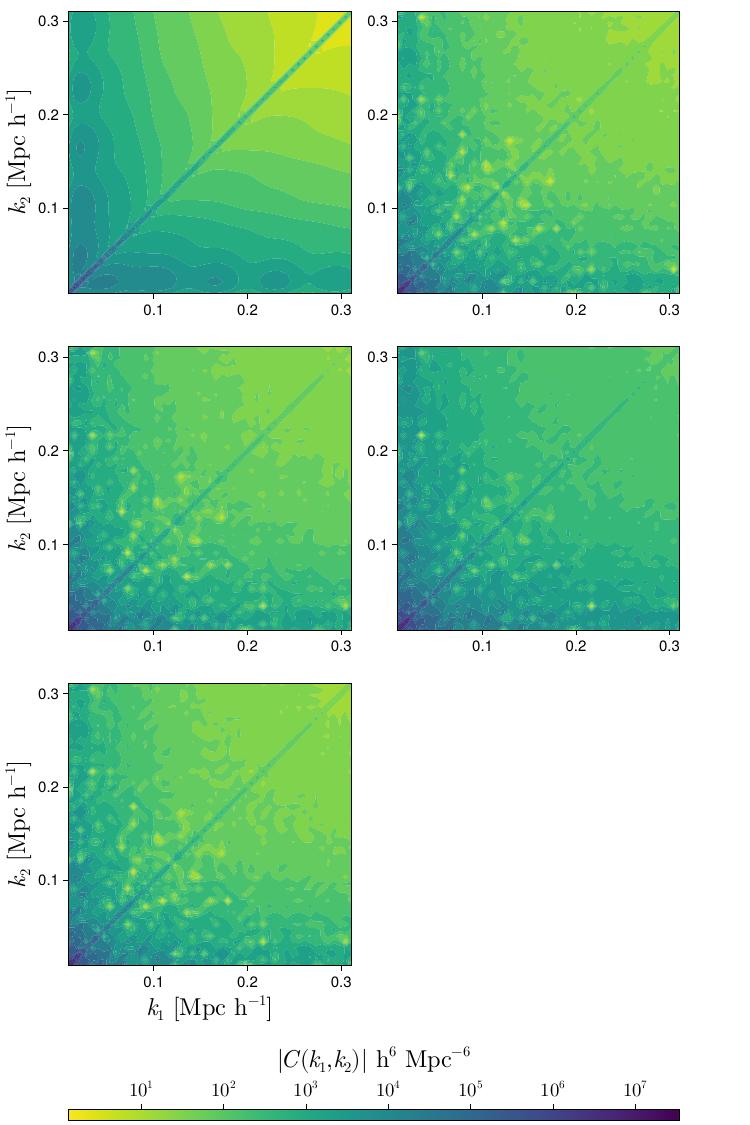}\hspace{0.5cm}
    \includegraphics[width=7.2cm]{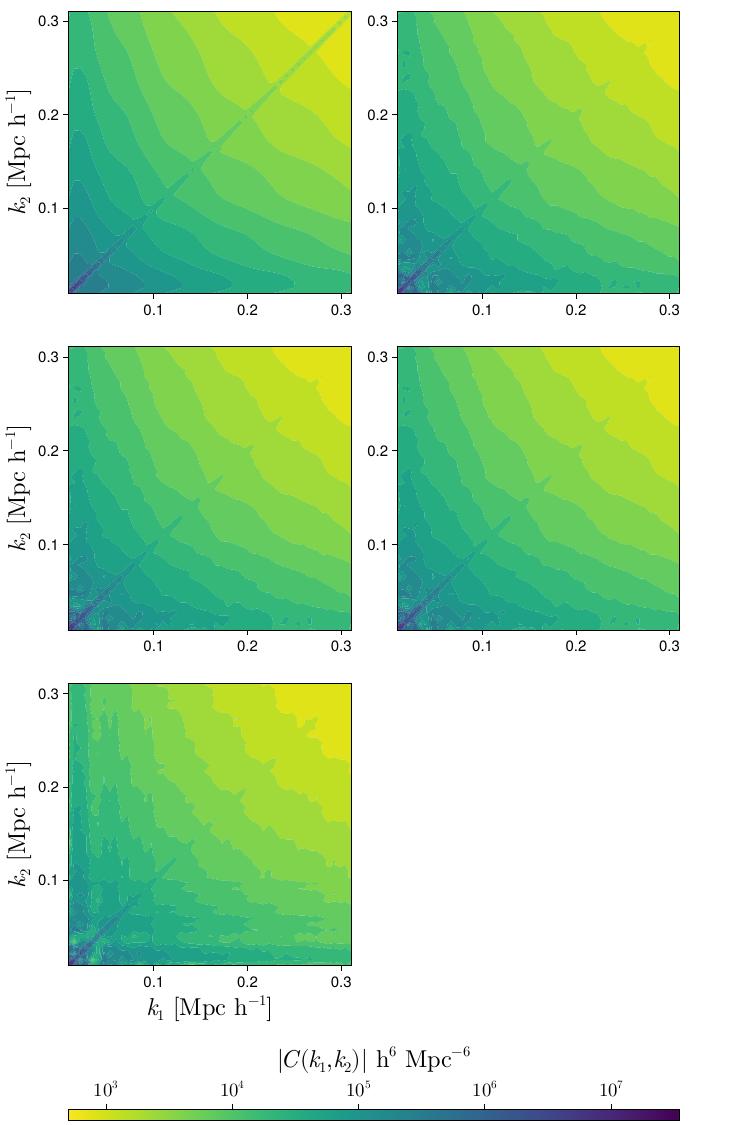}}
  \centering{
    \includegraphics[width=7.2cm]{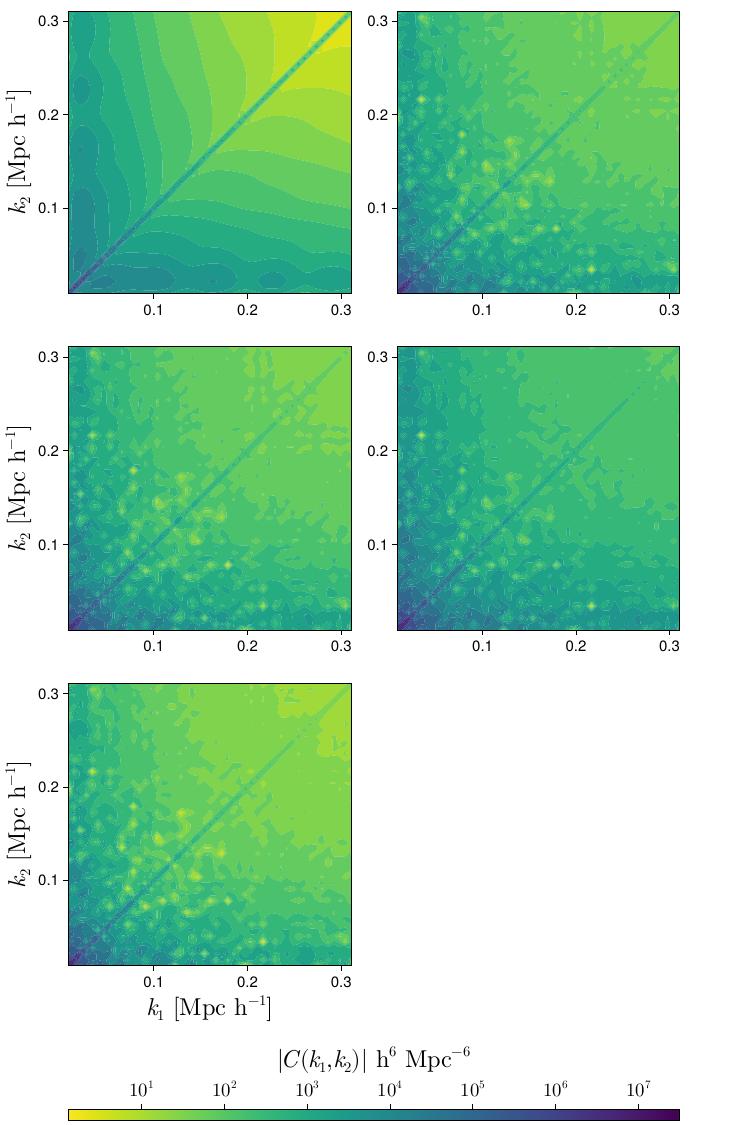}\hspace{0.5cm}
    \includegraphics[width=7.2cm]{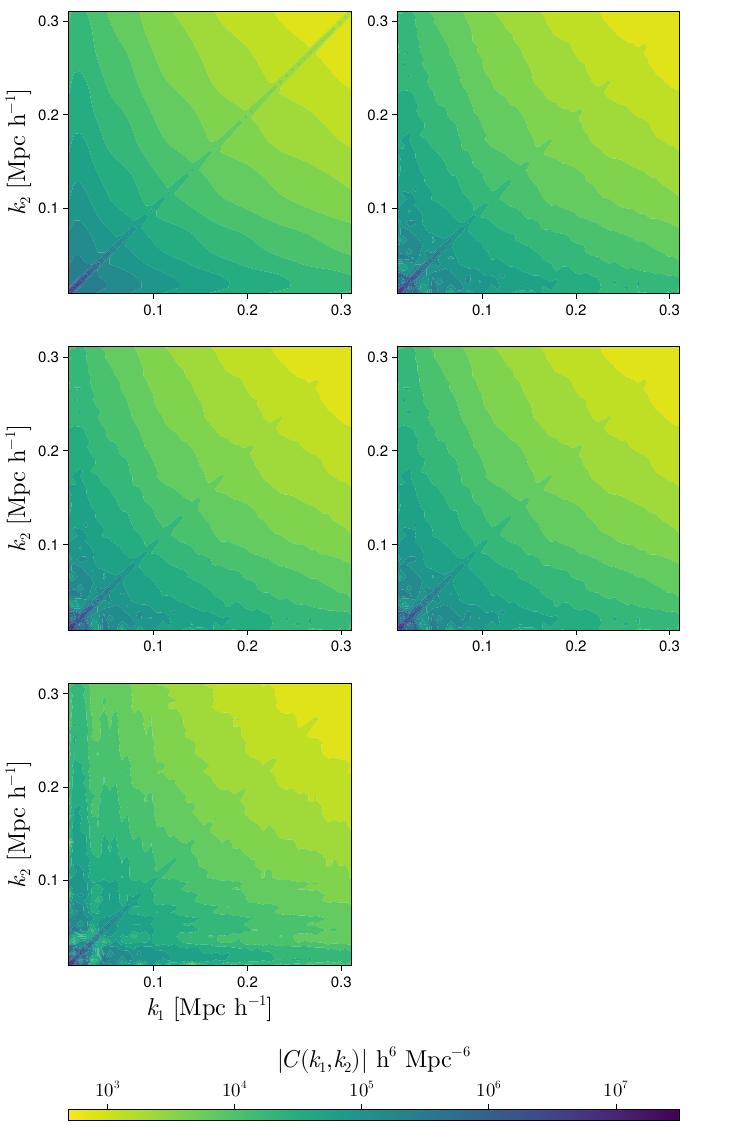}}
  \caption{The same as \Fig{fig:COm}, but this time we show the
    cosmological dependence of the covariance matrix on the baryon
    density parameter $\Omega_{\mathrm{b}}$. The top and bottom sets
    of panels show results for the $\Omega_{\mathrm{b}}^+$ and
    $\Omega_{\mathrm{b}}^-$ variations in cosmology,
    respectively.\label{fig:COb}}
\end{figure*}


\begin{figure*}
  \centering{
    \includegraphics[width=7.2cm]{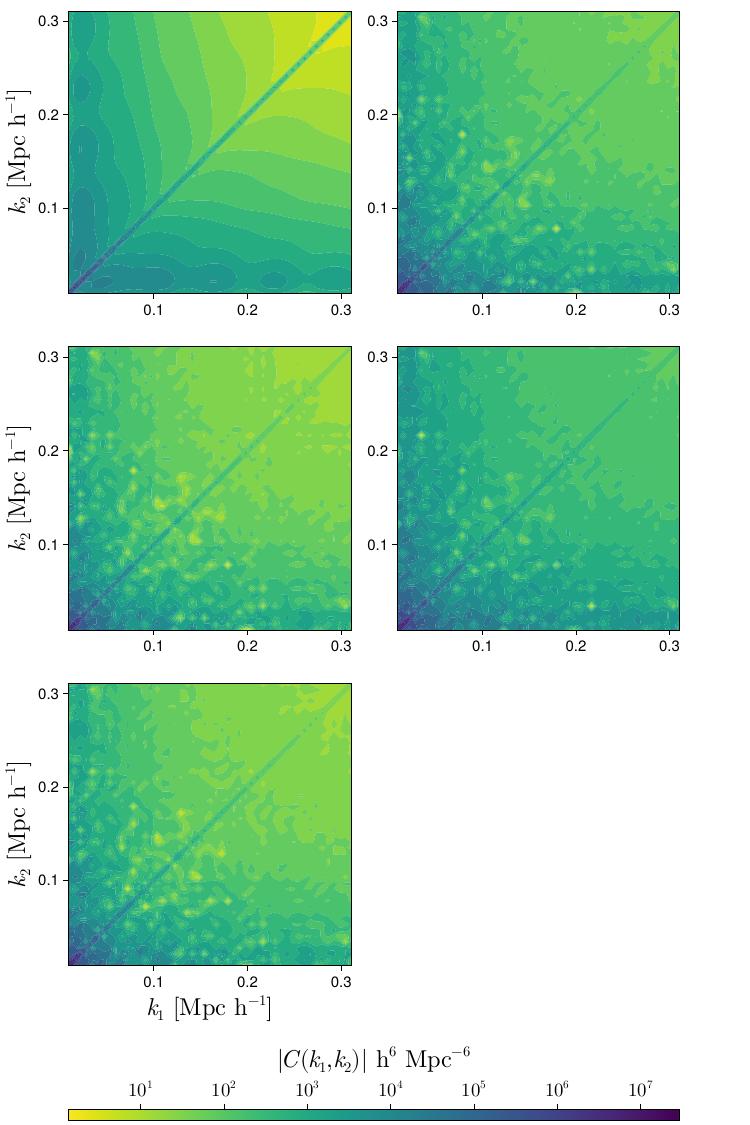}\hspace{0.5cm}
    \includegraphics[width=7.2cm]{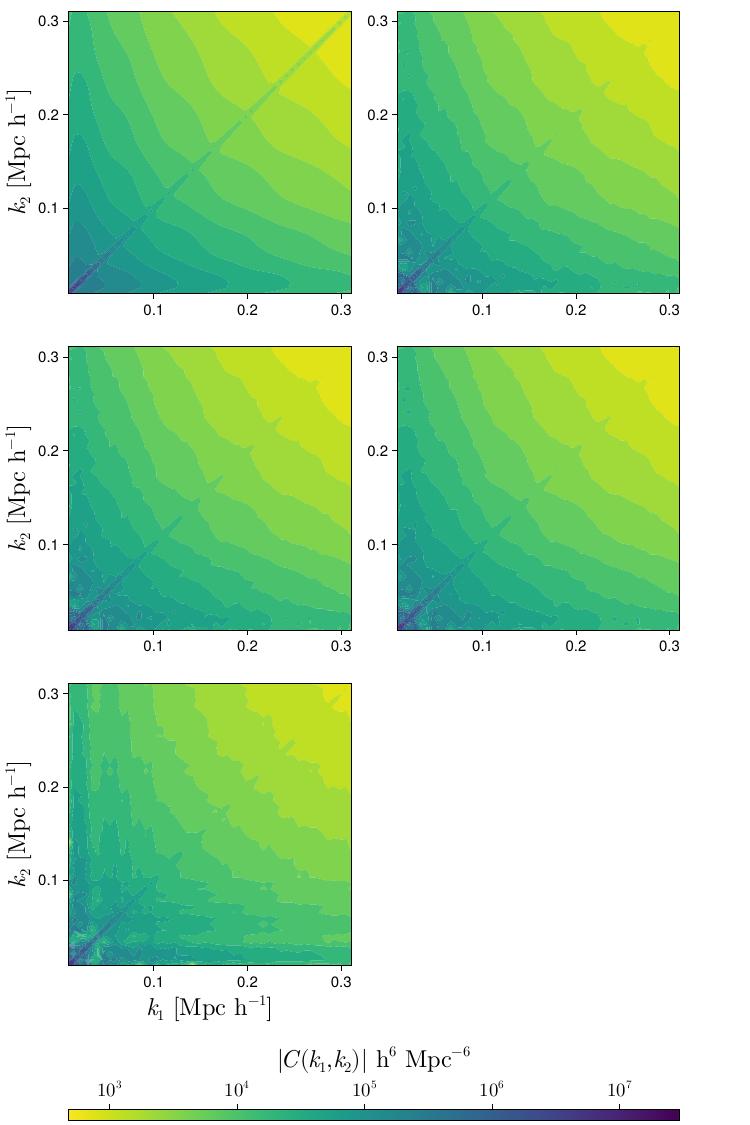}}
  \centering{
    \includegraphics[width=7.2cm]{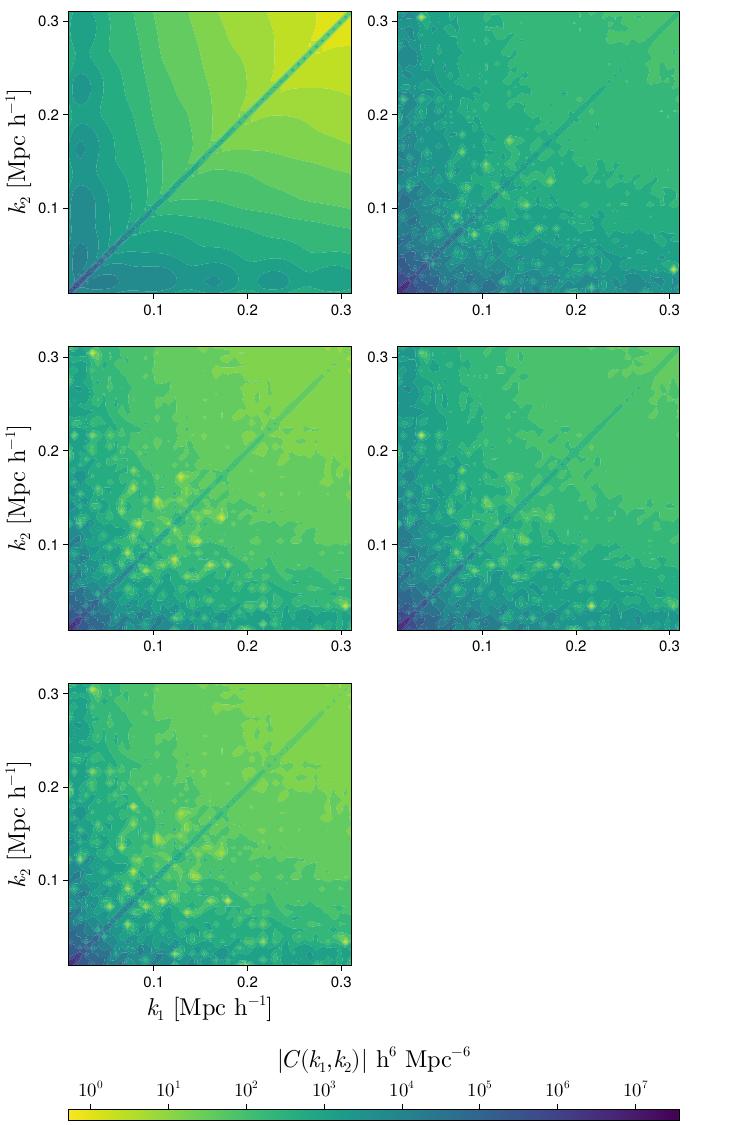}\hspace{0.5cm}
    \includegraphics[width=7.2cm]{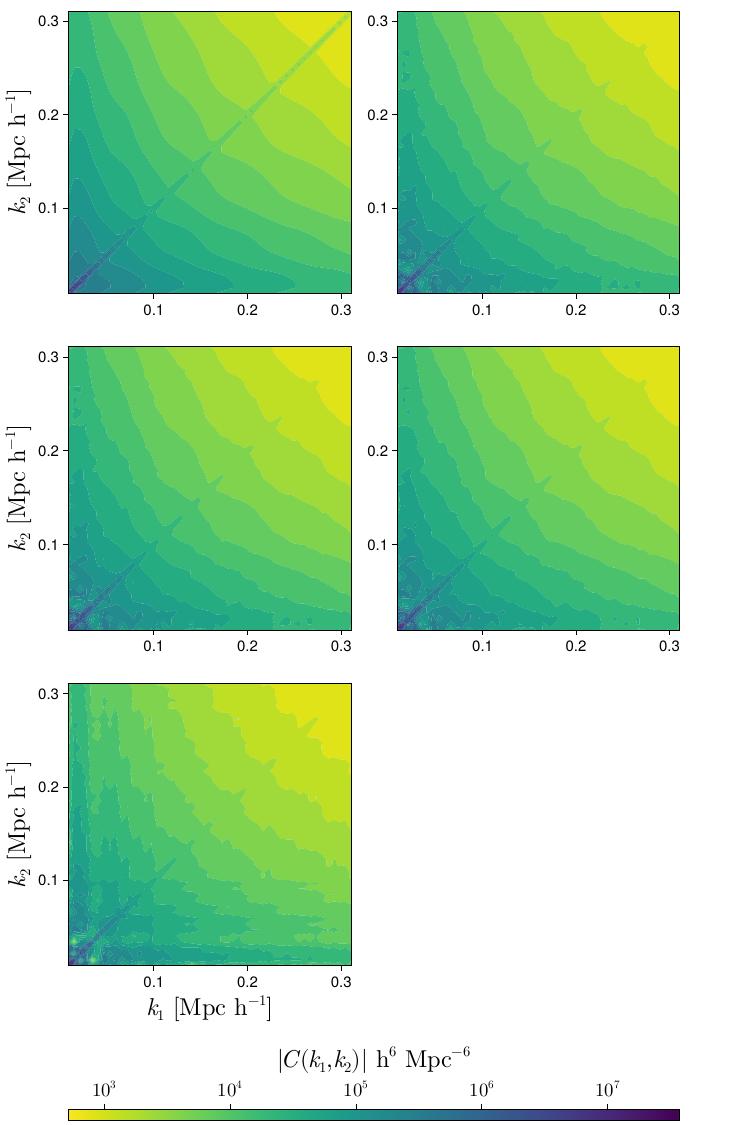}}
  \caption{The same as \Fig{fig:COm}, but this time we show the
    cosmological dependence of the covariance matrix on the primordial
    scalar index $n_{\mathrm{s}}$. The top and bottom sets of panels
    show results for the $n_{\mathrm{s}}^+$ and $n_{\mathrm{s}}^-$
    variations in cosmology, respectively.\label{fig:Cns}}
\end{figure*}


\begin{figure*}
  \centering{
    \includegraphics[width=7.2cm]{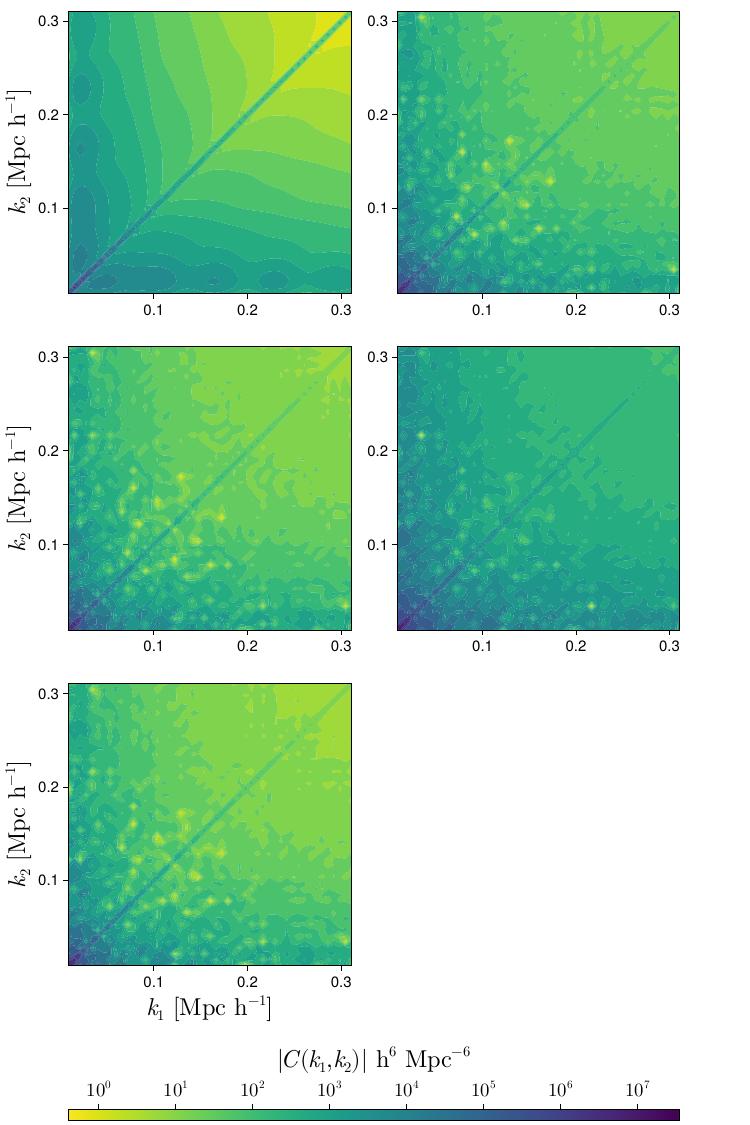}\hspace{0.5cm}
    \includegraphics[width=7.2cm]{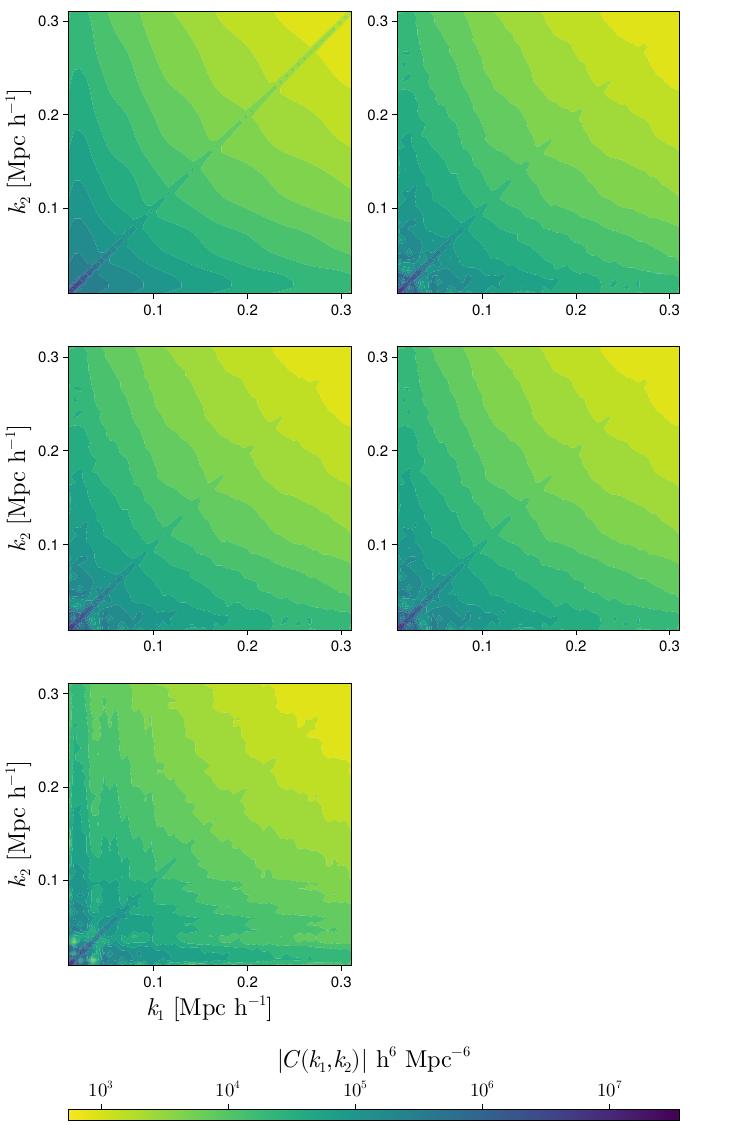}}
  \centering{
    \includegraphics[width=7.2cm]{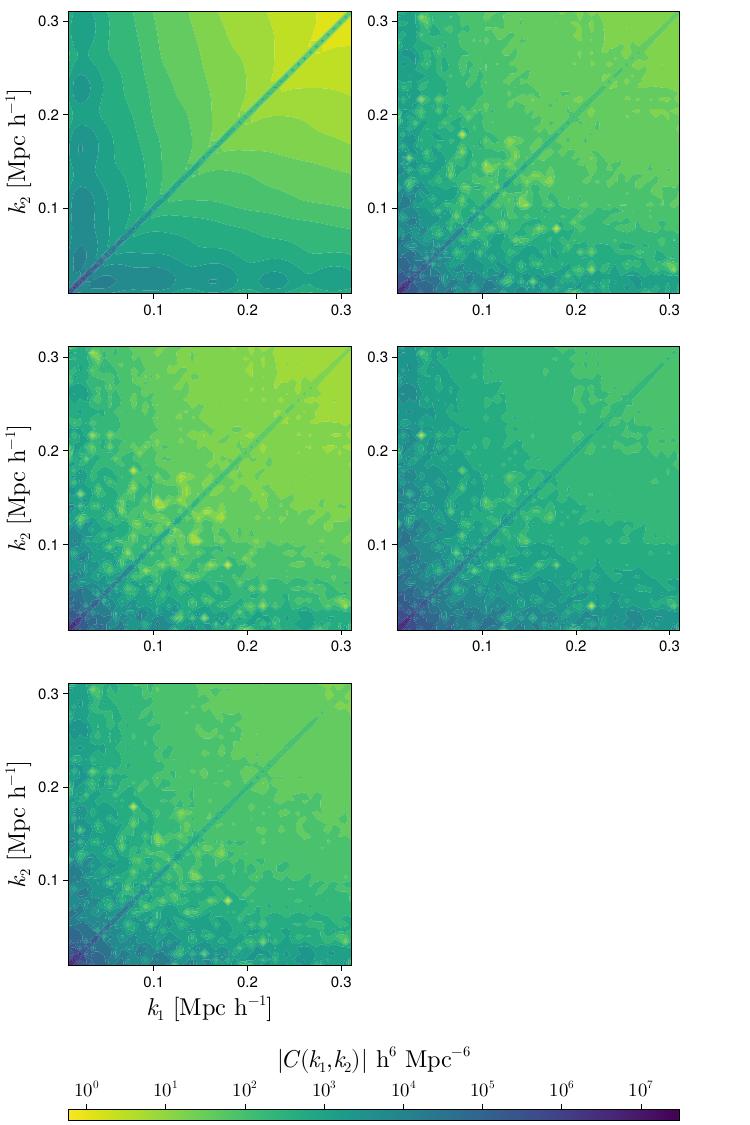}\hspace{0.5cm}
    \includegraphics[width=7.2cm]{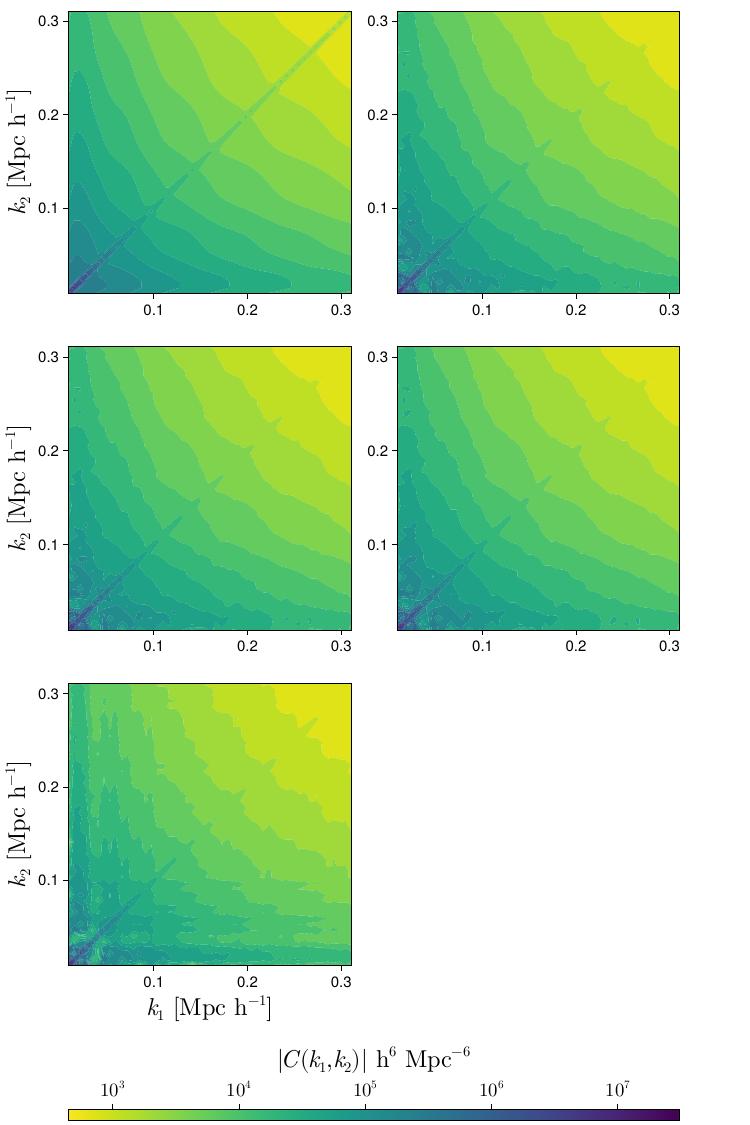}}
  \caption{The same as \Fig{fig:COm}, but this time we show the
    cosmological dependence of the covariance matrix on
    $\sigma_8$. The top and bottom sets of panels show results for the
    $\sigma_8^+$ and $\sigma_8^-$ variations in cosmology,
    respectively.\label{fig:Cs8}}
\end{figure*}


\begin{figure*}
  \centering{
    \includegraphics[width=7.2cm]{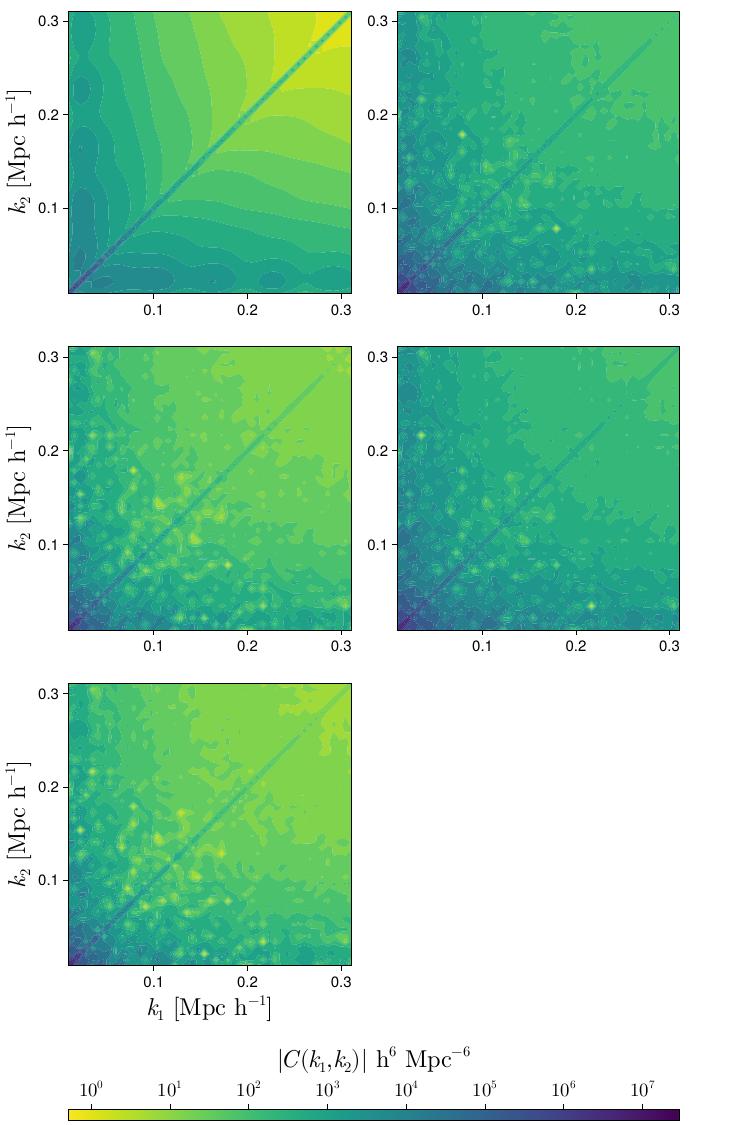}\hspace{0.5cm}
    \includegraphics[width=7.2cm]{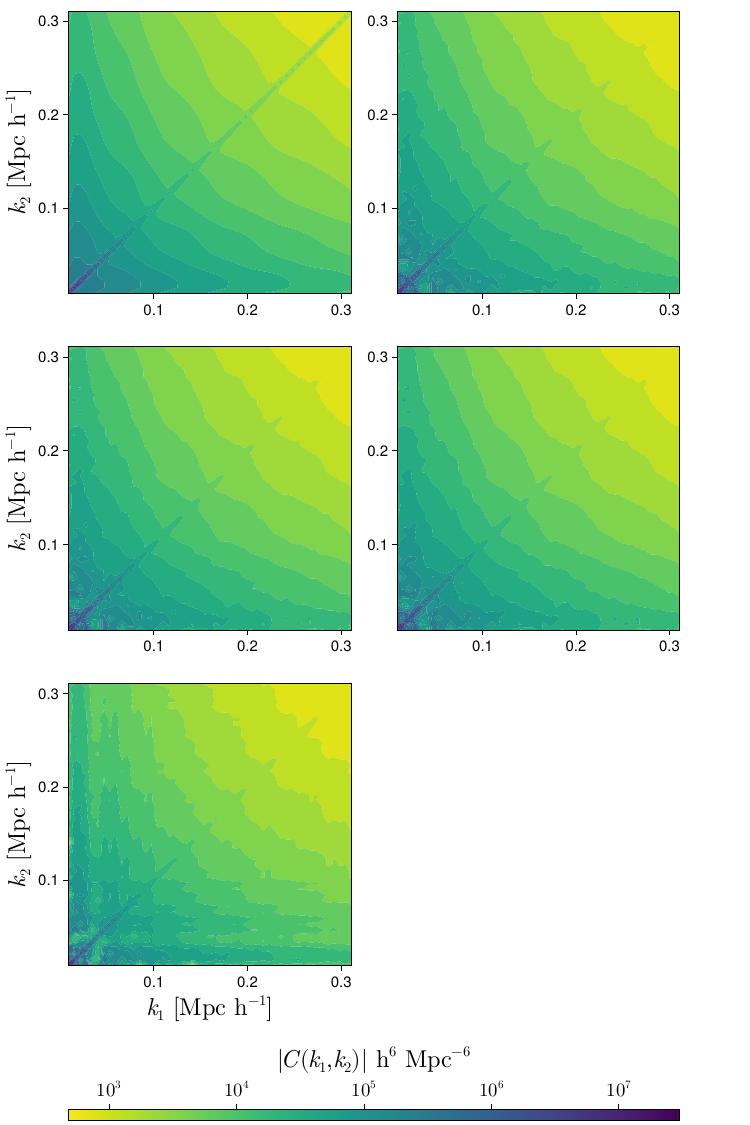}}
  \centering{
    \includegraphics[width=7.2cm]{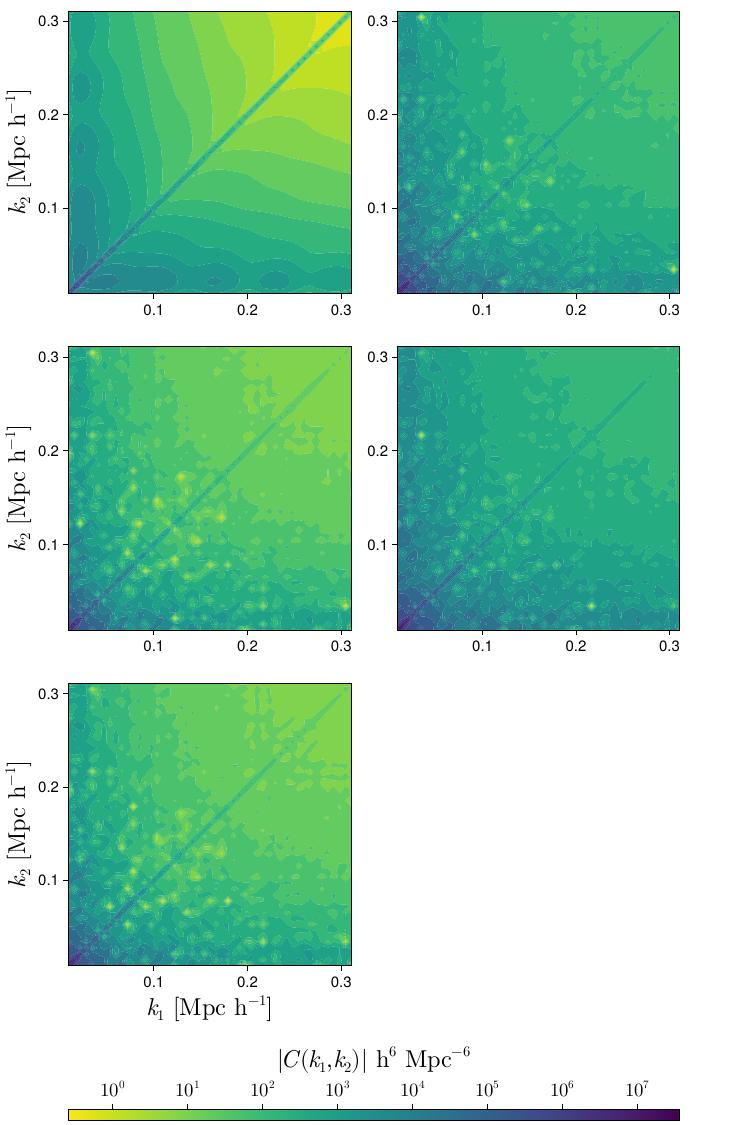}\hspace{0.5cm}
    \includegraphics[width=7.2cm]{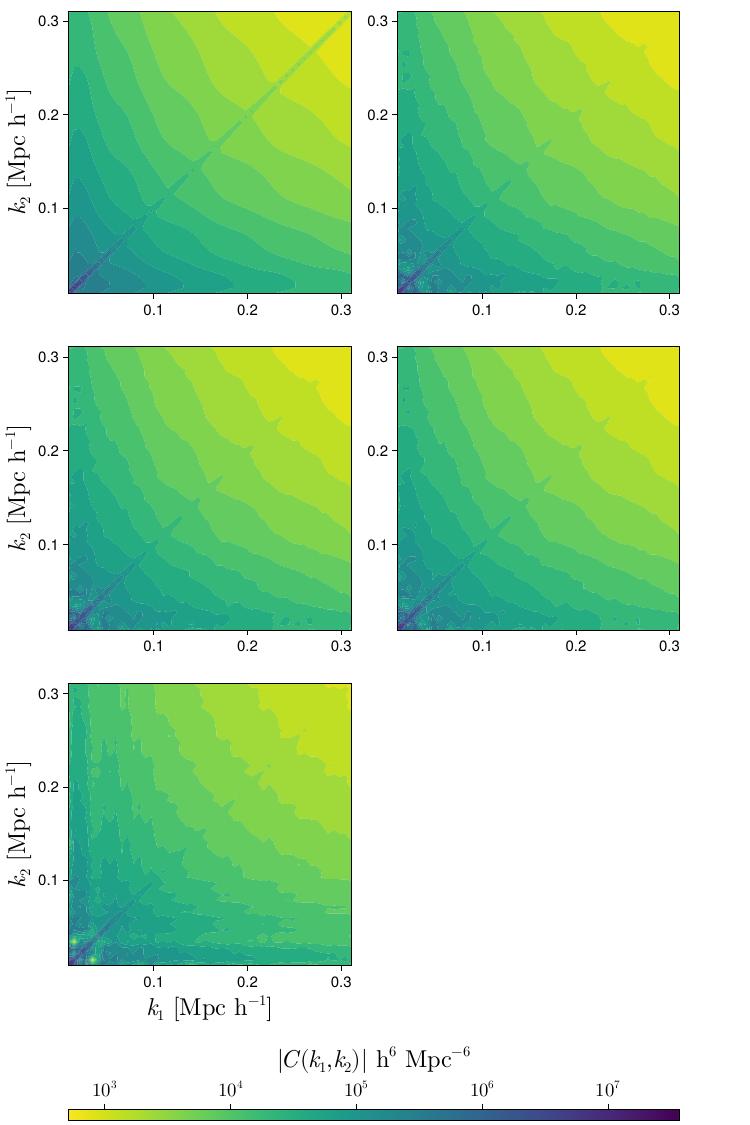}}
  \caption{The same as \Fig{fig:COm}, but this time we show the
    cosmological dependence of the covariance matrix on the
    dimensionless Hubble parameter $h$. The top and bottom sets of
    panels show results for the $h^+$ and $h^-$ variations in
    cosmology, respectively.\label{fig:Ch}}
\end{figure*}

\begin{figure*}
  \centering{
    \includegraphics[width=7.2cm]{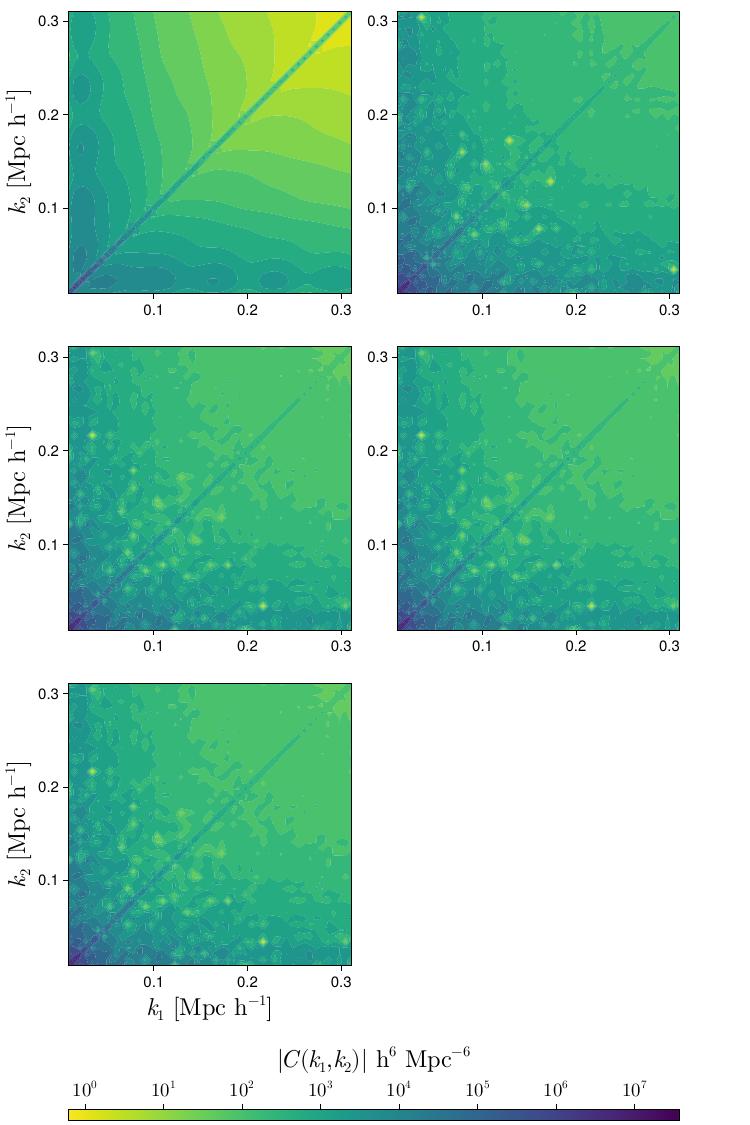}\hspace{0.5cm}
    \includegraphics[width=7.2cm]{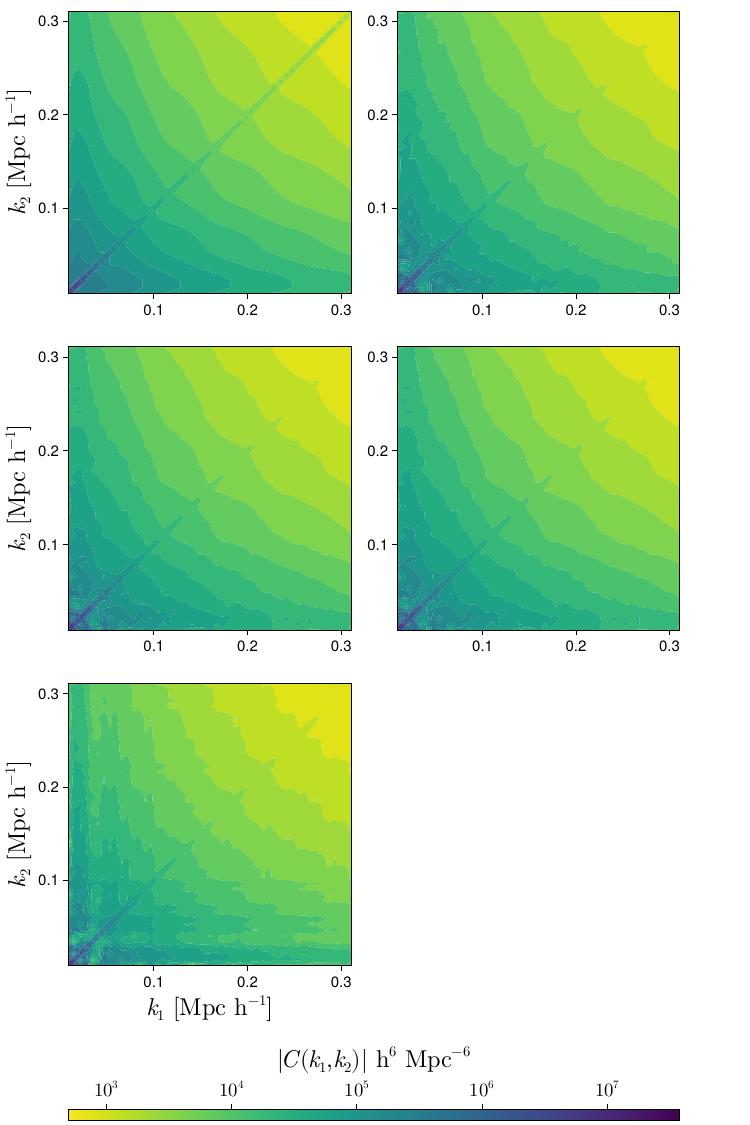}}
  \centering{
    \includegraphics[width=7.2cm]{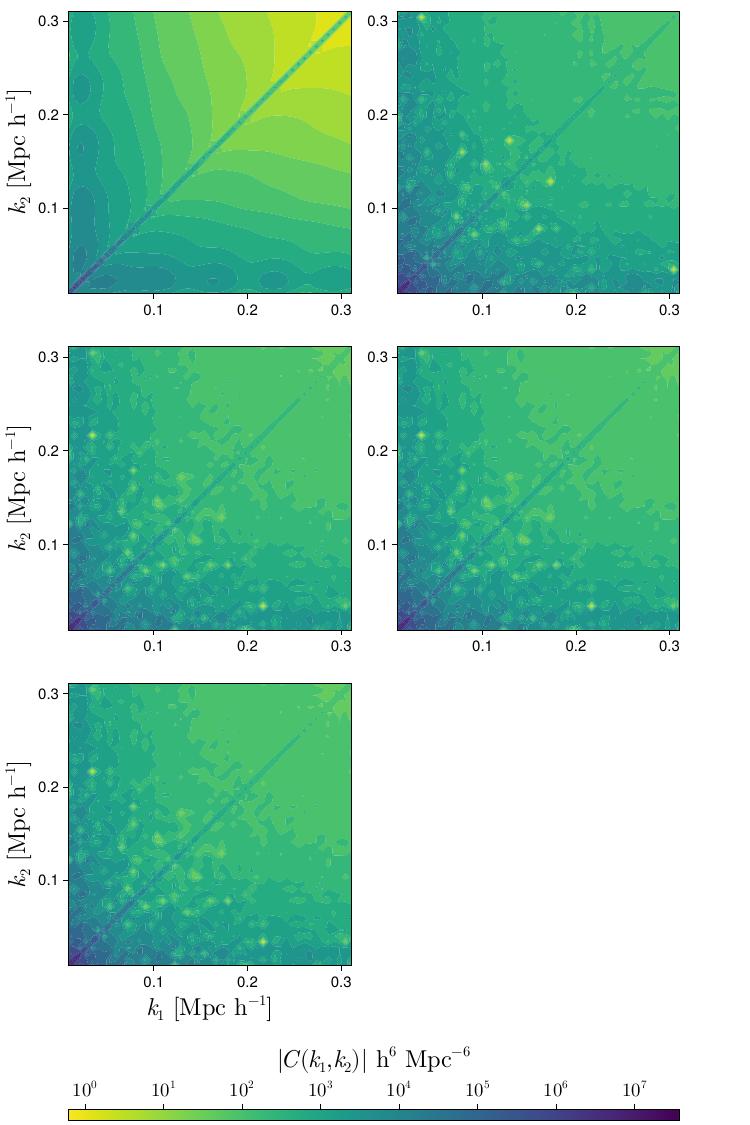}\hspace{0.5cm}
    \includegraphics[width=7.2cm]{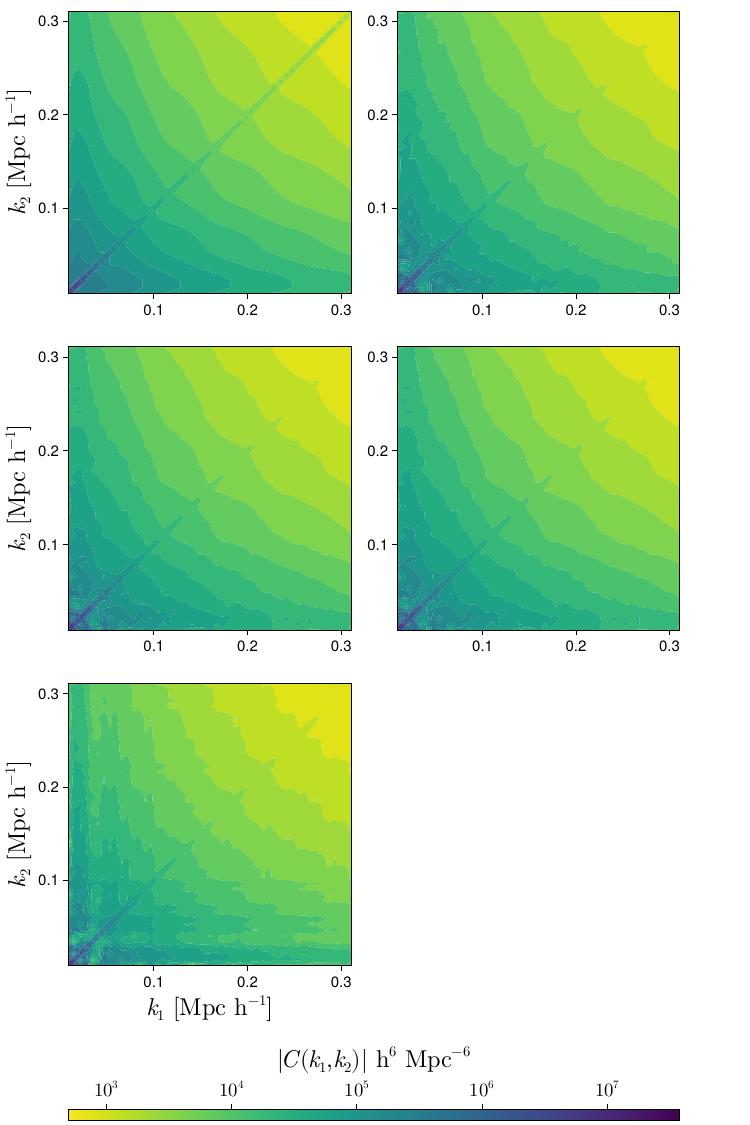}}
  \caption{The same as \Fig{fig:COm}, but this time we show the
    cosmological dependence of the covariance matrix on the
    dimensionless Hubble parameter $w_{0}$. The top and bottom sets of
    panels show results for the $w_{0}^+$ and $w_{0}^-$ variations in
    cosmology, respectively.\label{fig:Ch}}
\end{figure*}


\section{Cosmology dependence of the covariance}\label{sec:reconstruction}


\subsection{Approximating the covariance matrices}

We now turn to the issue of predicting the covariance matrix as a
function of cosmology using our Taylor expansion model. We note that there are three cases that we will
investigate.

\begin{itemize}
\item Case I: we define $C_{\parallel}$ as in \Eqn{eq:Cparallel}. We take
  $\widetilde{\Ymat}$ as given by \Eqn{eq:Ycovmodel1} and we model the
  full response matrix $\Rmat_{\alpha}$ using \Eqn{eq:fullR}.  This
  is the most complete estimator of the covariance matrix, but risks
  including sources of noise that contribute to the results through
  the estimated derivatives and the noisy precision matrix for the
  fiducial model.

\item
  Case II: we define $C_{\parallel,\mathrm{a}}$ by once again making
  use of \Eqn{eq:Cparallel} and \Eqn{eq:Ycovmodel1}, but this time we
  insert the approximate response matrix, as given by \Eqn{eq:aR} when
  computing $\widetilde{\Ymat}$, where we hold the correlation matrix
  constant.  This is less complete than the full $C_{\parallel}$, but
  avoids a potential source of noise in the derivative of the
  correlation matrix and is valid under the assumption that the
  correlation matrix does not vary significantly with the
  parameter under study.
  
\item Case III: we define $C_{\parallel,\mathrm{R}=0}$, found by
  setting the response matrix $\Rmat_{\alpha}$ to zero in
  $\widetilde{\Ymat}$. This provides the simplest and fastest
  estimator for the non-linear covariance matrix and removes all
  sources of noise from the derivatives of the measured and modelled
  matrices with respect to their parameters, but is only valid under
  the assumption that the accuracy of the model is not itself a
  function of the parameters under study, as that
  dependence is precisely what the response matrix measures.
\end{itemize}


\subsection{Validating the model}

In order to validate our model we compare the accuracies of the
different forms of the reconstructed covariance matrix by generating
predictions for $C_{\parallel}$, $C_{\parallel,\mathrm{a}}$, and
$C_{\parallel,\mathrm{R}=0}$ for the fiducial model and each of the
12 variations.  In the event that a given definition of $C_{\parallel}$
is perfect, it would recreate the non-linear covariance matrix
exactly.  In practice, we expect it to contain small errors due to a
combination of numerical errors and the use of a first order Taylor
expansion as an approximation for the true $\Ymat$ matrix.

\vspace{0.2cm}
\noindent $\bullet$ {\bf Fiducial model}: In the bottom row of
\Fig{fig:Cfid} we show the reconstructed covariance matrices for the
fiducial cosmology. The left two columns exclude super-sample
covariance and the right two panels include it. For these cases, the
bottom left panel gives $C_{\parallel}$ and the bottom right gives
$C_{\parallel,\mathrm{a}}$, respectively.  As can be seen, in this
case, both reconstructed matrices recover the fiducial covariance
matrix to high precision. This is reassuring, but entirely
unsurprising, since $\Delta\theta_{\alpha}=0$ and hence
$\widetilde{\Ymat(\bm\theta_0)}=\Ymat(\bm\theta_0)$, and the Taylor
expansion is turned off.

\vspace{0.2cm}
\noindent $\bullet$ {\bf Cosmological variations}: we now turn to the
more interesting case and examine how well the approach can
reconstruct the covariance matrices of the variations in cosmology
simulated in the Quijote runs. In
Figures~\ref{fig:COm}-\ref{fig:Ch} we show the results for variations
in the parameters $\{\Omega_{\rm m},\Omega_{\rm b},n_{\rm
  s},\sigma_8,h\}$. The pairs of panels on the left and right show the
results for the case without and with super-sample covariance,
respectively. In each figure, the top row of panels show results for
the positive variation in the parameter in question and the bottom
sets show the results for the negative variation. In each set of five
panels: the top left panel shows the SPT model covariance matrix; the
top right panel shows the non-linear covariance matrix measured from
the 500 simulations in the Quijote suite; the centre left panel shows
$C_{\parallel}$, reconstructed with the full response matrix; the
centre right panel shows $C_{\parallel,\mathrm{a}}$, where the
correlation matrix is assumed to be independent of cosmology; the
bottom panel shows $C_{\parallel,\mathrm{R}=0}$, where the response
matrix has been set to zero.

We see that, overall, the models perform very well. For the
reconstructions $C_{\parallel}$ and $C_{\parallel,\mathrm{a}}$, we
notice that there are only small differences with and without the SSC
correction. This suggests that the approximation that the correlation
matrix is roughly constant over variations in the cosmological
parameters is a reasonable one. For the case
$C_{\parallel,\mathrm{R}=0}$ we see that this, in some cases, is also
an adequate description. There are however some more noticeable
deviations from the true measured matrix, especially for the cases
with SSC corrections.  Here we see deviations at low wavemodes,
indicating a strong parameter dependent disparity between SPT and
non-linear covariances on large scales. The figures indicate that the
full response matrix provides the most accurate reconstruction of the
original non-linear covariance matrix.

In Fig.~\ref{fig:ROm}-\ref{fig:Rh} we show the reduced covariance matrix
for all cosmologies under study in the context of SPT, non-linear simulation results, and both the full and approximate forms of the reconstructed covariance matrix.  Since the Gaussian terms dominate the diagonal, this serves primarily to highlight the shape of the off-diagonal terms and to permit comparison of these shapes between the various models and the non-linear results.  As can clearly be seen in all cases, the reconstructed matrices are substantially more accurate to the non-linear data than the SPT matrix, both with and without SSC corrections.  Indeed, the SPT model fails to capture a substantial amount of the off-diagonal terms at $k\sim 0.15~h~\mathrm{Mpc}^{-1}$ which are captured most accurately with the full response matrix and to a lesser extent, at least in some cases, with the approximate response matrix.


\begin{figure*}
  \centering
    \includegraphics[width=0.33\textwidth]{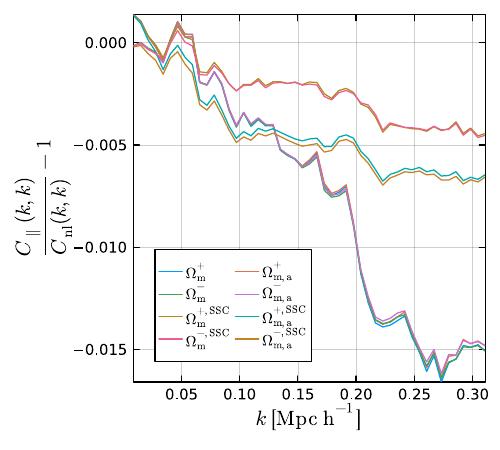}
    \includegraphics[width=0.33\textwidth]{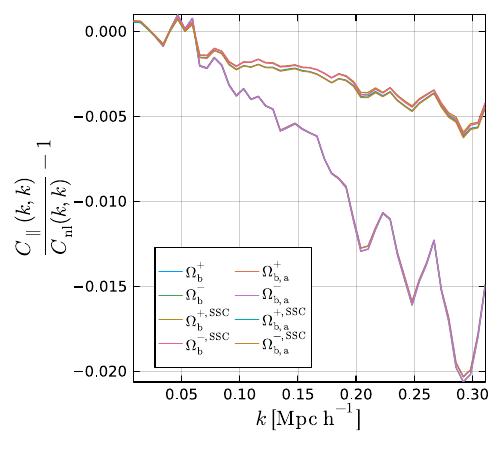}
    \includegraphics[width=0.33\textwidth]{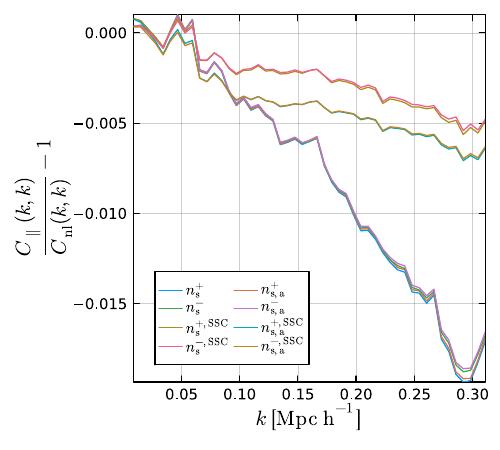}
    \includegraphics[width=0.33\textwidth]{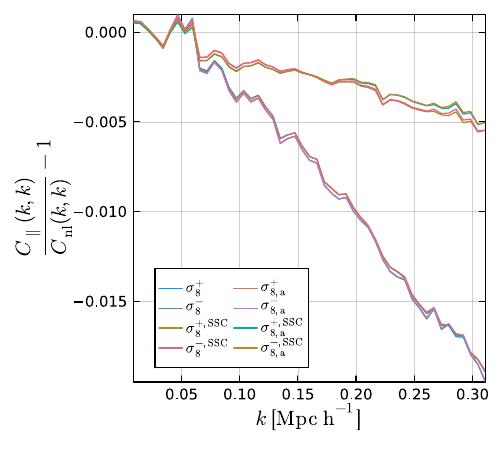}
    \includegraphics[width=0.33\textwidth]{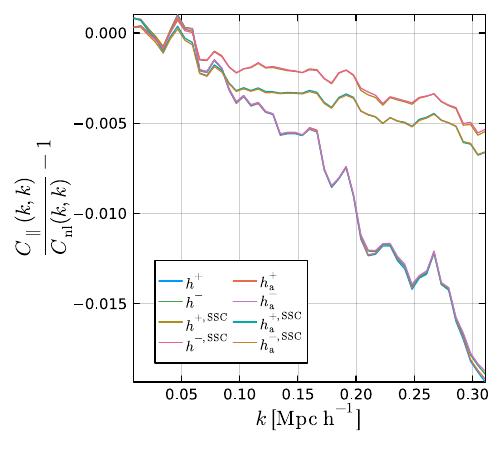}
    \includegraphics[width=0.33\textwidth]{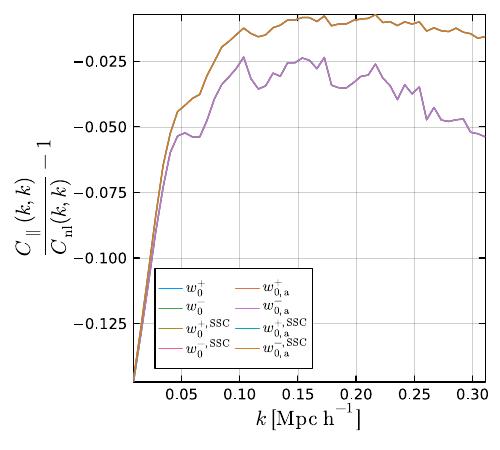}
  \caption{Each of the panels shows the residuals of the ratios of the
    diagonal elements of the reconstructed covariance matrix with the
    non-linear covariance matrices. Going clockwise from the top right
    panel we show results for the positive and negative variations in
    $\Omega_{\rm m}$, $\Omega_{\rm b}$, $n_{\rm s}$, $\sigma_8$, 
    $h$, and $w_{0}$. In each panel we show the results for the case
    $C_{\parallel}$ and $C_{\parallel,\mathrm{a}}$. }
  \label{fig:Drats}
\end{figure*}
\begin{figure*}
\centering
    \begin{minipage}{0.48\textwidth}\includegraphics[width=\textwidth]{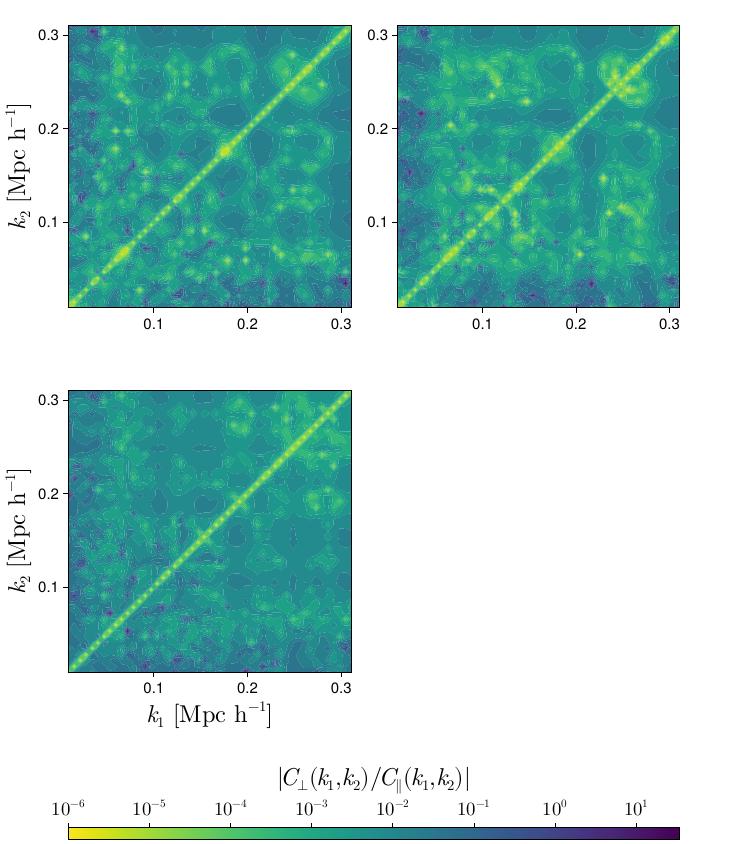}\end{minipage}
    \begin{minipage}{0.48\textwidth}\includegraphics[width=\textwidth]{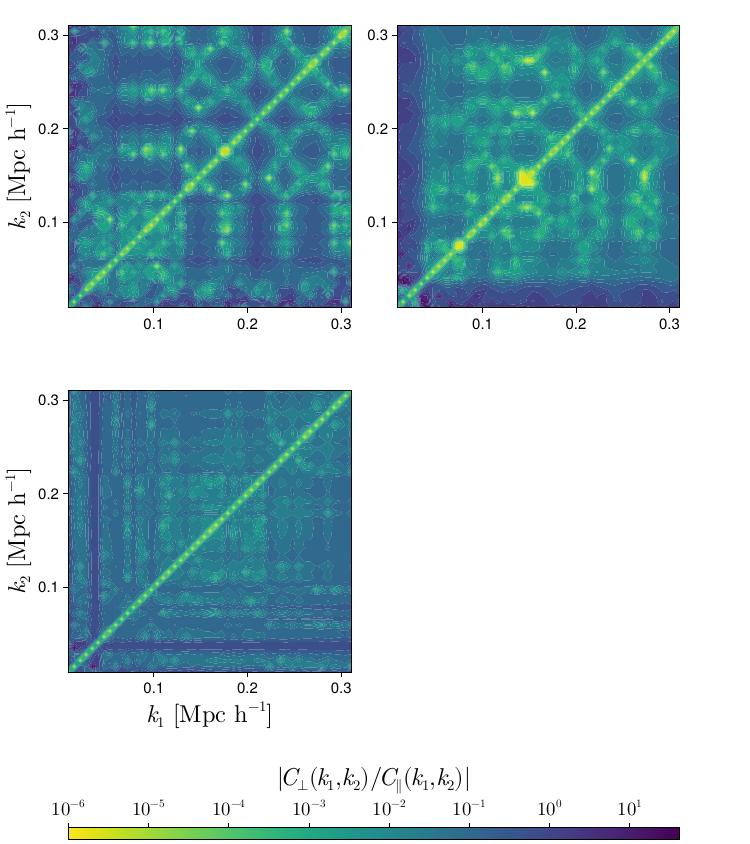}\end{minipage}
    \caption{The elementwise ratios of $C_{\perp}$ and $C_{\parallel}$ for the reduced case of $\Omega_{\mathrm{m}}$ without (left hand panel) and with (right hand panel) SSC corrections.  In both cases, the top left image shows the calculation made with the full response matrix, the top right image shows the calculation made with the constant correlation matrix approximation, and the bottom image shows the calculation made with the fiducial response matrix approximation.}
    \label{fig:CparatOmm}
\end{figure*}
\begin{figure*}
\centering
    \begin{minipage}{0.48\textwidth}\includegraphics[width=\textwidth]{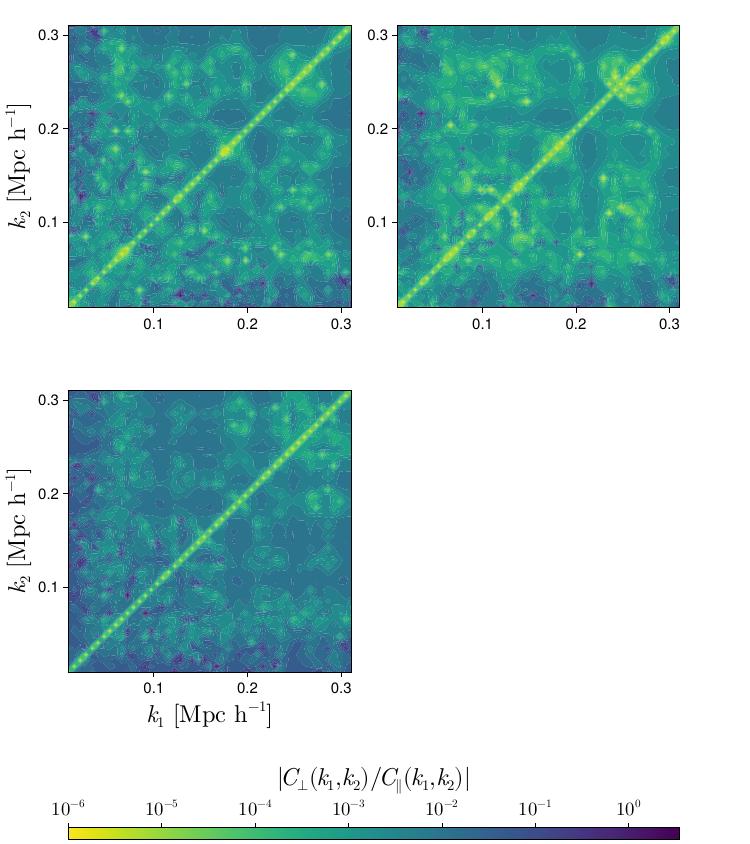}\end{minipage}
    \begin{minipage}{0.48\textwidth}\includegraphics[width=\textwidth]{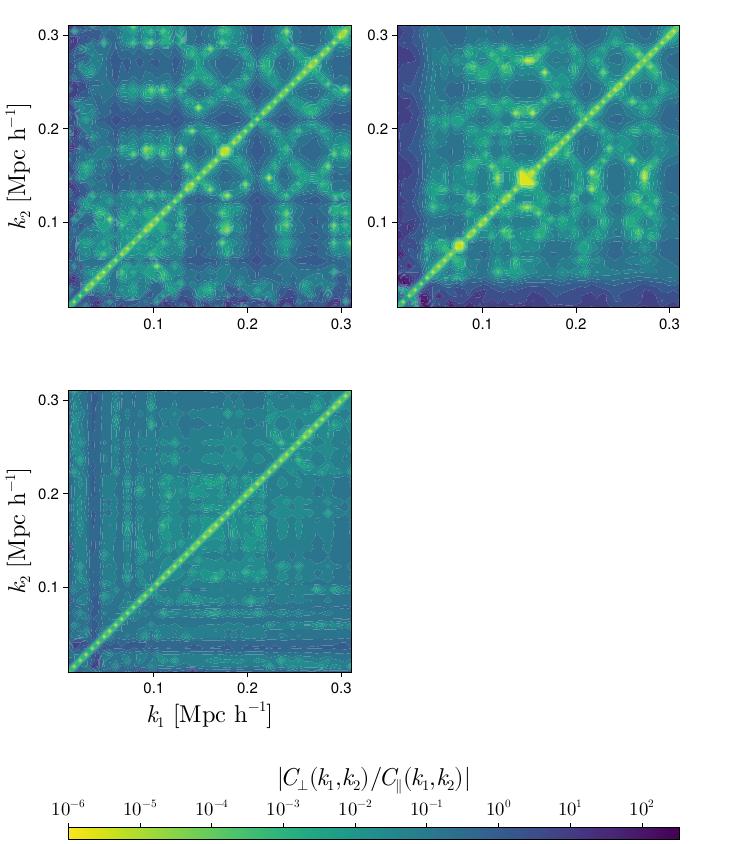}\end{minipage}
    \caption{The elementwise ratios of $C_{\perp}$ and $C_{\parallel}$ for the increased case of $\Omega_{\mathrm{m}}$ without (left hand panel) and with (right hand panel) SSC correction.  In both cases, the top left image shows the calculation made with the full response matrix, the top right image shows the calculation made with the constant correlation matrix approximation, and the bottom image shows the calculation made with the fiducial response matrix approximation.}
    \label{fig:CparatOmp}
\end{figure*}
\begin{figure*}
\centering
    \begin{minipage}{0.48\textwidth}\includegraphics[width=\textwidth]{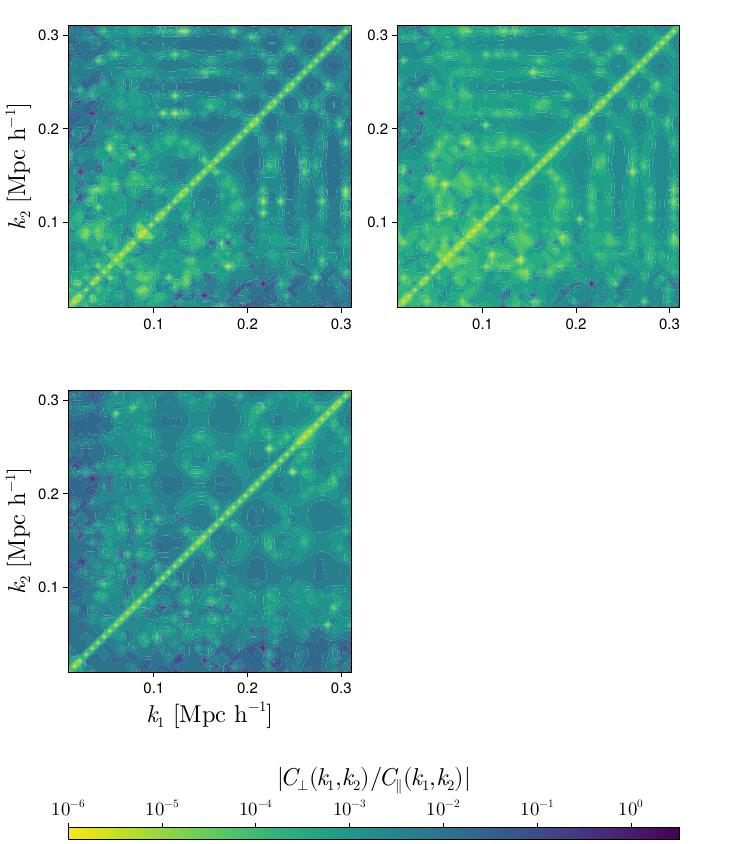}\end{minipage}
    \begin{minipage}{0.48\textwidth}\includegraphics[width=\textwidth]{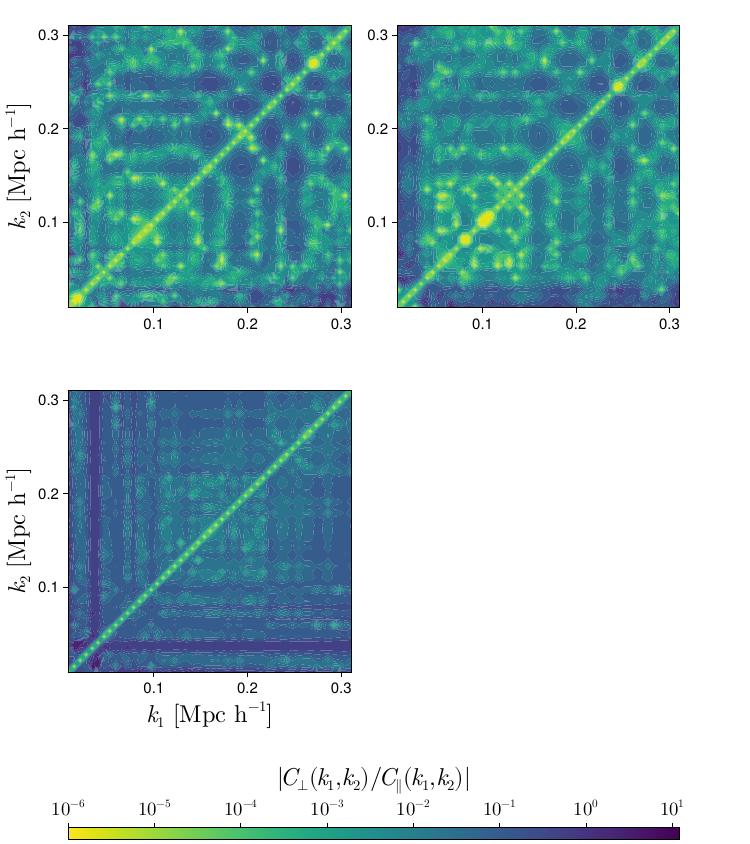}\end{minipage}
    \caption{The elementwise ratios of $C_{\perp}$ and $C_{\parallel}$ for the reduced case of $\Omega_{\mathrm{m}}$ without (left hand panel) and with (right hand panel) SSC correction.  In both cases, the top left image shows the calculation made with the full response matrix, the top right image shows the calculation made with the constant correlation matrix approximation, and the bottom image shows the calculation made with the fiducial response matrix approximation.}
    \label{fig:CparatObm}
\end{figure*}
\begin{figure*}
\centering
    \begin{minipage}{0.48\textwidth}\includegraphics[width=\textwidth]{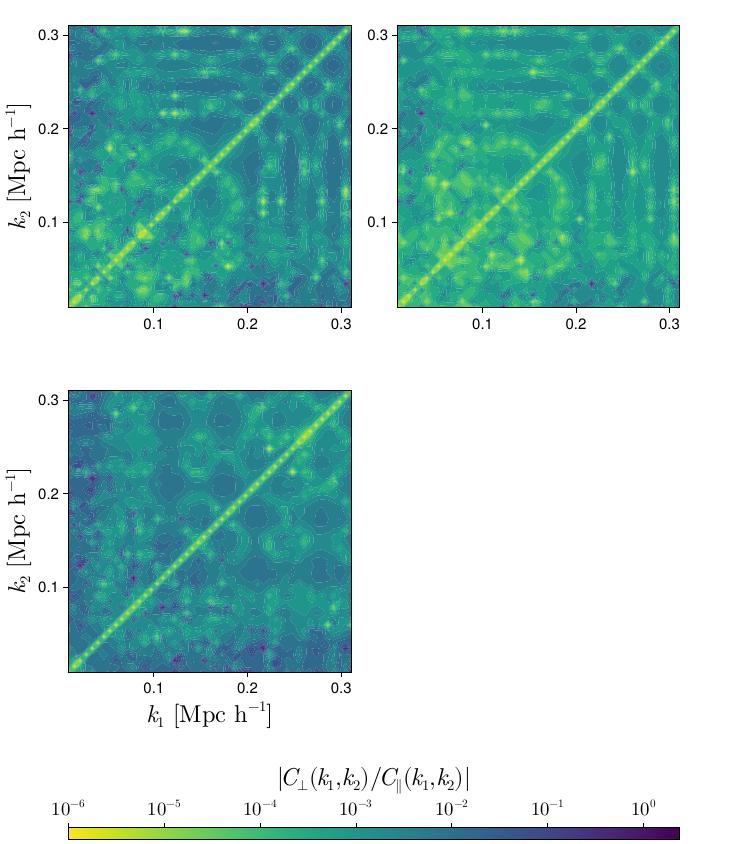}\end{minipage}
    \begin{minipage}{0.48\textwidth}\includegraphics[width=\textwidth]{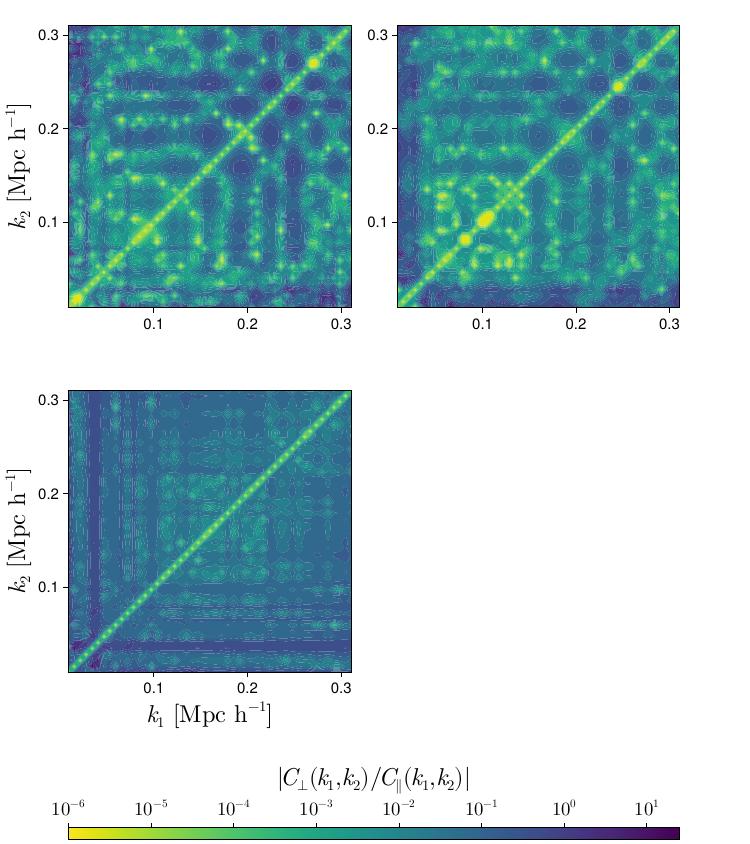}\end{minipage}
    \caption{The elementwise ratios of $C_{\perp}$ and $C_{\parallel}$ for the increased case of $\Omega_{\mathrm{b}}$ without (left hand panel) and with (right hand panel) SSC correction.  In both cases, the top left image shows the calculation made with the full response matrix, the top right image shows the calculation made with the constant correlation matrix approximation, and the bottom image shows the calculation made with the fiducial response matrix approximation.}
    \label{fig:CparatObp}
\end{figure*}
\begin{figure*}
\centering
    \begin{minipage}{0.48\textwidth}\includegraphics[width=\textwidth]{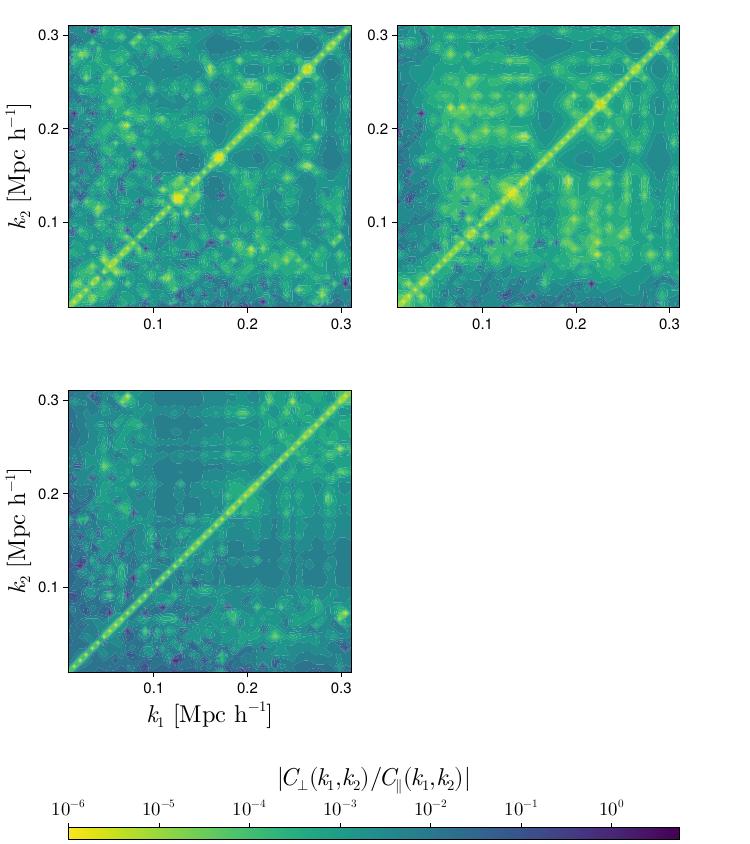}\end{minipage}
    \begin{minipage}{0.48\textwidth}\includegraphics[width=\textwidth]{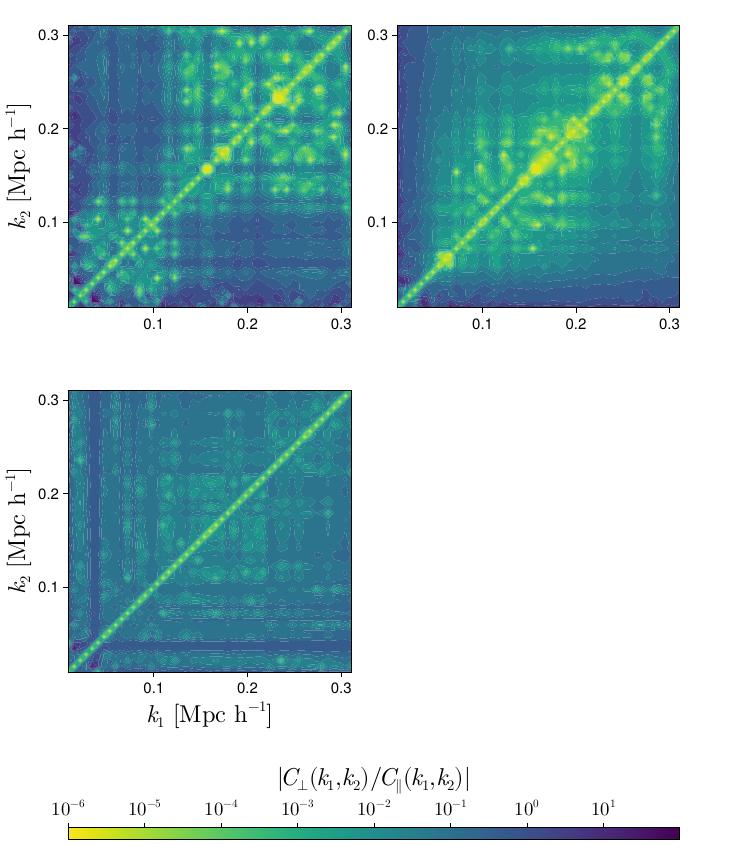}\end{minipage}
    \caption{The elementwise ratios of $C_{\perp}$ and $C_{\parallel}$ for the reduced case of $n_{s}$ without (left hand panel) and with (right hand panel) SSC correction.  In both cases, the top left image shows the calculation made with the full response matrix, the top right image shows the calculation made with the constant correlation matrix approximation, and the bottom image shows the calculation made with the fiducial response matrix approximation.}
    \label{fig:CparatOmm}
\end{figure*}
\begin{figure*}
\centering
    \begin{minipage}{0.48\textwidth}\includegraphics[width=\textwidth]{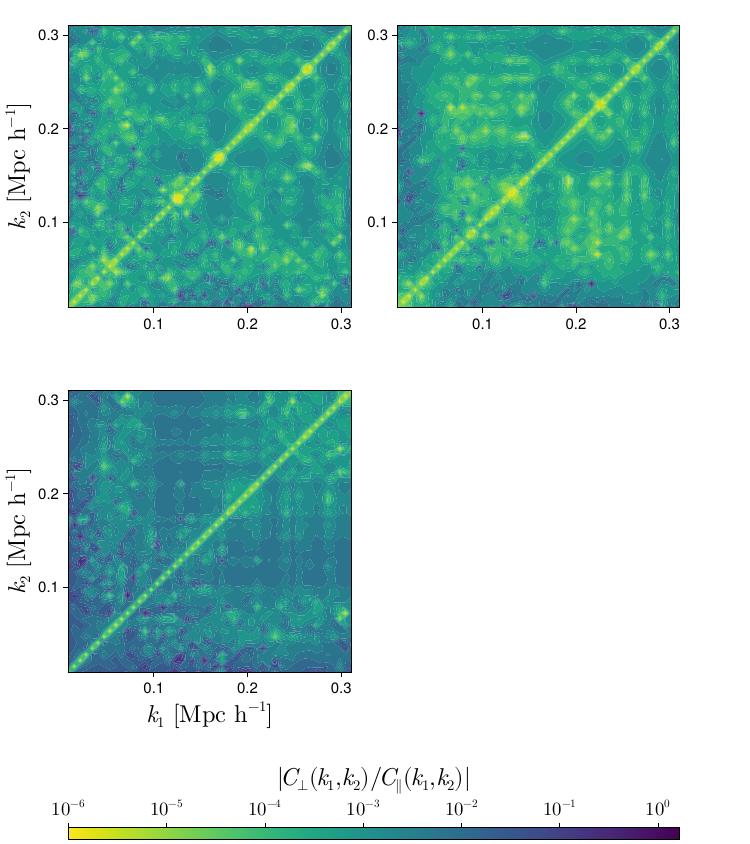}\end{minipage}
    \begin{minipage}{0.48\textwidth}\includegraphics[width=\textwidth]{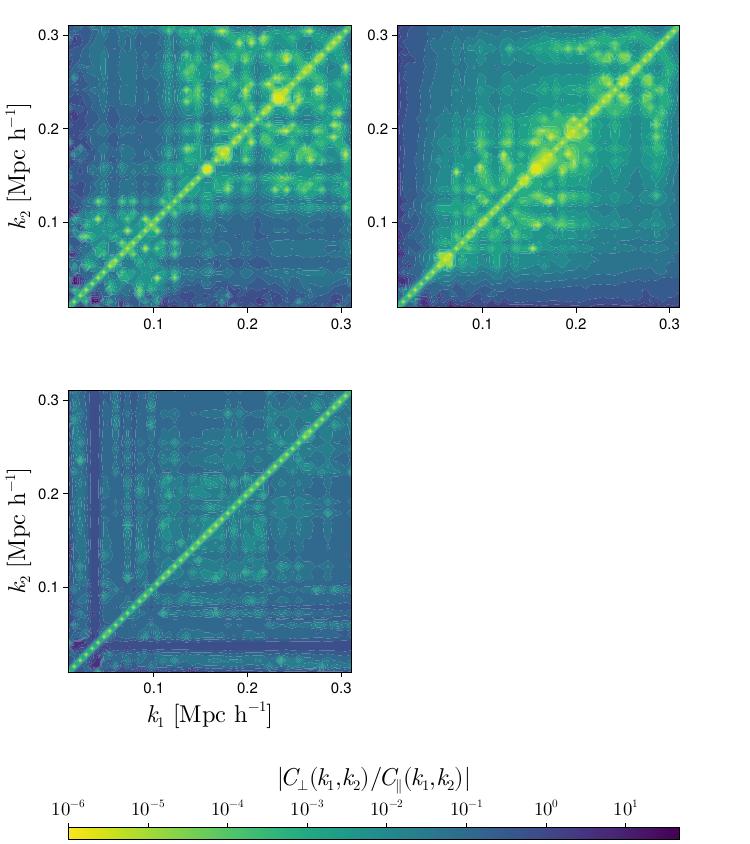}\end{minipage}
    \caption{The elementwise ratios of $C_{\perp}$ and $C_{\parallel}$ for the increased case of $n_{s}$ without (left hand panel) and with (right hand panel) SSC correction.  In both cases, the top left image shows the calculation made with the full response matrix, the top right image shows the calculation made with the constant correlation matrix approximation, and the bottom image shows the calculation made with the fiducial response matrix approximation.}
    \label{fig:CparatOmp}
\end{figure*}
\begin{figure*}
\centering
    \begin{minipage}{0.48\textwidth}\includegraphics[width=\textwidth]{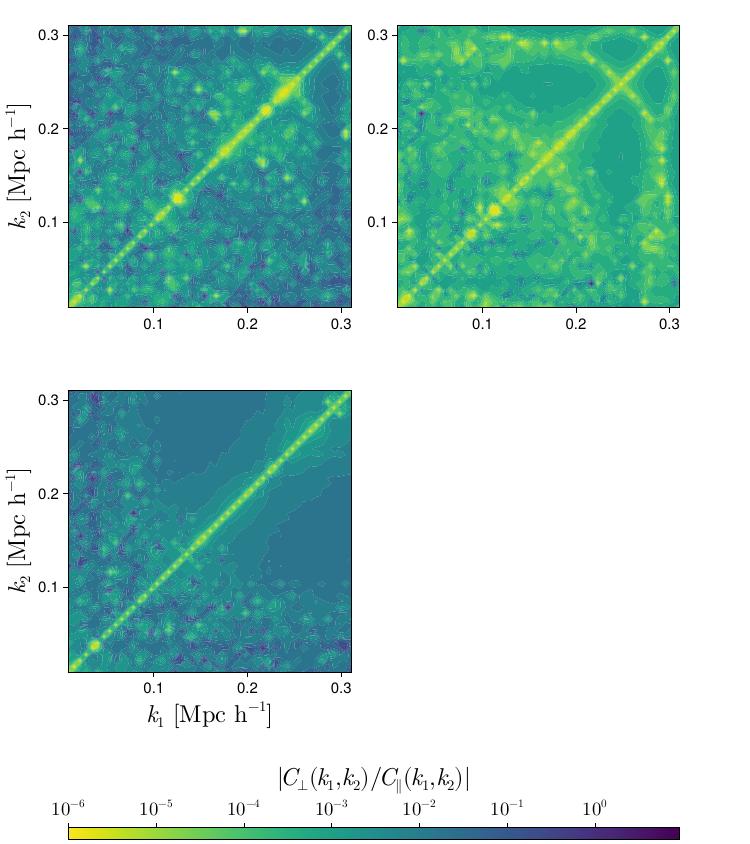}\end{minipage}
    \begin{minipage}{0.48\textwidth}\includegraphics[width=\textwidth]{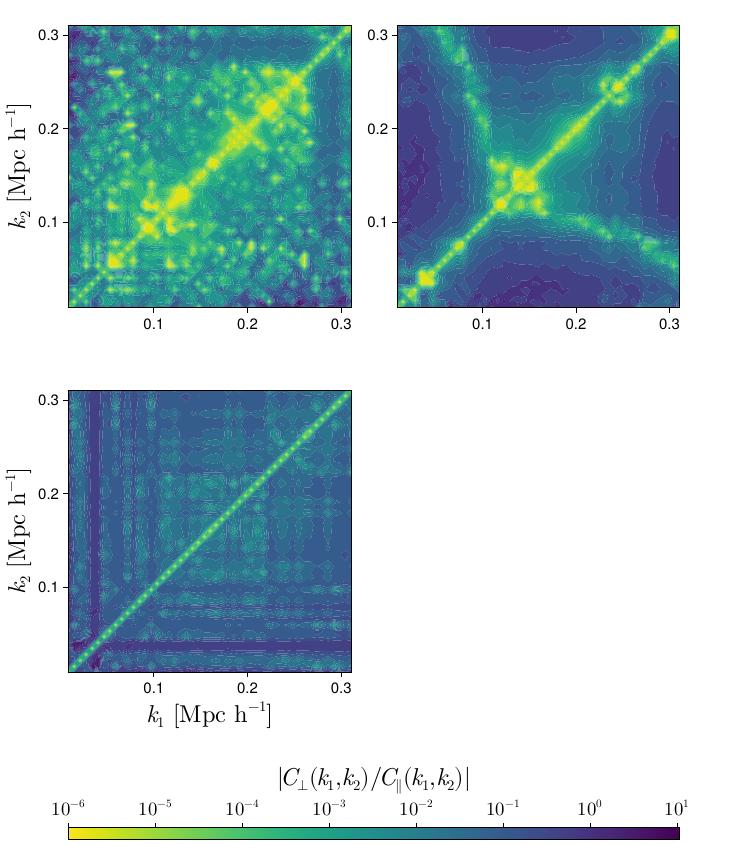}\end{minipage}
    \caption{The elementwise ratios of $C_{\perp}$ and $C_{\parallel}$ for the reduced case of $\sigma_{8}$ without (left hand panel) and with (right hand panel) SSC correction.  In both cases, the top left image shows the calculation made with the full response matrix, the top right image shows the calculation made with the constant correlation matrix approximation, and the bottom image shows the calculation made with the fiducial response matrix approximation.}
    \label{fig:Cparats8m}
\end{figure*}
\begin{figure*}
\centering
    \begin{minipage}{0.48\textwidth}\includegraphics[width=\textwidth]{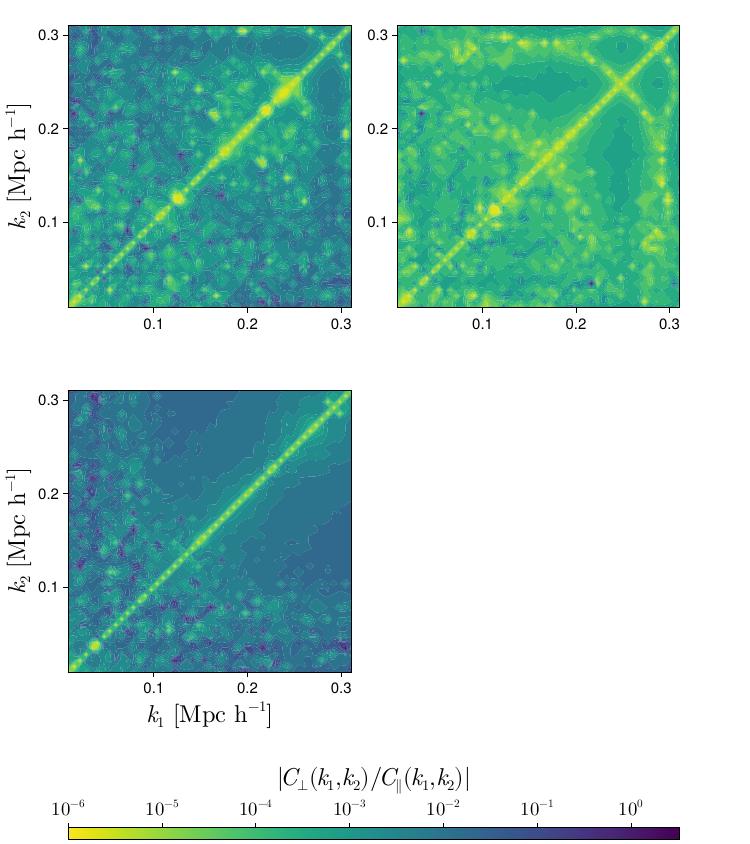}\end{minipage}
    \begin{minipage}{0.48\textwidth}\includegraphics[width=\textwidth]{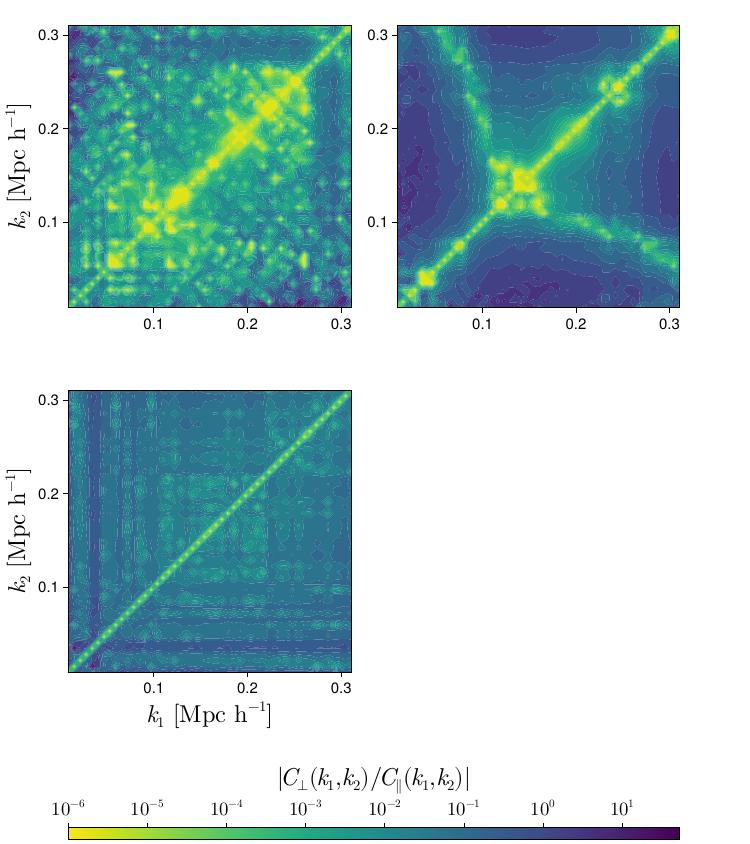}\end{minipage}
    \caption{The elementwise ratios of $C_{\perp}$ and $C_{\parallel}$ for the increased case of $\sigma_{8}$ without (left hand panel) and with (right hand panel) SSC correction.  In both cases, the top left image shows the calculation made with the full response matrix, the top right image shows the calculation made with the constant correlation matrix approximation, and the bottom image shows the calculation made with the fiducial response matrix approximation.}
    \label{fig:Cparats8p}
\end{figure*}
\begin{figure*}
\centering
    \begin{minipage}{0.48\textwidth}\includegraphics[width=\textwidth]{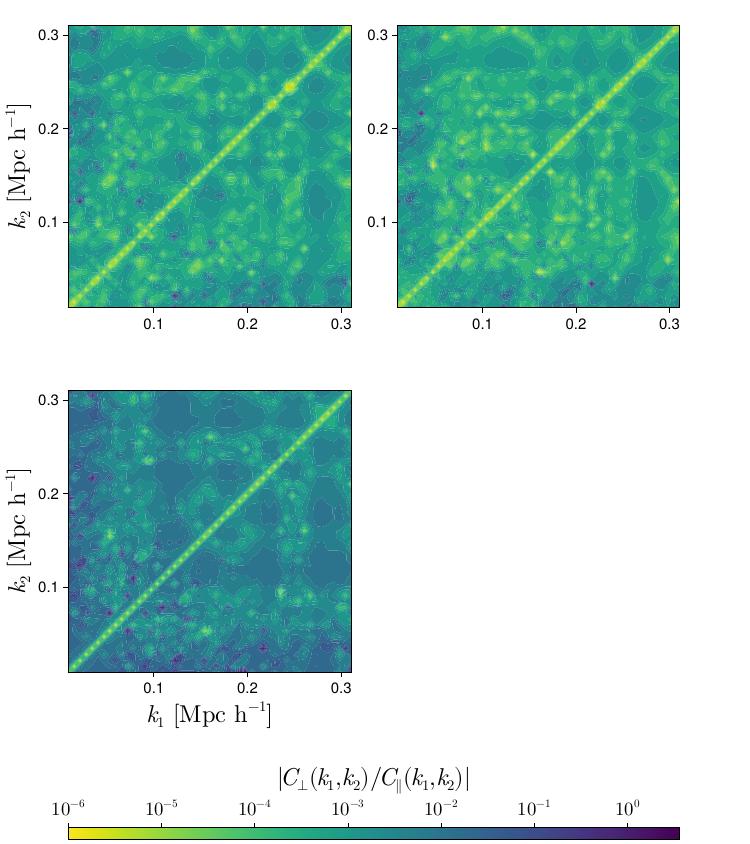}\end{minipage}
    \begin{minipage}{0.48\textwidth}\includegraphics[width=\textwidth]{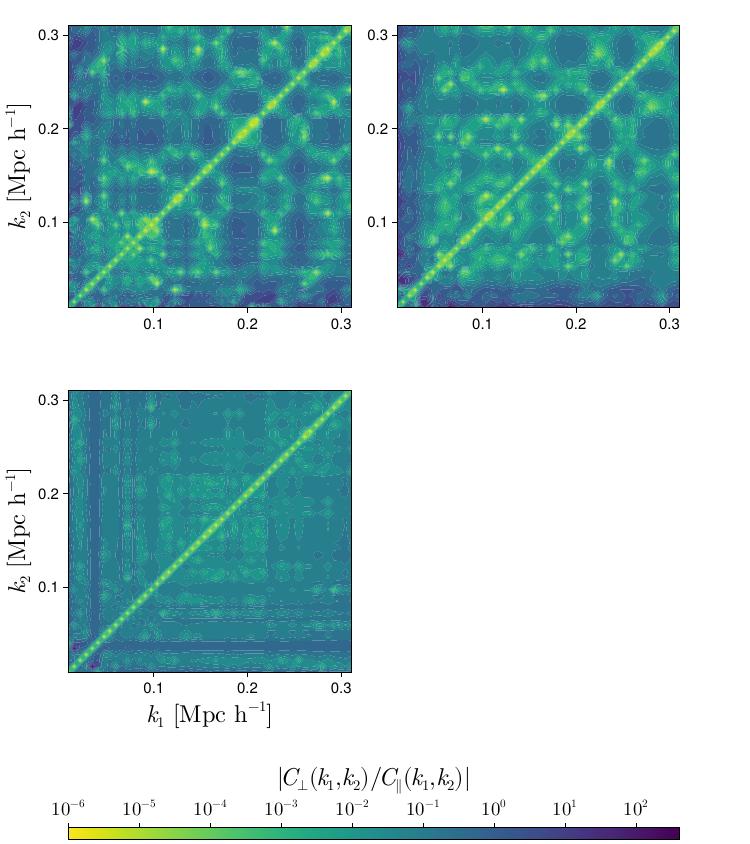}\end{minipage}
    \caption{The elementwise ratios of $C_{\perp}$ and $C_{\parallel}$ for the reduced case of $h$ without (left hand panel) and with (right hand panel) SSC correction.  In both cases, the top left image shows the calculation made with the full response matrix, the top right image shows the calculation made with the constant correlation matrix approximation, and the bottom image shows the calculation made with the fiducial response matrix approximation.}
    \label{fig:Cparathm}
\end{figure*}
\begin{figure*}
\centering
    \begin{minipage}{0.48\textwidth}\includegraphics[width=\textwidth]{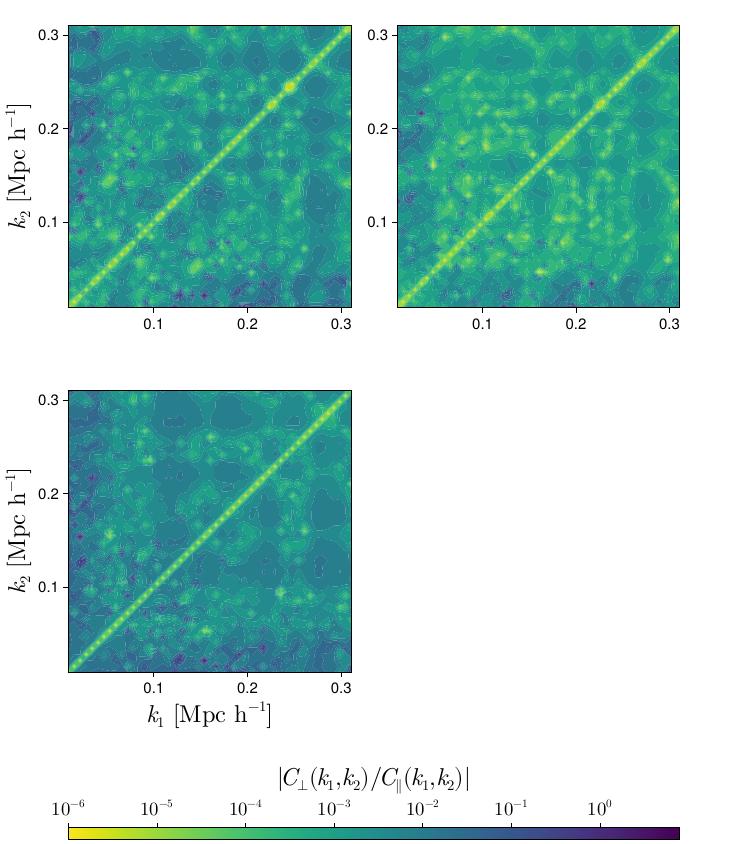}\end{minipage}
    \begin{minipage}{0.48\textwidth}\includegraphics[width=\textwidth]{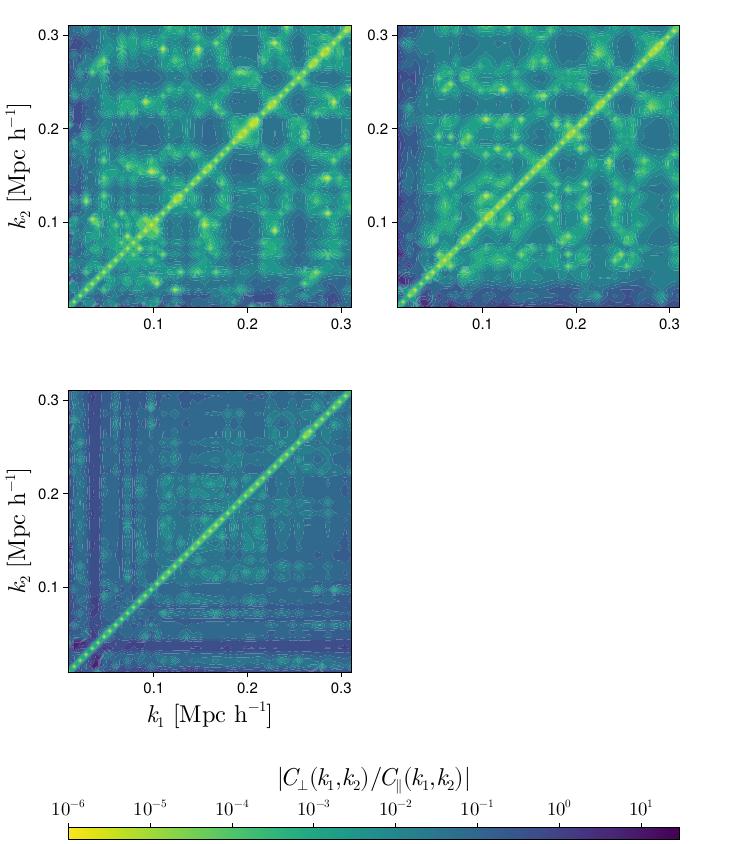}\end{minipage}
    \caption{The elementwise ratios of $C_{\perp}$ and $C_{\parallel}$ for the increased case of $h$ without (left hand panel) and with (right hand panel) SSC correction.  In both cases, the top left image shows the calculation made with the full response matrix, the top right image shows the calculation made with the constant correlation matrix approximation, and the bottom image shows the calculation made with the fiducial response matrix approximation.}
    \label{fig:Cparathp}
\end{figure*}

\begin{figure*}
\centering
    \begin{minipage}{0.48\textwidth}\includegraphics[width=\textwidth]{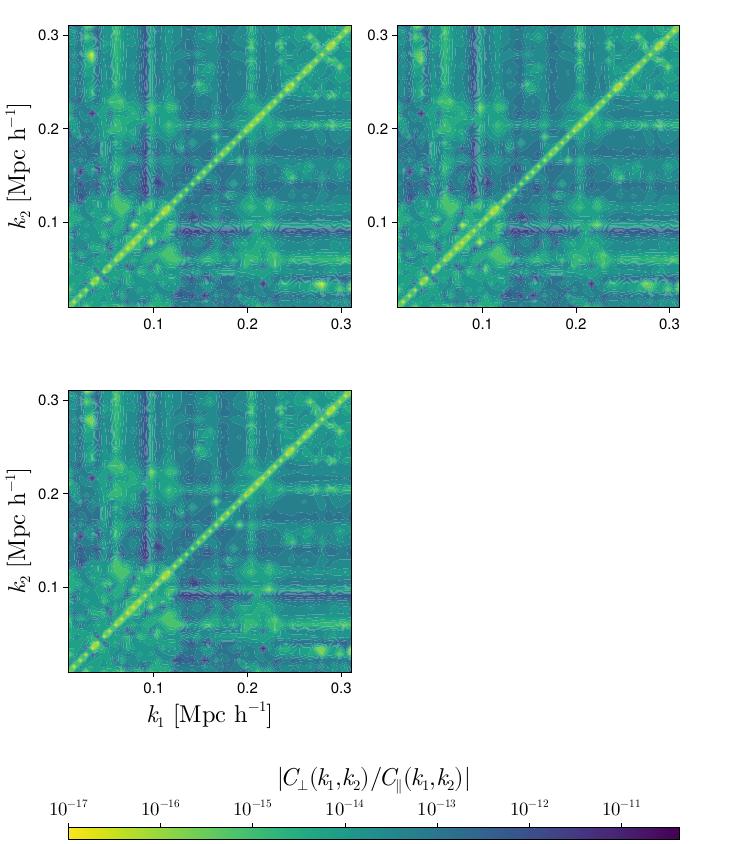}\end{minipage}
    \begin{minipage}{0.48\textwidth}\includegraphics[width=\textwidth]{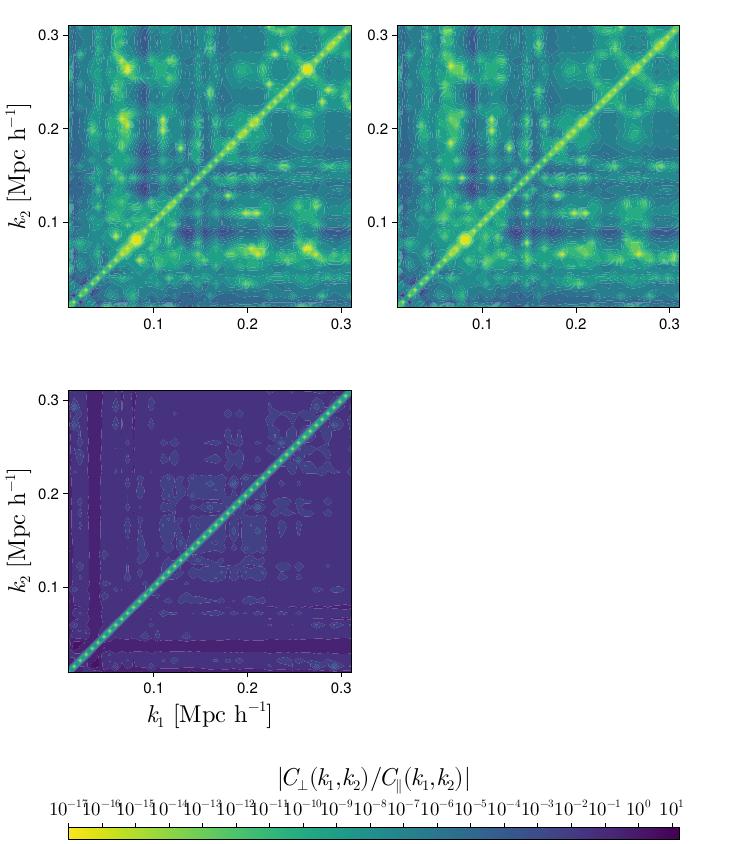}\end{minipage}
    \caption{The elementwise ratios of $C_{\perp}$ and $C_{\parallel}$ for the reduced case of $w0$ without (left hand panel) and with (right hand panel) SSC correction.  In both cases, the top left image shows the calculation made with the full response matrix, the top right image shows the calculation made with the constant correlation matrix approximation, and the bottom image shows the calculation made with the fiducial response matrix approximation.}
    \label{fig:Cparatw0m}
\end{figure*}

\begin{figure*}
\centering
    \begin{minipage}{0.48\textwidth}\includegraphics[width=\textwidth]{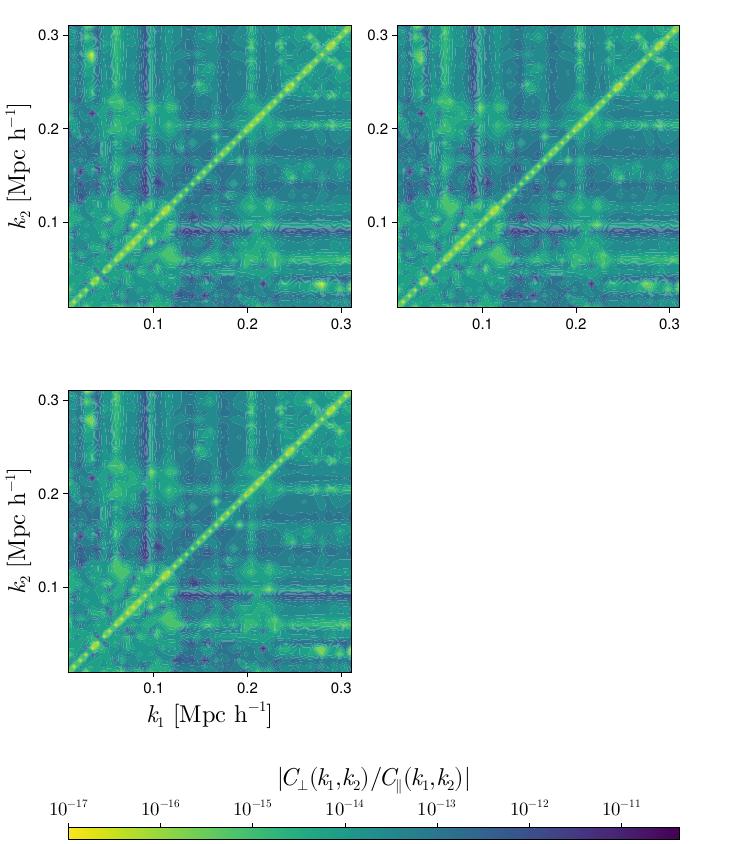}\end{minipage}
    \begin{minipage}{0.48\textwidth}\includegraphics[width=\textwidth]{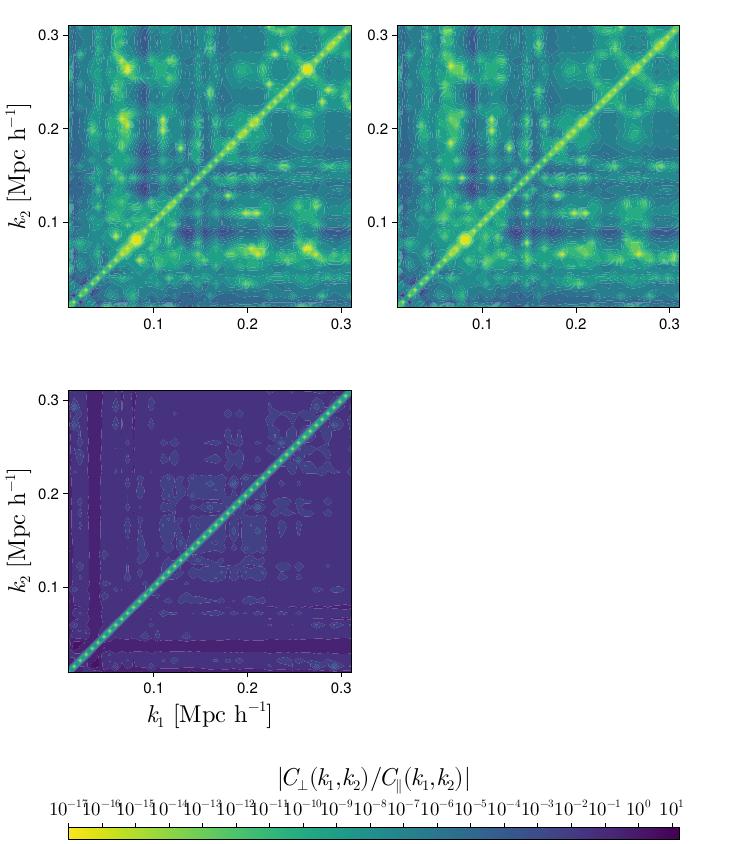}\end{minipage}
    \caption{The elementwise ratios of $C_{\perp}$ and $C_{\parallel}$ for the increased case of $w_{0}$ without (left hand panel) and with (right hand panel) SSC correction.  In both cases, the top left image shows the calculation made with the full response matrix, the top right image shows the calculation made with the constant correlation matrix approximation, and the bottom image shows the calculation made with the fiducial response matrix approximation.}
    \label{fig:Cparatw0p}
\end{figure*}

\subsection{Residuals of the variances}

In \Fig{fig:Drats} we show the residual of the ratios of the diagonal
elements of $C_{\parallel}$ and $C_{\parallel,\mathrm{a}}$ with
respect to the true measured nonlinear covariance matrices from the
Quijote simulations. These plots show quantitatively the
deviations between the measurements and the model.  As can be seen,
the residual is $\lesssim6\%$ in all cases for $k\lesssim0.3\,\kMpc$,
indicating a high level of accuracy in our model. On larger scales,
$k\lesssim0.15\,\kMpc$, the model predictions are $\lesssim1\%$, with
the exception of the variations in $h$, which show slightly larger
deviations. Furthermore, we see in the cases of the parameters
$\Omega_{\mathrm{m}}$, $\Omega_{\mathrm{b}}$, and $n_{\mathrm{s}}$,
the full response matrix consistently gives more accurate results than
the approximate response matrix, though both achieve a good level of
accuracy up to the limit of the range of wavemodes that we study.
This indicates that the correlation matrix has a non-negligible rate
of change with respect to these parameters.  In the cases of
$\sigma_{8}$ and $h$, there is not such a clear disparity between the
full and approximate response matrix results, indicating that the
correlation matrix has a minimal dependence on those parameters
compared to the rate of change of the diagonal.


\subsection{Reconstructed matrix accuracy}

As noted in \S\ref{sec:model} one of the deficiencies of our approach
is that the modelled covariance matrix $\Cov$ is not guaranteed to be
symmetric. We force this symmetry by constructing $\Cov_{\parallel}$
and $\Cov_{\perp}$. 

Another test of the accuracy of our method
then is to compute the ratios of the Frobenius norms of the
perpendicular and parallel matrices: $\| C_{\perp}\|_{\rm F} / \|
C_{\parallel}\|_{\rm F}$. Any large deviations from zero would
indicate a breakdown of the method and the invalidity of the Taylor
expansion approach. We recall that the Frobenius norm is defined as
\be \|\Cov\|_{\rm F} = \sqrt{\sum_{i}\sum_j |C_{ij}|^2} \ .\ee

These results are presented in Table~\ref{tab}. All of the results show that the reconstructed matrices
show no strong indication of a significant $\Cov_{\perp}$
contribution. In most cases, we see that $\|\Cov_{\parallel}\|_{\rm
  F}$ is multiple orders of magnitude greater than
$\|\Cov_{\perp}\|_{\rm F}$. We note that the reconstructions with the
response matrix set to zero (the elements labelled $R_{0}$ in the table)
yielded the least accurate results.  This indicates that the response
matrix is generally important for the capturing the true cosmology
dependence of the covariance matrix. Oddly, for $h$ in the presence of
SSC, actually setting $R_{0}=0$ appears to improve the accuracy of the
model. This may indicate issues with residual noise in this case.

As a further inspection in Figs.~\ref{fig:CparatOmm}-\ref{fig:Cparatw0p}
we present the element-wise ratios of
$C_{\perp}(k_{1},k_{2})/C_{\parallel}(k_{1},k_{2})$ for the range of
cosmological models that we consider.


\begin{table*}
\begin{tabular}{|c|c|c|c|c|c|c|}
\hline
Cosmology & Full & Approx & $R_{0}$ & Full SSC & Approx SSC & $R_{0}$ SSC\\\hline
$\Omega_{\mathrm{m}}^{-}$ &
$2.586 \times 10^{-5}$ &
$2.953 \times 10^{-5}$ &
$1.33 \times 10^{-3}$ &
$2.647 \times 10^{-5}$ &
$8.707 \times 10^{-5}$ &
$2.199 \times 10^{-2}$\\\hline
$\Omega_{\mathrm{m}}^{+}$ &
$5.464 \times 10^{-5}$ &
$6.096 \times 10^{-5}$ &
$1.345 \times 10^{-3}$ &
$5.816 \times 10^{-5}$ &
$1.011 \times 10^{-4}$ &
$2.265 \times 10^{-2}$\\\hline
$\Omega_{\mathrm{b}}^{-}$ &
$7.032 \times 10^{-6}$ &
$8.397 \times 10^{-6}$ &
$3.543 \times 10^{-4}$ &
$5.455 \times 10^{-6}$ &
$1.347 \times 10^{-5}$ &
$2.241 \times 10^{-2}$\\\hline
$\Omega_{\mathrm{b}}^{+}$ &
$6.89 \times 10^{-6}$ &
$8.135 \times 10^{-6}$ &
$3.423 \times 10^{-4}$ &
$5.374 \times 10^{-6}$ &
$1.34 \times 10^{-5}$ &
$2.215 \times 10^{-2}$\\\hline
$n_{s}^{-}$ &
$1.028 \times 10^{-5}$ &
$8.920 \times 10^{-6}$ &
$8.83 \times 10^{-4}$ &
$1.933 \times 10^{-5}$ &
$3.82 \times 10^{-5}$ &
$2.216 \times 10^{-2}$\\\hline
$n_{s}^{+}$ &
$7.944 \times 10^{-6}$ &
$9.436 \times 10^{-6}$ &
$9.164 \times 10^{-4}$ &
$1.89 \times 10^{-5}$ &
$3.686 \times 10^{-5}$ &
$2.243 \times 10^{-2}$\\\hline
$\sigma_{8}^{-}$ &
$4.236 \times 10^{-6}$ &
$5.231 \times 10^{-6}$ &
$1.309 \times 10^{-4}$ &
$3.165 \times 10^{-6}$ &
$8.335 \times 10^{-6}$ &
$2.14 \times 10^{-2}$\\\hline
$\sigma_{8}^{+}$ &
$4.747 \times 10^{-6}$ &
$5.878 \times 10^{-6}$ &
$1.347 \times 10^{-4}$ &
$3.636 \times 10^{-6}$ &
$8.484 \times 10^{-6}$ &
$2.318 \times 10^{-2}$\\\hline
$h^{-}$ &
$1.255 \times 10^{-5}$ &
$1.504 \times 10^{-5}$ &
$8.06 \times 10^{-4}$ &
$1.823 \times 10^{-5}$ &
$4.283 \times 10^{-5}$ &
$2.208 \times 10^{-2}$\\\hline
$h^{+}$ &
$1.924 \times 10^{-5}$ &
$2.245 \times 10^{-5}$ &
$8.149 \times 10^{-4}$ &
$2.488 \times 10^{-5}$ &
$4.54 \times 10^{-5}$ &
$2.25 \times 10^{-2}$\\\hline
$w_{0}^{+}$ &
$1.234 \times 10^{-15}$ &
$1.234 \times 10^{-15}$ &
$1.237 \times 10^{-15}$ &
$3.096 \times 10^{-15}$ &
$3.093 \times 10^{-15}$ &
$2.228 \times 10^{-2}$\\\hline
$w_{0}^{-}$ &
$1.238 \times 10^{-15}$ &
$1.239 \times 10^{-15}$ &
$1.236 \times 10^{-15}$ &
$3.106 \times 10^{-15}$ &
$3.108 \times 10^{-15}$ &
$2.228 \times 10^{-2}$\\\hline

\end{tabular}
\caption{A table of the ratios of the norms of the estimated
  errors $C_{\perp}$ and the estimated covariance matrices
  $C_{\parallel}$ for our assorted reconstructed cosmologies.  We note that the vanishingly small numbers for $w_{0}$ are the result of the $w_{0}$ simulations being normalised at z=0, making them similar enough to the fiducial present time cosmology that we could use the fiducial SPT covariance matrices for them.}
\label{tab}
\end{table*}


Given all of these results, we conclude that the full response matrix
is necessary for the most accurate model of the non-linear covariance
matrix.  This indicates that the correlation matrix exhibits a greater
parameter dependence than previously assumed, since the only
difference between the full and approximate response matrices is the
presence of the derivative of the correlation matrix with respect to
the parameter under study.

We note that we developed the approximate response matrix to remove
the risk of noise without signal from the derivative of the
correlation matrix.  However, we have found that our model exhibits
minimal noise even with these terms present.  Thus, given that they
improve accuracy without substantially increasing noise, we conclude
that the correct way to model future covariance matrices is using the
full response matrix.


\section{Conclusions and discussion}\label{sec:discussion}

Upcoming surveys like Euclid and LSST will provide unprecedented volumes of data for the study of cosmology and fundamental physics. Testing the
$\Lambda$CDM model and the wide array of alternates and extensions, such as
a time varying equation of state for dark energy, will require us to
make optimal use of this data. Establishing an accurate and precise
method for exploring this menagerie of physical models is essential for
us to converge on the correct theoretical interpretation of the
data. 

In this paper, we have presented a new method for computation of the
cosmological dependence of the covariance matrix of the matter power
spectrum. This enables the rapid evaluation of the covariance
matrix without recourse to running new simulations beyond those used to establish the model.  It
therefore represents an important step towards the development of
likelihood modelling codes that include an accurate model for the
variations in the errors on cosmology.

In \S\ref{sec:introduction} and \S\ref{sec:background} we outlined the
deficiency associated current likelihood analysis involving
clustering probes, reviewed the current literature on the subject, and outlined
some key elements of the relevant theory.

In \S\ref{sec:model} we developed our proposed modelling of the
cosmology dependence. Specifically, we proposed that if one projects the
true covariance onto a reasonably good precision matrix, scaling out
the model matrix, this quantity can be well described by a
low-order matrix Taylor expansion. We termed the first-order
correction from the Taylor expansion the response
matrix. Qualitatively speaking, this describes the extent to which
the accuracy of a model varies as a function of the parameters being
considered. We showed that this response can be measured from a suite
of simulations.

In \S\ref{sec:simulations} we present the Quijote suite of $N$-body
simulations that we used to explore the cosmological dependence of the
covariance matrices. Here we also presented a basic analysis of the
simulations. 

In \S\ref{sec:results} we explore the predictions from SPT and
included SSC corrections. We also presented an analysis of
the approximate models for the projection of the model covariance on
to the true covariance matrix. We noted that this was close to being a
diagonal matrix, but that it had off-diagonal structure, which
informed us that the SPT covariance was not accurate enough for
modelling the covariance matrix alone.

In \S\ref{sec:reconstruction} We have used this method with three
different variants: one in which we employ the full response matrix,
one in which we employ a simplified approximate response matrix, and
one in which the response matrix is set to zero and all parameter
dependence comes from an interpolation over SPT models.  Applying the
method to recreate the covariance matrices we used as inputs for our
model, we were able to show that they recreated them to several
percent precision out to at least $k\approx
0.3~\mathrm{h}~\mathrm{Mpc}^{-1}$.  We found that, in all cosmologies
considered, all three methods worked well in the case that SSC
corrections were omitted, with the method using the full response matrix generally
being the most effective. The inclusion of SSC corrections resulted in
the reconstructions with the response set to zero becoming more
inaccurate, indicating that the SSC corrections substantially alter
the accuracy of the model in this case. We also performed a series of
other validation tests that the model was able to pass at a high
level.

While our validation tests appear to have been successful, it would
have been preferable to be able to run ensembles of simulations at
independent points in parameters space. This, however, owing to the
computational demands was not feasible with our resources.  In future
work, we will look to generate such samples for improved validation
analysis.

This work has been done in the context of the clustering of
collisionless matter, sampled by discrete particles in $N$-body
simulations.  This is a rather idealised case, and to model galaxy
clustering we would need to extend this framework to include galaxy
biasing and redshift space distortions. In addition, the survey
systematics such as the mask and selection effects will need to be incorporated. On the other hand, for modelling weak lensing
statistics, we would need to show that the model can be made to work up to $k\gtrsim1\kMpc$. We leave it to
a future work to explore such extensions.

\section{Acknowledgements}
Robert E. Smith would like to acknowledge the help of Francisco Villaescusa-Navarro, who provided information that was used to correctly analyse the data and generate non-linear covariance matrices from the Quijote simulations.  The authors would also like to acknowledge the help of the Science and Technology Facilities Council (STFC), who provided funding for this research through a consolidated grant.


\bibliographystyle{mnras}
\bibliography{bibliography}

@article{Dore:2014cca,
    author = "Dor\'e, Olivier and others",
    title = "{Cosmology with the SPHEREX All-Sky Spectral Survey}",
    eprint = "1412.4872",
    archivePrefix = "arXiv",
journal= "",
    primaryClass = "astro-ph.CO",
    month = "12",
    year = "2014"
}

@article{Steele:2020tak,
    author = "Steele, Theodore and Baldauf, Tobias",
    title = "{Precise Calibration of the One-Loop Bispectrum in the Effective Field Theory of Large Scale Structure}",
    eprint = "2009.01200",
    archivePrefix = "arXiv",
    primaryClass = "astro-ph.CO",
    doi = "10.1103/PhysRevD.103.023520",
    journal = "Phys. Rev. D",
    volume = "103",
    number = "2",
    pages = "023520",
    year = "2021"
}

@article{Steele:2021lnz,
    author = "Steele, Theodore and Baldauf, Tobias",
    title = "{Precise Calibration of the One-Loop Trispectrum in the Effective Field Theory of Large Scale Structure}",
    eprint = "2101.10289",
    archivePrefix = "arXiv",
    primaryClass = "astro-ph.CO",
    doi = "10.1103/PhysRevD.103.103518",
    journal = "Phys. Rev. D",
    volume = "103",
    number = "10",
    pages = "103518",
    year = "2021"
}

@ARTICLE{2019ApJ...873..111I,
       author = {Ivezi{\'c}, {\v{Z}eljko} and others},
        title = "{LSST: From Science Drivers to Reference Design and Anticipated Data Products}",
      journal = {\apj},
         year = 2019,
        month = mar,
       volume = {873},
       number = {2},
          eid = {111},
        pages = {111},
          doi = {10.3847/1538-4357/ab042c},
archivePrefix = {arXiv},
       eprint = {0805.2366},
 primaryClass = {astro-ph},
       adsurl = {https://ui.adsabs.harvard.edu/abs/2019ApJ...873..111I}
}

@article{Wang:2021moa,
    author = "Wang, Zhaoyu and others",
    title = "{The clustering of galaxies in the DESI imaging legacy surveys DR8: I. The luminosity and color dependent intrinsic clustering}",
    eprint = "2106.14159",
    archivePrefix = "arXiv",
    primaryClass = "astro-ph.GA",
    doi = "10.1007/s11433-021-1707-6",
    journal = "Sci. China Phys. Mech. Astron.",
    volume = "64",
    number = "8",
    pages = "289811",
    year = "2021"
}

@article{Richard:2019dwt,
    author = "Richard, Johan and others",
    title = "{4MOST Consortium Survey 8: Cosmology Redshift Survey (CRS)}",
    eprint = "1903.02474",
    archivePrefix = "arXiv",
    primaryClass = "astro-ph.CO",
    doi = "10.18727/0722-6691/5127",
    journal = "The Messenger",
    volume = "175",
    pages = "50--53",
    year = "2019"
}

@article{interloperbias,
author = {Pullen, Anthony and Hirata, Christopher and Doré, Olivier and Raccanelli, Alvise},
year = {2015},
month = {07},
pages = {},
title = {Interloper bias in future large-scale structure surveys},
volume = {68},
journal = {Publications of the Astronomical Society of Japan},
doi = {10.1093/pasj/psv118}
}

@article{Bernardeauetal2002,
    author = "Bernardeau, F. and Colombi, S. and Gaztanaga, E. and Scoccimarro, R.",
    title = "{Large scale structure of the universe and cosmological perturbation theory}",
    eprint = "astro-ph/0112551",
    archivePrefix = "arXiv",
    reportNumber = "SACLAY-T01-142",
    doi = "10.1016/S0370-1573(02)00135-7",
    journal = "Phys. Rept.",
    volume = "367",
    pages = "1--248",
    year = "2002"
}

@article{TakadaHu2013,
    author = "Takada, Masahiro and Hu, Wayne",
    title = "{Power Spectrum Super-Sample Covariance}",
    eprint = "1302.6994",
    archivePrefix = "arXiv",
    primaryClass = "astro-ph.CO",
    doi = "10.1103/PhysRevD.87.123504",
    journal = "Phys. Rev. D",
    volume = "87",
    number = "12",
    pages = "123504",
    year = "2013"
}

@article{Lietal2014,
    author = "Li, Yin and Hu, Wayne and Takada, Masahiro",
    title = "{Super-Sample Covariance in Simulations}",
    eprint = "1401.0385",
    archivePrefix = "arXiv",
    primaryClass = "astro-ph.CO",
    doi = "10.1103/PhysRevD.89.083519",
    journal = "Phys. Rev. D",
    volume = "89",
    number = "8",
    pages = "083519",
    year = "2014"
}

@article{Villaescusa-Navarroetal2020,
    author = "Villaescusa-Navarro, Francisco and others",
    title = "{The Quijote simulations}",
    eprint = "1909.05273",
    archivePrefix = "arXiv",
    primaryClass = "astro-ph.CO",
    doi = "10.3847/1538-4365/ab9d82",
    journal = "Astrophys. J. Suppl.",
    volume = "250",
    number = "1",
    pages = "2",
    year = "2020"
}

@article{Mohammedetal2017,
    author = "Mohammed, Irshad and Seljak, Uros and Vlah, Zvonimir",
    title = "{Perturbative approach to covariance matrix of the matter power spectrum}",
    eprint = "1607.00043",
    archivePrefix = "arXiv",
    primaryClass = "astro-ph.CO",
    reportNumber = "FERMILAB-PUB-16-247-A",
    doi = "10.1093/mnras/stw3196",
    journal = "Mon. Not. Roy. Astron. Soc.",
    volume = "466",
    number = "1",
    pages = "780--797",
    year = "2017"
}

@ARTICLE{Takahashietal2009,
       author = {{Takahashi}, Ryuichi and {Yoshida}, Naoki and {Takada}, Masahiro and {Matsubara}, Takahiko and {Sugiyama}, Naoshi and {Kayo}, Issha and {Nishizawa}, Atsushi J. and {Nishimichi}, Takahiro and {Saito}, Shun and {Taruya}, Atsushi},
        title = "{Simulations of Baryon Acoustic Oscillations. II. Covariance Matrix of the Matter Power Spectrum}",
      journal = {\apj},
         year = 2009,
        month = jul,
       volume = {700},
       number = {1},
        pages = {479-490},
          doi = {10.1088/0004-637X/700/1/479},
archivePrefix = {arXiv},
       eprint = {0902.0371},
 primaryClass = {astro-ph.CO},
       adsurl = {https://ui.adsabs.harvard.edu/abs/2009ApJ...700..479T}
}

@ARTICLE{SellentinHeavens2016,
       author = {{Sellentin}, Elena and {Heavens}, Alan F.},
        title = "{Parameter inference with estimated covariance matrices}",
      journal = {\mnras},
         year = 2016,
        month = feb,
       volume = {456},
       number = {1},
        pages = {L132-L136},
          doi = {10.1093/mnrasl/slv190},
archivePrefix = {arXiv},
       eprint = {1511.05969},
 primaryClass = {astro-ph.CO},
       adsurl = {https://ui.adsabs.harvard.edu/abs/2016MNRAS.456L.132S}
}

@ARTICLE{TaylorWatts2001,
       author = {{Taylor}, A.~N. and {Watts}, P.~I.~R.},
        title = "{Parameter information from non-linear cosmological fields}",
      journal = {\mnras},
         year = 2001,
        month = dec,
       volume = {328},
       number = {4},
        pages = {1027-1038},
          doi = {10.1046/j.1365-8711.2001.04874.x},
archivePrefix = {arXiv},
       eprint = {astro-ph/0010014},
 primaryClass = {astro-ph},
       adsurl = {https://ui.adsabs.harvard.edu/abs/2001MNRAS.328.1027T}
}

@ARTICLE{EuclidPaper1_2025,
       author = {{Euclid Collaboration} and {Mellier}, Y. and {Abdurro'uf} and {Acevedo Barroso}, J.~A. and {Ach{\'u}carro}, A. and {Adamek}, J. and {Adam}, R. and {Addison}, G.~E. and {Aghanim}, N. and {Aguena}, M. and {Ajani}, V. and {Akrami}, Y. and {Al-Bahlawan}, A. and {Alavi}, A. and {Albuquerque}, I.~S. and {Alestas}, G. and {Alguero}, G. and {Allaoui}, A. and {Allen}, S.~W. and {Allevato}, V. and {Alonso-Tetilla}, A.~V. and {Altieri}, B. and {Alvarez-Candal}, A. and {Alvi}, S. and {Amara}, A. and {Amendola}, L. and {Amiaux}, J. and {Andika}, I.~T. and {Andreon}, S. and {Andrews}, A. and {Angora}, G. and {Angulo}, R.~E. and {Annibali}, F. and {Anselmi}, A. and {Anselmi}, S. and {Arcari}, S. and {Archidiacono}, M. and {Aric{\`o}}, G. and {Arnaud}, M. and {Arnouts}, S. and {Asgari}, M. and {Asorey}, J. and {Atayde}, L. and {Atek}, H. and {Atrio-Barandela}, F. and {Aubert}, M. and {Aubourg}, E. and {Auphan}, T. and {Auricchio}, N. and {Aussel}, B. and {Aussel}, H. and {Avelino}, P.~P. and {Avgoustidis}, A. and {Avila}, S. and {Awan}, S. and {Azzollini}, R. and {Baccigalupi}, C. and {Bachelet}, E. and {Bacon}, D. and {Baes}, M. and {Bagley}, M.~B. and {Bahr-Kalus}, B. and {Balaguera-Antolinez}, A. and {Balbinot}, E. and {Balcells}, M. and {Baldi}, M. and {Baldry}, I. and {Balestra}, A. and {Ballardini}, M. and {Ballester}, O. and {Balogh}, M. and {Ba{\~n}ados}, E. and {Barbier}, R. and {Bardelli}, S. and {Baron}, M. and {Barreiro}, T. and {Barrena}, R. and {Barriere}, J.-C. and {Barros}, B.~J. and {Barthelemy}, A. and {Bartolo}, N. and {Basset}, A. and {Battaglia}, P. and {Battisti}, A.~J. and {Baugh}, C.~M. and {Baumont}, L. and {Bazzanini}, L. and {Beaulieu}, J.-P. and {Beckmann}, V. and {Belikov}, A.~N. and {Bel}, J. and {Bellagamba}, F. and {Bella}, M. and {Bellini}, E. and {Benabed}, K. and {Bender}, R. and {Benevento}, G. and {Bennett}, C.~L. and {Benson}, K. and {Bergamini}, P. and {Bermejo-Climent}, J.~R. and {Bernardeau}, F. and {Bertacca}, D. and {Berthe}, M. and {Berthier}, J. and {Bethermin}, M. and {Beutler}, F. and {Bevillon}, C. and {Bhargava}, S. and {Bhatawdekar}, R. and {Bianchi}, D. and {Bisigello}, L. and {Biviano}, A. and {Blake}, R.~P. and {Blanchard}, A. and {Blazek}, J. and {Blot}, L. and {Bosco}, A. and {Bodendorf}, C. and {Boenke}, T. and {B{\"o}hringer}, H. and {Boldrini}, P. and {Bolzonella}, M. and {Bonchi}, A. and {Bonici}, M. and {Bonino}, D. and {Bonino}, L. and {Bonvin}, C. and {Bon}, W. and {Booth}, J.~T. and {Borgani}, S. and {Borlaff}, A.~S. and {Borsato}, E. and {Bose}, B. and {Botticella}, M.~T. and {Boucaud}, A. and {Bouche}, F. and {Boucher}, J.~S. and {Boutigny}, D. and {Bouvard}, T. and {Bouwens}, R. and {Bouy}, H. and {Bowler}, R.~A.~A. and {Bozza}, V. and {Bozzo}, E. and {Branchini}, E. and {Brando}, G. and {Brau-Nogue}, S. and {Brekke}, P. and {Bremer}, M.~N. and {Brescia}, M. and {Breton}, M.-A. and {Brinchmann}, J. and {Brinckmann}, T. and {Brockley-Blatt}, C. and {Brodwin}, M. and {Brouard}, L. and {Brown}, M.~L. and {Bruton}, S. and {Bucko}, J. and {Buddelmeijer}, H. and {Buenadicha}, G. and {Buitrago}, F. and {Burger}, P. and {Burigana}, C. and {Busillo}, V. and {Busonero}, D. and {Cabanac}, R. and {Cabayol-Garcia}, L. and {Cagliari}, M.~S. and {Caillat}, A. and {Caillat}, L. and {Calabrese}, M. and {Calabro}, A. and {Calderone}, G. and {Calura}, F. and {Camacho Quevedo}, B. and {Camera}, S. and {Campos}, L. and {Ca{\~n}as-Herrera}, G. and {Candini}, G.~P. and {Cantiello}, M. and {Capobianco}, V. and {Cappellaro}, E. and {Cappelluti}, N. and {Cappi}, A. and {Caputi}, K.~I. and {Cara}, C. and {Carbone}, C. and {Cardone}, V.~F. and {Carella}, E. and {Carlberg}, R.~G. and {Carle}, M. and {Carminati}, L. and {Caro}, F. and {Carrasco}, J.~M. and {Carretero}, J. and {Carrilho}, P. and {Carron Duque}, J. and {Carry}, B.},
        title = "{Euclid: I. Overview of the Euclid mission}",
      journal = {\aap},
         year = 2025,
        month = may,
       volume = {697},
          eid = {A1},
        pages = {A1},
          doi = {10.1051/0004-6361/202450810},
archivePrefix = {arXiv},
       eprint = {2405.13491},
 primaryClass = {astro-ph.CO},
       adsurl = {https://ui.adsabs.harvard.edu/abs/2025A&A...697A...1E}
}

@ARTICLE{Carron2013,
       author = {{Carron}, J.},
        title = "{On the assumption of Gaussianity for cosmological two-point statistics and parameter dependent covariance matrices}",
      journal = {\aap},
         year = 2013,
        month = mar,
       volume = {551},
          eid = {A88},
        pages = {A88},
          doi = {10.1051/0004-6361/201220538},
archivePrefix = {arXiv},
       eprint = {1204.4724},
 primaryClass = {astro-ph.CO},
       adsurl = {https://ui.adsabs.harvard.edu/abs/2013A&A...551A..88C}
}

@ARTICLE{Heavens2009,
       author = {{Heavens}, Alan},
        title = "{Statistical techniques in cosmology}",
      journal = {arXiv e-prints},
         year = 2009,
        month = jun,
          eid = {arXiv:0906.0664},
        pages = {arXiv:0906.0664},
          doi = {10.48550/arXiv.0906.0664},
archivePrefix = {arXiv},
       eprint = {0906.0664},
 primaryClass = {astro-ph.CO},
       adsurl = {https://ui.adsabs.harvard.edu/abs/2009arXiv0906.0664H}
}

@ARTICLE{Eifleretal2009,
       author = {{Eifler}, T. and {Schneider}, P. and {Hartlap}, J.},
        title = "{Dependence of cosmic shear covariances on cosmology. Impact on parameter estimation}",
      journal = {\aap},
         year = 2009,
        month = aug,
       volume = {502},
       number = {3},
        pages = {721-731},
          doi = {10.1051/0004-6361/200811276},
archivePrefix = {arXiv},
       eprint = {0810.4254},
 primaryClass = {astro-ph},
       adsurl = {https://ui.adsabs.harvard.edu/abs/2009A&A...502..721E}
}

@ARTICLE{GilMarinetal2015,
       author = {{Gil-Mar{\'\i}n}, H{\'e}ctor and {Nore{\~n}a}, Jorge and {Verde}, Licia and {Percival}, Will J. and {Wagner}, Christian and {Manera}, Marc and {Schneider}, Donald P.},
        title = "{The power spectrum and bispectrum of SDSS DR11 BOSS galaxies - I. Bias and gravity}",
      journal = {\mnras},
         year = 2015,
        month = jul,
       volume = {451},
       number = {1},
        pages = {539-580},
          doi = {10.1093/mnras/stv961},
archivePrefix = {arXiv},
       eprint = {1407.5668},
 primaryClass = {astro-ph.CO},
       adsurl = {https://ui.adsabs.harvard.edu/abs/2015MNRAS.451..539G}
}

@ARTICLE{NovellMasotetal2025,
       author = {{Novell-Masot}, S. and {Gil-Mar{\'\i}n}, H. and {Verde}, L. and {Aguilar}, J. and {Ahlen}, S. and {Bailey}, S. and {BenZvi}, S. and {Bianchi}, D. and {Brooks}, D. and {Buckley-Geer}, E. and {Carnero Rosell}, A. and {Chaussidon}, E. and {Claybaugh}, T. and {Cole}, S. and {Cuceu}, A. and {Dawson}, K.~S. and {de la Macorra}, A. and {Demina}, R. and {Dey}, A. and {Dey}, B. and {Doel}, P. and {Ferraro}, S. and {Font-Ribera}, A. and {Forero-Romero}, J.~E. and {Gazta{\~n}aga}, E. and {Gontcho A Gontcho}, S. and {Gonzalez-Morales}, A.~X. and {Gutierrez}, G. and {Herrera-Alcantar}, H.~K. and {Honscheid}, K. and {Howlett}, C. and {Juneau}, S. and {Kehoe}, R. and {Kirkby}, D. and {Kisner}, T. and {Kremin}, A. and {Lamman}, C. and {Landriau}, M. and {Le Guillou}, L. and {Levi}, M.~E. and {Magneville}, C. and {Manera}, M. and {Meisner}, A. and {Miquel}, R. and {Moustakas}, J. and {Mu{\~n}oz-Guti{\'e}rrez}, A. and {Myers}, A.~D. and {Nadathur}, S. and {Niz}, G. and {Noriega}, H.~E. and {Percival}, W.~J. and {Poppett}, C. and {Prada}, F. and {P{\'e}rez-R{\`a}fols}, I. and {Ross}, A.~J. and {Rossi}, G. and {Samushia}, L. and {Sanchez}, E. and {Schlegel}, D. and {Schubnell}, M. and {Seo}, H. and {Silber}, J. and {Sprayberry}, D. and {Tarl{\'e}}, G. and {Vargas-Maga{\~n}a}, M. and {Weaver}, B.~A. and {Zarrouk}, P. and {Zhou}, R. and {Zou}, H.},
        title = "{Full-Shape analysis of the power spectrum and bispectrum of DESI DR1 LRG and QSO samples}",
      journal = {\jcap},
         year = 2025,
        month = jun,
       volume = {2025},
       number = {6},
          eid = {005},
        pages = {005},
          doi = {10.1088/1475-7516/2025/06/005},
archivePrefix = {arXiv},
       eprint = {2503.09714},
 primaryClass = {astro-ph.CO},
       adsurl = {https://ui.adsabs.harvard.edu/abs/2025JCAP...06..005N}
}

@ARTICLE{Kilbingeretal2013,
       author = {{Kilbinger}, Martin and {Fu}, Liping and {Heymans}, Catherine and {Simpson}, Fergus and {Benjamin}, Jonathan and {Erben}, Thomas and {Harnois-D{\'e}raps}, Joachim and {Hoekstra}, Henk and {Hildebrandt}, Hendrik and {Kitching}, Thomas D. and {Mellier}, Yannick and {Miller}, Lance and {Van Waerbeke}, Ludovic and {Benabed}, Karim and {Bonnett}, Christopher and {Coupon}, Jean and {Hudson}, Michael J. and {Kuijken}, Konrad and {Rowe}, Barnaby and {Schrabback}, Tim and {Semboloni}, Elisabetta and {Vafaei}, Sanaz and {Velander}, Malin},
        title = "{CFHTLenS: combined probe cosmological model comparison using 2D weak gravitational lensing}",
      journal = {\mnras},
         year = 2013,
        month = apr,
       volume = {430},
       number = {3},
        pages = {2200-2220},
          doi = {10.1093/mnras/stt041},
archivePrefix = {arXiv},
       eprint = {1212.3338},
 primaryClass = {astro-ph.CO},
       adsurl = {https://ui.adsabs.harvard.edu/abs/2013MNRAS.430.2200K}
}

@ARTICLE{Byunetal2017,
       author = {{Byun}, Joyce and {Eggemeier}, Alexander and {Regan}, Donough and {Seery}, David and {Smith}, Robert E.},
        title = "{Towards optimal cosmological parameter recovery from compressed bispectrum statistics}",
      journal = {\mnras},
         year = 2017,
        month = oct,
       volume = {471},
       number = {2},
        pages = {1581-1618},
          doi = {10.1093/mnras/stx1681},
archivePrefix = {arXiv},
       eprint = {1705.04392},
 primaryClass = {astro-ph.CO},
       adsurl = {https://ui.adsabs.harvard.edu/abs/2017MNRAS.471.1581B}
}

@ARTICLE{Sefusattietal2006,
       author = {{Sefusatti}, Emiliano and {Crocce}, Mart{\'\i}n and {Pueblas}, Sebasti{\'a}n and {Scoccimarro}, Rom{\'a}n},
        title = "{Cosmology and the bispectrum}",
      journal = {\prd},
         year = 2006,
        month = jul,
       volume = {74},
       number = {2},
          eid = {023522},
        pages = {023522},
          doi = {10.1103/PhysRevD.74.023522},
archivePrefix = {arXiv},
       eprint = {astro-ph/0604505},
 primaryClass = {astro-ph},
       adsurl = {https://ui.adsabs.harvard.edu/abs/2006PhRvD..74b3522S}
}

@ARTICLE{Labatieetal2012,
       author = {{Labatie}, A. and {Starck}, J.~L. and {Lachi{\`e}ze-Rey}, M.},
        title = "{Effect of Model-dependent Covariance Matrix for Studying Baryon Acoustic Oscillations}",
      journal = {\apj},
         year = 2012,
        month = dec,
       volume = {760},
       number = {2},
          eid = {97},
        pages = {97},
          doi = {10.1088/0004-637X/760/2/97},
archivePrefix = {arXiv},
       eprint = {1210.0878},
 primaryClass = {astro-ph.CO},
       adsurl = {https://ui.adsabs.harvard.edu/abs/2012ApJ...760...97L}
}

@ARTICLE{Feldmanetal1994,
       author = {{Feldman}, Hume A. and {Kaiser}, Nick and {Peacock}, John A.},
        title = "{Power-Spectrum Analysis of Three-dimensional Redshift Surveys}",
      journal = {\apj},
         year = 1994,
        month = may,
       volume = {426},
        pages = {23},
          doi = {10.1086/174036},
archivePrefix = {arXiv},
       eprint = {astro-ph/9304022},
 primaryClass = {astro-ph},
       adsurl = {https://ui.adsabs.harvard.edu/abs/1994ApJ...426...23F}
}

@ARTICLE{MeiksinWhite1999,
       author = {{Meiksin}, A. and {White}, Martin},
        title = "{The growth of correlations in the matter power spectrum}",
      journal = {\mnras},
         year = 1999,
        month = oct,
       volume = {308},
       number = {4},
        pages = {1179-1184},
          doi = {10.1046/j.1365-8711.1999.02825.x},
archivePrefix = {arXiv},
       eprint = {astro-ph/9812129},
 primaryClass = {astro-ph},
       adsurl = {https://ui.adsabs.harvard.edu/abs/1999MNRAS.308.1179M}
}

@ARTICLE{Scoccimarroetal1999,
       author = {{Scoccimarro}, Rom{\'a}n and {Zaldarriaga}, Matias and {Hui}, Lam},
        title = "{Power Spectrum Correlations Induced by Nonlinear Clustering}",
      journal = {\apj},
         year = 1999,
        month = dec,
       volume = {527},
       number = {1},
        pages = {1-15},
          doi = {10.1086/308059},
archivePrefix = {arXiv},
       eprint = {astro-ph/9901099},
 primaryClass = {astro-ph},
       adsurl = {https://ui.adsabs.harvard.edu/abs/1999ApJ...527....1S}
}

@ARTICLE{NeyrinckSzapudi2008,
       author = {{Neyrinck}, Mark C. and {Szapudi}, Istv{\'a}n},
        title = "{Baryon oscillations in galaxy and matter power-spectrum covariance matrices}",
      journal = {\mnras},
         year = 2008,
        month = mar,
       volume = {384},
       number = {3},
        pages = {1221-1230},
          doi = {10.1111/j.1365-2966.2007.12803.x},
archivePrefix = {arXiv},
       eprint = {0710.3586},
 primaryClass = {astro-ph},
       adsurl = {https://ui.adsabs.harvard.edu/abs/2008MNRAS.384.1221N}
}

@ARTICLE{Smith2009,
       author = {{Smith}, Robert E.},
        title = "{Covariance of cross-correlations: towards efficient measures for large-scale structure}",
      journal = {\mnras},
         year = 2009,
        month = dec,
       volume = {400},
       number = {2},
        pages = {851-865},
          doi = {10.1111/j.1365-2966.2009.15490.x},
archivePrefix = {arXiv},
       eprint = {0810.1960},
 primaryClass = {astro-ph},
       adsurl = {https://ui.adsabs.harvard.edu/abs/2009MNRAS.400..851S}
}

@ARTICLE{WadekarScoccimarro2020,
       author = {{Wadekar}, Digvijay and {Scoccimarro}, Rom{\'a}n},
        title = "{Galaxy power spectrum multipoles covariance in perturbation theory}",
      journal = {\prd},
         year = 2020,
        month = dec,
       volume = {102},
       number = {12},
          eid = {123517},
        pages = {123517},
          doi = {10.1103/PhysRevD.102.123517},
archivePrefix = {arXiv},
       eprint = {1910.02914},
 primaryClass = {astro-ph.CO},
       adsurl = {https://ui.adsabs.harvard.edu/abs/2020PhRvD.102l3517W}
}

@ARTICLE{Blotetal2015,
       author = {{Blot}, L. and {Corasaniti}, P.~S. and {Alimi}, J.-M. and {Reverdy}, V. and {Rasera}, Y.},
        title = "{Matter power spectrum covariance matrix from the DEUS-PUR {\ensuremath{\Lambda}}CDM simulations: mass resolution and non-Gaussian errors}",
      journal = {\mnras},
         year = 2015,
        month = jan,
       volume = {446},
       number = {2},
        pages = {1756-1764},
          doi = {10.1093/mnras/stu2190},
archivePrefix = {arXiv},
       eprint = {1406.2713},
 primaryClass = {astro-ph.CO},
       adsurl = {https://ui.adsabs.harvard.edu/abs/2015MNRAS.446.1756B}
}

@ARTICLE{WhitePadmanabhan2015,
       author = {{White}, Martin and {Padmanabhan}, Nikhil},
        title = "{Including parameter dependence in the data and covariance for cosmological inference}",
      journal = {\jcap},
         year = 2015,
        month = dec,
       volume = {2015},
       number = {12},
        pages = {058-058},
          doi = {10.1088/1475-7516/2015/12/058},
archivePrefix = {arXiv},
       eprint = {1508.00566},
 primaryClass = {astro-ph.CO},
       adsurl = {https://ui.adsabs.harvard.edu/abs/2015JCAP...12..058W}
}

@ARTICLE{Blotetal2021,
       author = {{Blot}, Linda and {Corasaniti}, Pier-Stefano and {Rasera}, Yann and {Agarwal}, Shankar},
        title = "{Cosmological model parameter dependence of the matter power spectrum covariance from the DEUS-PUR Cosmo simulations}",
      journal = {\mnras},
         year = 2021,
        month = jan,
       volume = {500},
       number = {2},
        pages = {2532-2542},
          doi = {10.1093/mnras/staa3444},
archivePrefix = {arXiv},
       eprint = {2007.14984},
 primaryClass = {astro-ph.CO},
       adsurl = {https://ui.adsabs.harvard.edu/abs/2021MNRAS.500.2532B}
}

@BOOK{Peacock1999,
       author = {{Peacock}, John A.},
        title = "{Cosmological Physics}",
         year = 1999,
       adsurl = {https://ui.adsabs.harvard.edu/abs/1999coph.book.....P}
}

@BOOK{Peebles1980,
       author = {{Peebles}, P.~J.~E.},
        title = "{The large-scale structure of the universe}",
         year = 1980,
       adsurl = {https://ui.adsabs.harvard.edu/abs/1980lssu.book.....P}
}

@ARTICLE{Sirko2005,
       author = {{Sirko}, Edwin},
        title = "{Initial Conditions to Cosmological N-Body Simulations, or, How to Run an Ensemble of Simulations}",
      journal = {\apj},
         year = 2005,
        month = nov,
       volume = {634},
       number = {2},
        pages = {728-743},
          doi = {10.1086/497090},
archivePrefix = {arXiv},
       eprint = {astro-ph/0503106},
 primaryClass = {astro-ph},
       adsurl = {https://ui.adsabs.harvard.edu/abs/2005ApJ...634..728S}
}

@ARTICLE{Scoccimarro1998,
       author = {{Scoccimarro}, Roman},
        title = "{Transients from initial conditions: a perturbative analysis}",
      journal = {\mnras},
         year = 1998,
        month = oct,
       volume = {299},
       number = {4},
        pages = {1097-1118},
          doi = {10.1046/j.1365-8711.1998.01845.x},
archivePrefix = {arXiv},
       eprint = {astro-ph/9711187},
 primaryClass = {astro-ph},
       adsurl = {https://ui.adsabs.harvard.edu/abs/1998MNRAS.299.1097S}
}

@ARTICLE{Springel2005,
       author = {{Springel}, Volker},
        title = "{The cosmological simulation code GADGET-2}",
      journal = {\mnras},
         year = 2005,
        month = dec,
       volume = {364},
       number = {4},
        pages = {1105-1134},
          doi = {10.1111/j.1365-2966.2005.09655.x},
archivePrefix = {arXiv},
       eprint = {astro-ph/0505010},
 primaryClass = {astro-ph},
       adsurl = {https://ui.adsabs.harvard.edu/abs/2005MNRAS.364.1105S}
}

@ARTICLE{Lewisetal2000,
       author = {{Lewis}, Antony and {Challinor}, Anthony and {Lasenby}, Anthony},
        title = "{Efficient Computation of Cosmic Microwave Background Anisotropies in Closed Friedmann-Robertson-Walker Models}",
      journal = {\apj},
         year = 2000,
        month = aug,
       volume = {538},
       number = {2},
        pages = {473-476},
          doi = {10.1086/309179},
archivePrefix = {arXiv},
       eprint = {astro-ph/9911177},
 primaryClass = {astro-ph},
       adsurl = {https://ui.adsabs.harvard.edu/abs/2000ApJ...538..473L}
}

@MISC{Pylians,
    author = {{Villaescusa-Navarro}, Francisco},
    title = "{Pylians: Python libraries for the analysis of numerical simulations}",
    howpublished = {Astrophysics Source Code Library, record ascl:1811.008},
    year = 2018,
    month = nov,
    eid = {ascl:1811.008},
    pages = {ascl:1811.008},
    archivePrefix = {ascl},
    eprint = {1811.008},
    adsurl = {https://ui.adsabs.harvard.edu/abs/2018ascl.soft11008V}
}

@ARTICLE{Smithetal2007,
       author = {{Smith}, Robert E. and {Scoccimarro}, Rom{\'a}n and {Sheth}, Ravi K.},
        title = "{Scale dependence of halo and galaxy bias: Effects in real space}",
      journal = {\prd},
         year = 2007,
        month = mar,
       volume = {75},
       number = {6},
          eid = {063512},
        pages = {063512},
          doi = {10.1103/PhysRevD.75.063512},
archivePrefix = {arXiv},
       eprint = {astro-ph/0609547},
 primaryClass = {astro-ph},
       adsurl = {https://ui.adsabs.harvard.edu/abs/2007PhRvD..75f3512S}
}

@ARTICLE{Goroffetal1986,
       author = {{Goroff}, M.~H. and {Grinstein}, B. and {Rey}, S.-J. and {Wise}, M.~B.},
        title = "{Coupling of modes of cosmological mass density fluctuations}",
      journal = {\apj},
         year = 1986,
        month = dec,
       volume = {311},
        pages = {6-14},
          doi = {10.1086/164749},
       adsurl = {https://ui.adsabs.harvard.edu/abs/1986ApJ...311....6G}
}


\newpage

\onecolumn
\appendix


\section{Standard Perturbation Theory}\label{app:SPT}

In this appendix, we provide a brief introduction to Standard Perturbation
Theory (SPT).  For a more in-depth review we refer the reader to
\citet{Bernardeauetal2002}. 

This model begins with the assumption that gravitating matter can be treated as a set of
collisionless particles that behave as a pressureless fluid.  This is a
reasonable description of dark matter, up to shell-crossing at which
point we expect multi-streaming in phase space. For baryonic fluids
and neutrinos, additional infrastructure must be added which we will not discuss here.  

Under this approximation, we can
describe matter via the collisionless Boltzmann equation
\citep{Peebles1980}
\begin{equation}
  \frac{\mathrm{D}f}{\mathrm{D}t}=\frac{\partial f}{\partial t}
  +\frac{\mathbf{p}}{ma^{2}}\cdot\frac{df}{d\mathbf{x}}-m\frac{d\phi}{d\mathbf{x}}\frac{df}{d\mathbf{p}}=0~,
    \label{eq:Boltzmann}
\end{equation}
where $f$ is the phase space distribution function, $\bx$ is the
comoving position coordinate, ${\bf p}=ma^2\dot{\bx}$ is the canonical
momentum, $a$ is the universal expansion factor and $\phi(\bx)$ is the
peculiar gravitational potential
\be \phi(\bx) = -Ga^2\int\dx'\frac{\rho(\bx')-\rhob}{|\bx'-\bx|}\ ,\ee
where $\rho{\bx}-\rhob$ is the difference between the matter density
at position $\bx$ and the universal background.  Note that here we are assuming an expanding background FLRW model and
the Newtonian gauge.  

Taking the first two moments of this equation
gives us forms for the matter and momentum densities provide the equations of motion
for a collisionless perfect fluid:
\begin{align}
    \dot{\rho}+3H\rho+\frac{1}{a}\partial_{i}(\rho v^{i})=0~,\label{eq:continuity}\\
    \dot{v}^{i}+Hv^{i}+\frac{1}{a}v^{j}\partial_{j}v^{i}+\frac{1}{a}\partial_{i}\phi=0~,\label{eq:Euler}
\end{align}
where ${\bf v}=a\dot{\bx}$ is the proper peculiar velocity and
$H\equiv \dot{a}/a$. Defining the density contrast,
$\delta=\rho/\bar{\rho}-1$, Eq.~\eqref{eq:continuity} becomes
\begin{equation}
    \partial_{\eta}\delta+\nabla\cdot\left[\left(1+\delta\right)v^{i}\right]=0
    \label{eq:continuity2}
\end{equation}
where the conformal time is defined by $ad\eta\equiv dt$.

We may now take the derivative of Eq.~\eqref{eq:Euler} and define the
velocity divergence field, $\theta=\partial_{i}v^{i}$ to obtain
\begin{equation}
\partial_{\eta}\theta+\mathcal{H}\theta+v^{i}\partial_{i}\theta+\partial_{i}v^{j}\partial_{j}v^{i}+\nabla^{2}\phi=0~.
    \label{eq:Euler2}
\end{equation}
Taking the Fourier transforms of Eqs.~\eqref{eq:continuity2} and
\eqref{eq:Euler2} gives us
\begin{align}
\partial_{\eta}\delta(\mathbf{k},\eta)+\theta(\mathbf{k},\eta)=\mathcal{S}_{\alpha}(\mathbf{k},\eta)~,\\
\partial_{\eta}\theta(\mathbf{k},eta)+\mathcal{H}\theta(\mathbf{k},\eta)+\frac{3}{2}\Omega_{\mathrm{m}}\mathcal{H}^{2}\delta(\mathbf{k},\eta)=\mathcal{S}_{\beta}(\mathbf{k},\eta)~,
\end{align}
where $\mathcal{S}_{\alpha}$ and $\mathcal{S}_{\beta}$ are the non-linear source terms defined as
\begin{align}
\mathcal{S}_{\alpha}(\mathbf{k},\eta)=-\int \frac{d^{3}\mathbf{q}}{(2\pi)^{3}}\alpha(\mathbf{q},\mathbf{k}-\mathbf{q})\theta(\mathbf{q},\eta)\delta(\mathbf{k}-\mathbf{q},\eta)~,\\
\mathcal{S}_{\beta}(\mathbf{k},\eta)=-\int \frac{d^{3}\mathbf{q}}{(2\pi)^{3}}\beta(\mathbf{q},\mathbf{k}-\mathbf{q})\theta(\mathbf{q},\eta)\theta(\mathbf{k}-\mathbf{q},\eta)~,
\end{align}
where the kernels 
\begin{align}
    \alpha(\mathbf{k}_{1},\mathbf{k}_{2})=\frac{\mathbf{k}_{1}\cdot(\mathbf{k}_{1}+\mathbf{k}_{2})}{k_{1}^{2}}~,\\
    \beta(\mathbf{k}_{1},\mathbf{k}_{2})=\frac{1}{2}(\mathbf{k}_{1}+\mathbf{k}_{2})^{2}\frac{\mathbf{k}_{1}\cdot\mathbf{k}_{2}}{k_{1}^{2}k_{2}^{2}}
\end{align}
capture the coupling between modes of different wavelengths.

We now assume EdS cosmology, which permits the separability of space and time components,
$\delta_{n}(\mathbf{k},a)=D_{n}(a)\delta_{n}(\mathbf{k})$ with $D_{n}(a)=a^{n}$,
making it a commonly used model for approximate calculations, while
growth factors for other cosmologies must be derived order by order.

We may now solve our equations of motion using a power series,
yielding the following expression for the $n$th order Fourier space
density field in terms of the linear field:
\begin{equation}
  \delta_{n}(\bk)=\int_{\mathbf{q}_{1}}\ldots\int_{\mathbf{q}_{n}}(2\pi)^{3}\delta^{\mathrm{D}}
  \left(\mathbf{k}-\sum_{i=1}^{n}\mathbf{q}_{i}\right)F_{n}(\mathbf{q}_{1},\ldots,\mathbf{q}_{n})
  \prod_{i=1}^{n}\delta_{1}(\mathbf{q}_{i})~, \label{eq:SPTden}
\end{equation}
where $\int_{\mathbf{q}}=\int\frac{d^{3}\mathbf{q}}{(2\pi)^3}$ and the
gravitational coupling kernels are given by \citep{Goroffetal1986}
\begin{align}
  F_{n}(\mathbf{k}_{1},\ldots,\mathbf{k}_{n})=
  &\sum_{m=1}^{n-1}\frac{G_{m}(\mathbf{k}_{1},\ldots,\mathbf{k}_{m})}{(2n+3)(n-1)}
  \bigg[(2n+1)\alpha(\kappa_{1}^{m},\kappa_{m+1}^{n})F_{n-m}(\mathbf{k}_{m+1},\ldots,\mathbf{k}_{n})
    +2\beta(\kappa_{1}^{m},\kappa_{m+1}^{n})G_{n-m}(\mathbf{k}_{m+1},\ldots,\mathbf{k}_{n})\bigg]~,\\
  G_{n}(\mathbf{k}_{1},\ldots,\mathbf{k}_{n})=
  &\sum_{m=1}^{n-1}\frac{G_{m}(\mathbf{k}_{1},
    \ldots,\mathbf{k}_{m})}{(2n+3)(n-1)}\bigg[3\alpha(\kappa_{1}^{m},\kappa_{m+1}^{n})
    F_{n-m}(\mathbf{k}_{m+1},\ldots,\mathbf{k}_{n})]+2n\beta(\kappa_{1}^{m},\kappa_{m+1}^{n})
  G_{n-m}(\mathbf{k}_{m+1},\ldots,\mathbf{k}_{n})\bigg]~,
\end{align}
where $\kappa_{i}^{j}=\sum_{l=i}^{j}\mathbf{k}_{l}$ and we in general
use the symmetrised forms for these kernels, $F_{n}^{s}$ and
$G_{n}^{s}$.

For the SPT kernels up to third order in the linear fields we have
$F^{\mathrm{s}}_{1}(\bq_1)=G^{\mathrm{s}}_{1}(\bq_1)=1$ and with the second
and third order kernels being:
\begin{align}
  F^{\mathrm{s}}_{2}(\bq_1,\bq_2) & = \frac{5}{7}+\frac{1}{2}\mu_{12}\left(\frac{q_1}{q_2}+\frac{q_2}{q_1}\right) +
  \frac{2}{7}\mu_{12}^2\ , \\
  G^{\mathrm{s}}_{2}(\bq_1,\bq_2) & = \frac{3}{7}+\frac{1}{2}\mu_{12}\left(\frac{q_1}{q_2}+\frac{q_2}{q_1}\right) +
  \frac{4}{7}\mu_{12}^2\ , \\
  F_{3}(\bq_1,\bq_2,\bq_3) & =
  \frac{1}{18}\left\{7\alpha(\bq_1,\bq_{23})F_{2}^{\mathrm{s}}(\bq_2,\bq_3)+2\beta(\bq_1,\bq_{23})G_{2}^{\rm (s)}(\bq_2,
  \bq_{3})\right\} 
  + \frac{1}{18}\left\{ G_2^{\rm (s)}(\bq_1,\bq_2)\left[7\alpha(\bq_{12},\bq_3)+2\beta(\bq_{12},\bq_3)\right]\right\}
  \label{eq:F3}\ ,\\  
  G_{3}(\bq_1,\bq_2,\bq_3) & =
  \frac{1}{18}\left\{3\alpha(\bq_1,\bq_{23})F_{2}^{\mathrm{s}}(\bq_2,\bq_3)+6\beta(\bq_1,\bq_{23})G_{2}^{\rm (s)}(\bq_2,
  \bq_{3})\right\} 
  + \frac{1}{18}\left\{ G_2^{\rm (s)}(\bq_1,\bq_2)\left[3\alpha(\bq_{12},\bq_3)+6\beta(\bq_{12},\bq_3)\right]\right\} \ ,
\end{align}
where $\bq_{ij}\equiv \bq_i+\bq_j$ and $\mu_{12}=\hat{\mathbf{q}}_{1}\cdot\hat{\mathbf{q}}_{2}$. Note that the third order kernels
require a final symmetrisation step:
\begin{align}
F^{\mathrm{s}}_{3}(\bq_1,\bq_2,\bq_3) & = \frac{1}{6}\left[F_{3}(\bq_1,\bq_2,\bq_3)+5\,{\rm perms.}\right] \ \label{eq:Fsym} , \\
G^{\mathrm{s}}_{3}(\bq_1,\bq_2,\bq_3) & = \frac{1}{6}\left[G_{3}(\bq_1,\bq_2,\bq_3)+5\,{\rm perms.}\right] \ .
\end{align}

These perturbative density fields are used to build correlators in
momentum space at various orders, which can be represented using a
Feynman diagram notation in which loops represent couplings between
large scale modes and those extending from large down to arbitrarily small scales.  Thus, we
generally refer to the leading order perturbative contributions to
correlators as tree-level and higher order contributions as $n$-loop
terms.  

Furthermore, we note that when constructing estimators for
covariance matrices from SPT correlation functions, we maintain a
consistent sum of the orders of all density fields across each
correlator.  So, when employing a power spectrum (two-point
correlator) up to two-loops, where we correlate fields whose orders
sum to 6, we will use for the same covariance matrix the one-loop
bispectrum (three-point correlator) and tree-level trispectrum
(four-point correlator), as these also correlate fields whose orders
sum to 6.


\subsection{The Power Spectrum}
From these power ordered density fields, we can construct correlation functions which are the expected values of products of fields.  The second order correlation function in Fourier space is called the power spectrum and is defined as $(2\pi)^{3}\delta^{\mathrm{D}}(k_{1}+k_{2})P_{ij}(k_{1},k_{2})\equiv\langle\delta_{i}(k_{1})\delta_{j}(k_{2})\rangle+(1-\delta^{\mathrm{K}}_{ij})\langle\delta_{j}(k_{1})\delta_{i}(k_{2})\rangle$ for Kronecker delta $\delta^{\mathrm{K}}_{ij}$, where we note that, due to isotropy, the power spectrum is dependent only upon the magnitude of its momentum vectors.  

The linear power spectrum, $P_{11}$, correlates two linear density fields and is the only tree-level power spectrum term.  The one-loop contributions to the power spectrum are 
\begin{align}
    P_{31}(k)=&6P_{11}(k)\int_{\mathbf{q}}F_{3}^{s}(\mathbf{k},\mathbf{q},-\mathbf{q})P_{11}(q)~,\\
    P_{22}(k)=&2\int_{\mathbf{q}}F_{2}^{s}(\mathbf{k}-\mathbf{q},\mathbf{q})^{2}P_{11}(\vert\mathbf{k}-\mathbf{q}\vert)P_{11}(q)~,
\end{align}
and the two-loop contributions are
\begin{align}
    P_{51}(k)=&30P_{11}(k)\int_{\mathbf{q}_{1}}\int_{\mathbf{q}_{2}}F_{5}^{s}(\mathbf{k},\mathbf{q}_{1},-\mathbf{q}_{1},\mathbf{q}_{2},-\mathbf{q}_{2})P_{11}(q_{1})P_{11}(q_{2})~,\\
    P_{42}(k)=&24\int_{\mathbf{q}_{1}}\int_{\mathbf{q}_{2}}F_{4}^{s}(\mathbf{q}_{1}-\mathbf{k},-\mathbf{q}_{1},\mathbf{q}_{2},-\mathbf{q}_{2})F_{2}^{s}(\mathbf{k}-\mathbf{q}_{1},\mathbf{q}_{1})P_{11}(q_{1})P_{11}(q_{2})P_{11}(\vert\mathbf{k}-\mathbf{q}_{1}\vert)~,\\
    P_{33,\mathrm{a}}(k)=&6\int_{\mathbf{q}_{1}}\int_{\mathbf{q}_{2}}F_{3}^{s}(\mathbf{q}_{1},\mathbf{q}_{2},\mathbf{k}-\mathbf{q}_{1}-\mathbf{q}_{2})F_{3}^{s}(-\mathbf{q}_{1},-\mathbf{q}_{2},-\mathbf{k}+\mathbf{q}_{1}+\mathbf{q}_{2})P_{11}(q_{1})P_{11}(q_{2})P_{11}(\vert\mathbf{k}-\mathbf{q}_{1}-\mathbf{q}_{2}\vert)~,\\
    P_{33,\mathrm{b}}(k)=&\frac{1}{4}\frac{P_{31}(k)^{2}}{P_{11}(k)}~,
\end{align}

The above terms can be summed to give an approximation of the non-linear power spectrum that would be measured in simulations, as $P_{\mathrm{nn}}\approx P_{11}+P_{31}+P_{22}+P_{51}+P_{42}+P_{33}$ where $P_{33}=P_{33,\mathrm{a}}+P_{33,\mathrm{b}}$.

\subsection{The Bispectrum}
The bispectrum is the Fourier space correlation function of three fields \cite{Steele:2020tak}, defined as $(2\pi)^{3}\delta^{\mathrm{D}}(\mathbf{k}_{1}+\mathbf{k}_{2}+\mathbf{k}_{3})B_{ijl}(\mathbf{k}_{1},\mathbf{k}_{2},\mathbf{k}_{3})\equiv\langle\delta_{i}(\mathbf{k}_{1})\delta_{j}(\mathbf{k}_{2})\delta_{l}(\mathbf{k}_{3})\rangle+\mathrm{perms.}$.  At tree level, we have
\begin{equation}
    B_{211}=2F_{2}^{s}(\mathbf{k}_{2},\mathbf{k}_{3})P_{11}(k_{2}),P_{11}(k_{3})+\mathrm{2~perms.}
\end{equation}
and at one-loop, the contributions are
\begin{align}
    B_{411}=&12\int_{\mathbf{q}}F_{4}^{s}(-\mathbf{k}_{2},-\mathbf{k}_{3},\mathbf{q},-\mathbf{q})P_{11}(k_{2})P_{11}(k_{3})P_{11}(q)+\mathrm{2~perms.}~,\\
    B_{321,\mathrm{a}}=&6\int_{\mathbf{q}}F_{3}^{s}(-\mathbf{k}_{3},-\mathbf{q},\mathbf{q}-\mathbf{k}_{2})F_{2}^{s}(\mathbf{q},\mathbf{k}_{2}-\mathbf{q})P_{11}(\vert\mathbf{k}_{2}-\mathbf{q}\vert)P_{11}(k_{3})P_{11}(q)+\mathrm{5~perms.}~,\\
    B_{321,\mathrm{b}}=&6\int_{\mathbf{q}}F_{3}^{s}(\mathbf{q},-\mathbf{q},\mathbf{k}_{3})F_{2}^{s}(\mathbf{k}_{2},\mathbf{k}_{3})P_{11}(k_{2})P_{11}(k_{3})P_{11}(q)+\mathrm{5~perms.}~,\\
    B_{222}=&8\int_{\mathbf{q}}F_{2}^{s}(-\mathbf{q},\mathbf{k}_{3}+\mathbf{q})F_{2}^{s}(\mathbf{k}_{3}+\mathbf{q},\mathbf{k}_{2}-\mathbf{q})F_{2}^{s}(\mathbf{k}_{2}-\mathbf{q},\mathbf{q})P_{11}(\vert\mathbf{k}_{2}-\mathbf{q}\vert)P_{11}(\vert\mathbf{k}_{3}+\mathbf{q}\vert)P_{11}(q)~.
\end{align}
These sum to give an approximation of the non-linear bispectrum: $B_{\mathrm{nnn}}\approx B_{211}+B_{411}+B_{321}+B_{222}$, where $B_{321}=B_{321,\mathrm{a}}+B_{321,\mathrm{b}}$.  Bispectra must obey the triangle condition that $\mathbf{k}_{1}+\mathbf{k}_{2}+\mathbf{k}_{3}=0$, such that we only need to explicitly specify two momentum vectors in order to be able to fully fit the function.

\subsection{The Trispectrum}
The trispectrum is the four-point Fourier space correlator \cite{Steele:2021lnz}, $(2\pi)^{3}\delta^{\mathrm{D}}(\mathbf{k}_{1}+\mathbf{k}_{2}+\mathbf{k}_{3}+\mathbf{k}_{4})T_{ijlm}(\mathbf{k}_{1},\mathbf{k}_{2},\mathbf{k}_{3},\mathbf{k}_{4})\equiv\langle\delta_{i}(\mathbf{k}_{1})\delta_{j}(\mathbf{k}_{2})\delta_{l}(\mathbf{k}_{3})\delta_{m}(\mathbf{k}_{4})\rangle+\mathrm{perms.}$.  Focusing only on the connected trispectrum, the tree-level contributions are given by
\begin{align}
    T_{3111}=&6F_{3}^{s}(\mathbf{k}_{2},\mathbf{k}_{3},\mathbf{k}_{4})P_{11}(k_{2})P_{11}(k_{3})P_{11}(k_{4})+\mathrm{3~perms.}~,\\
    T_{2211}=&4F_{2}^{s}(\mathbf{k}_{4},-\mathbf{k}_{3}-\mathbf{k}_{4})F_{2}^{s}(\mathbf{k}_{2},\mathbf{k}_{3}+\mathbf{k}_{4}) P_{11}(k_{2})P_{11}(k_{4})P_{11}(\vert\mathbf{k}_{3}+\mathbf{k}_{4}\vert)+\mathrm{11~perms.}~,
\end{align}
which sum to give the tree-level approximation of the non-linear trispectrum: $T_{\mathrm{nnnn}}\approx T_{3111}+T_{2211}$.

In general cases, trispectra can occupy a wide range of configurations as all four external momentum vectors can vary freely.  However, in the case of the covariance matrix of the power spectrum, we are only interested in trispectra with $\mathbf{k}_{3}=-\mathbf{k}_{1}$ and $\mathbf{k}_{4}=-\mathbf{k}_{2}$, simplifying calculations substantially.

\section{Simulation Parameters}
\label{app:sims}
For this work, we used the publicly available QUIJOTE simulation results.  In these simulations, $M_{\nu}=0$ and $\Omega_{\mathrm{k}}=0$, while 13 runs varied the parameters listed below.  In the fiducial case, every parameter has its central value.  In each of the other 12 runs, one of the parameters was varied to its higher or lower value while the other parameters were kept at their fiducial values.
\begin{align*}
    \Omega_{\mathrm{m}}&\in\{0.3075,0.3175,0.3275\}\\
    \Omega_{\mathrm{b}}&\in\{0.047,0.049,0.051\}\\
    n_{\mathrm{s}}&\in\{0.9424,0.9624,0.9824\}\\
    \sigma_{8}&\in\{0.819,0.834,0.849\}\\
    h&\in\{0.6511,0.6711,0.6911\}\\
    w_{0}&\in\{-1.05,-1,-0.95\}
\end{align*}

\section{Response Matrices}
\label{app:response}

In this appendix, we collect the full and approximate response and reduced response matrices for all of our cosmological permutations.

\begin{figure*}
\centering
    \begin{minipage}{0.48\textwidth}\includegraphics[width=\textwidth]{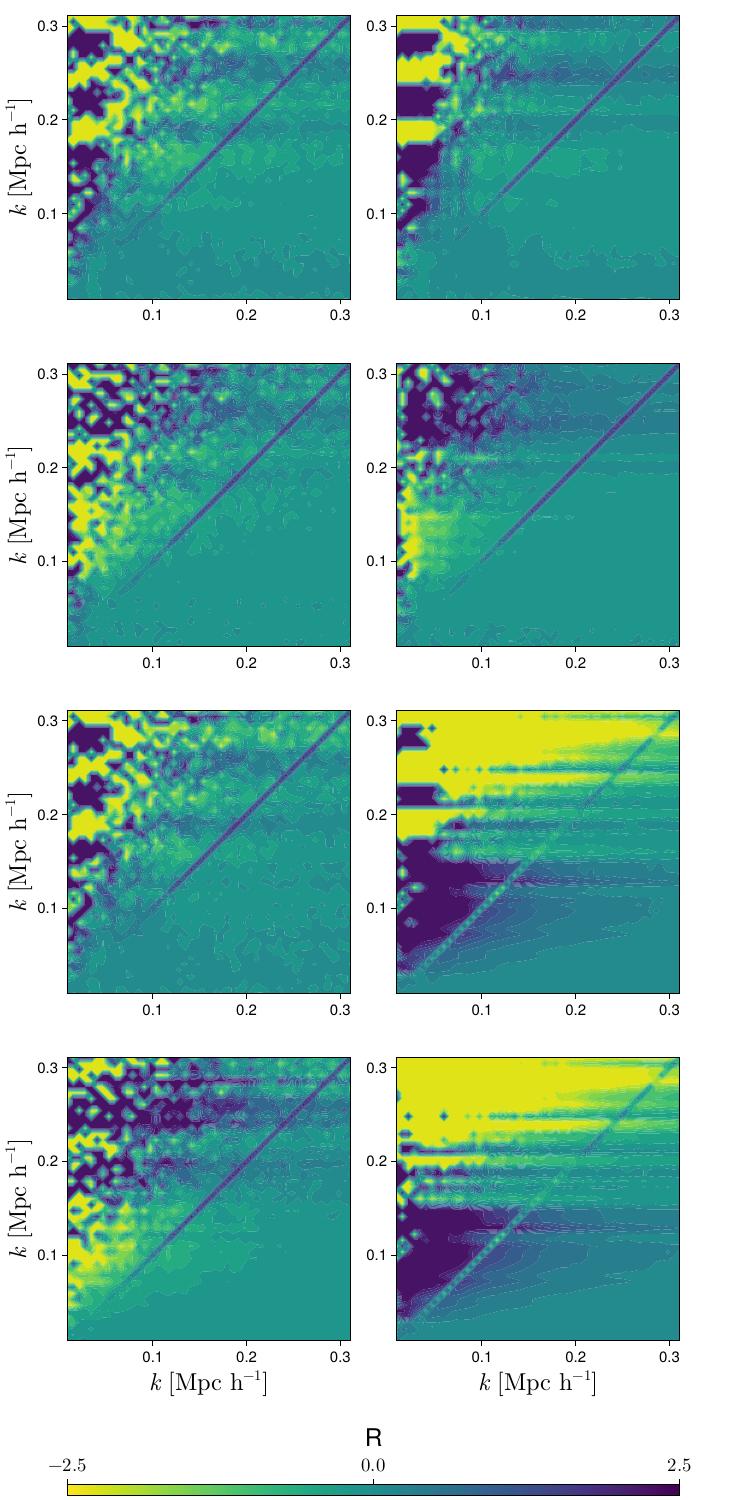}\end{minipage}
    \begin{minipage}{0.48\textwidth}\includegraphics[width=\textwidth]{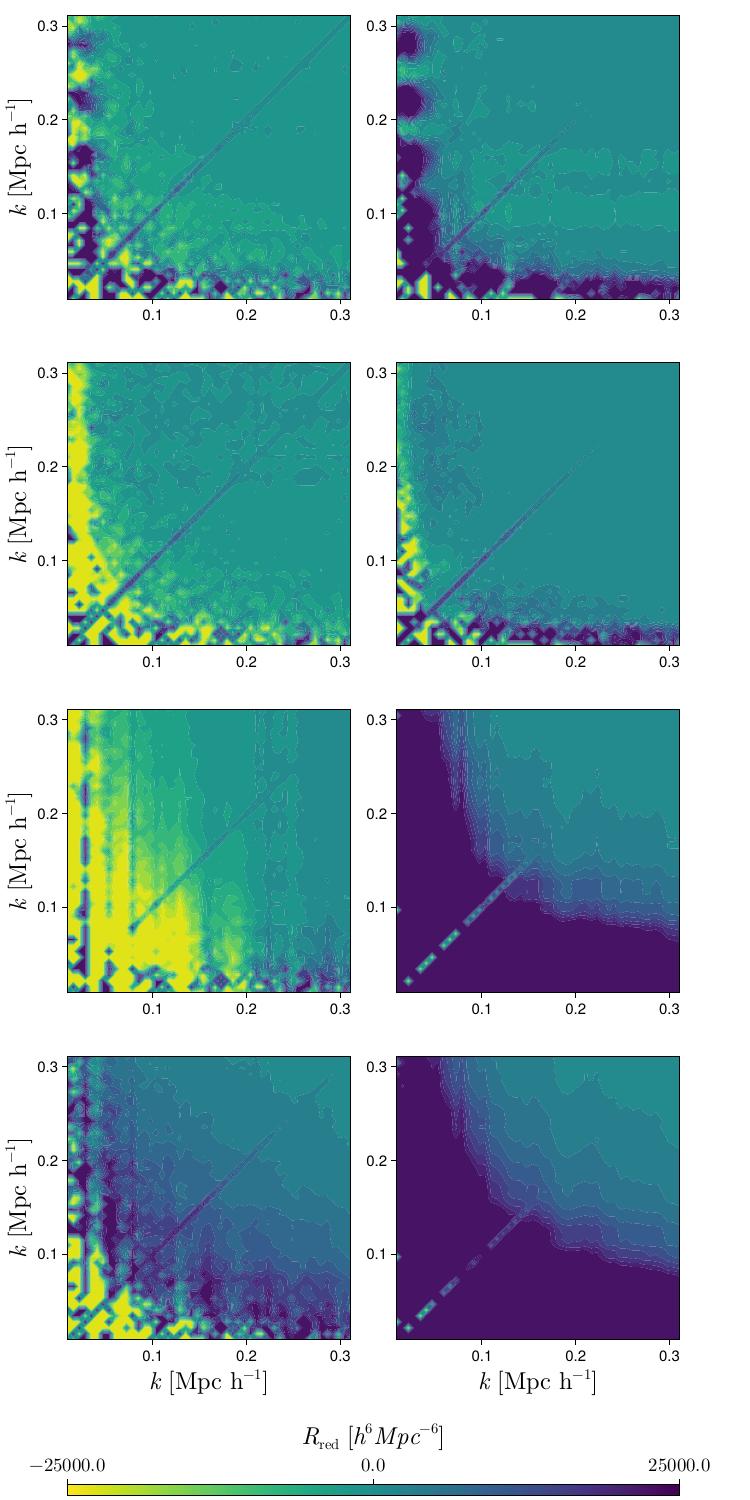}\end{minipage}
    \caption{The response matrices for $\Omega_{\mathrm{m}}$ (left hand pair of columns) and the corresponding reduced response matrices (right hand pair of columns).  In each pair of columns, the top row shows the non-Gaussian response matrices without SSC corrections, the second row show the Gaussian response matrices without SSC corrections, the third row show the non-Gaussian response matrices with SSC corrections, and the bottom row shows the Gaussian response matrices with SSC corrections, with the left hand column showing the results obtained with the full response matrix and the right hand column showing the results from the approximate form of the response matrix.}
\end{figure*}

\begin{figure*}
\centering
    \begin{minipage}{0.48\textwidth}\includegraphics[width=\textwidth]{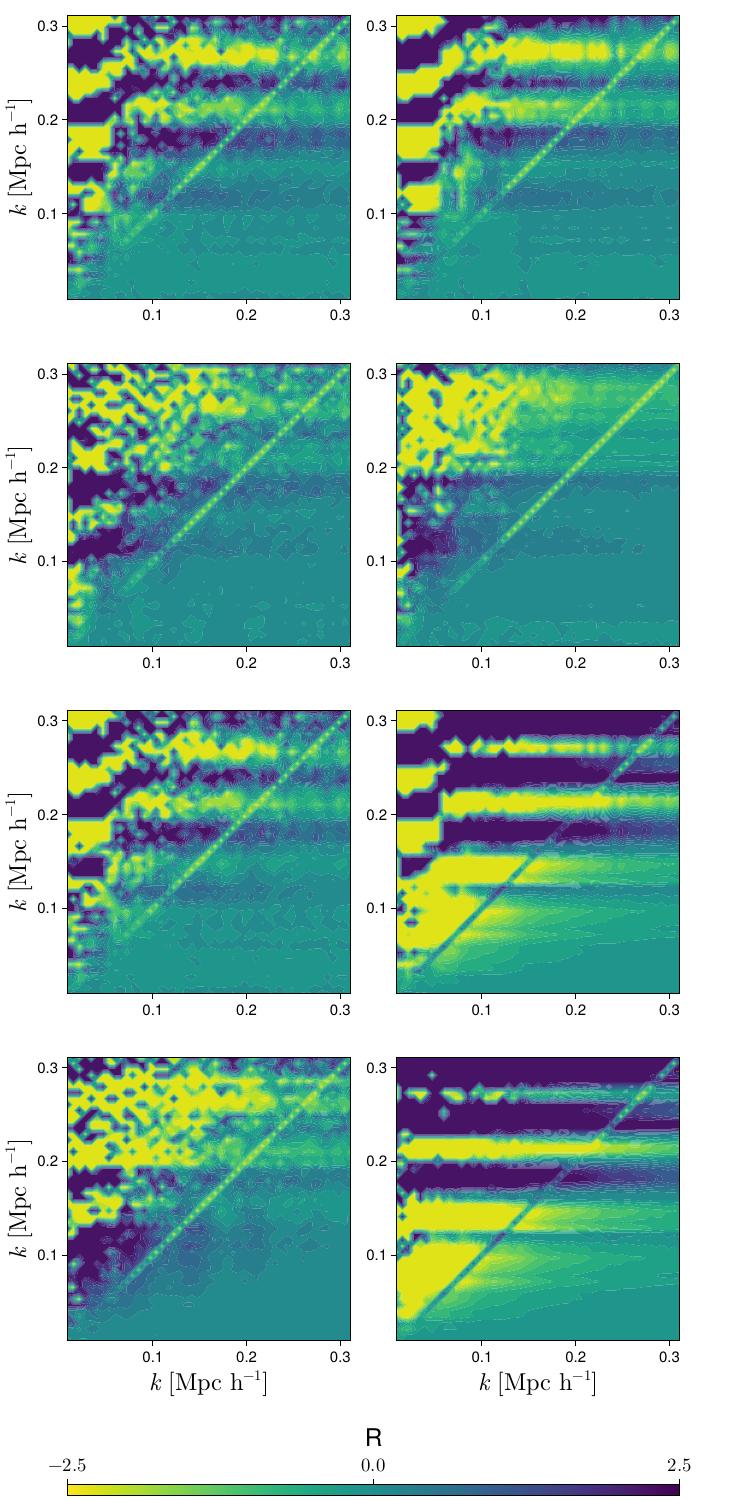}\end{minipage}
    \begin{minipage}{0.48\textwidth}\includegraphics[width=\textwidth]{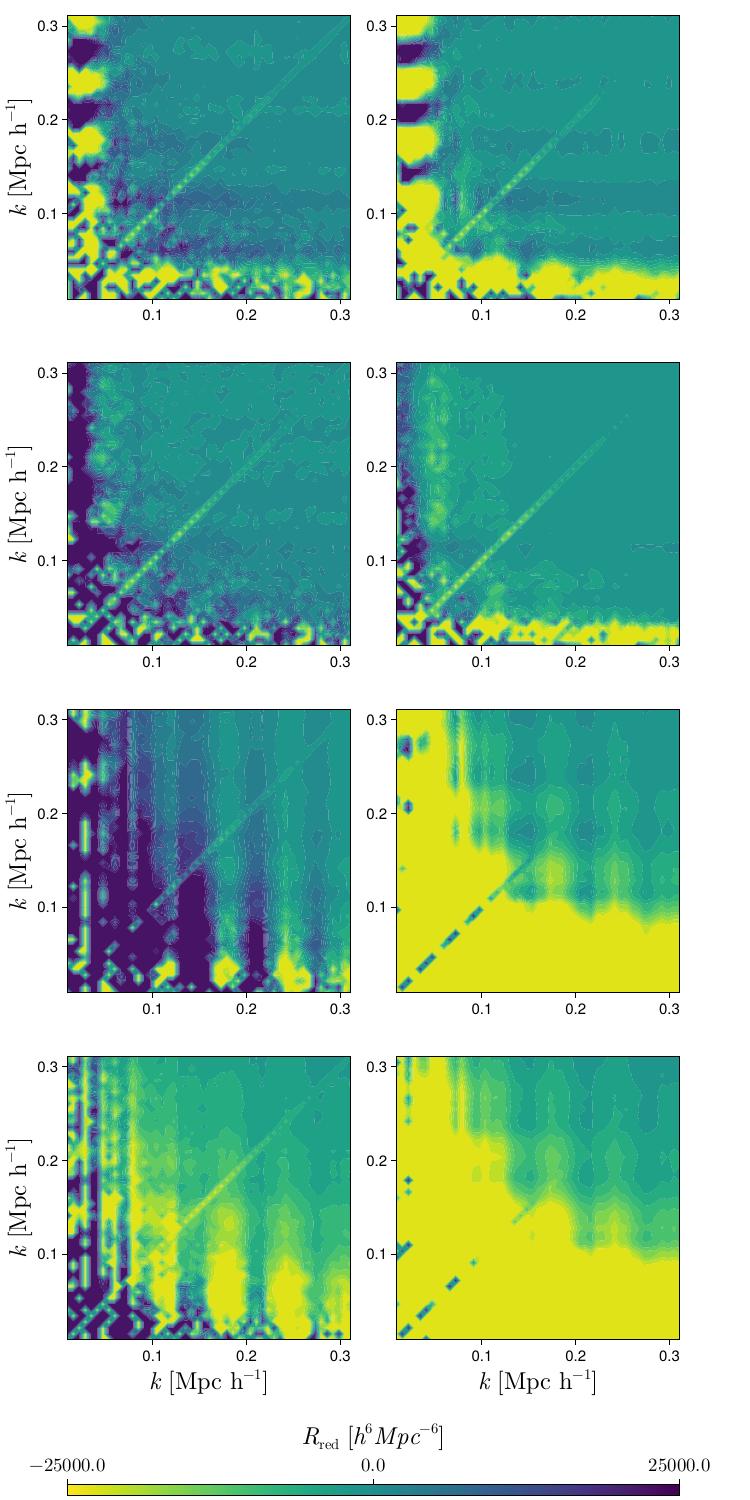}\end{minipage}
    \caption{The response matrices for $\Omega_{\mathrm{b}}$ (left hand pair of columns) and the corresponding reduced response matrices (right hand pair of columns).  In each pair of columns, the top row shows the non-Gaussian response matrices without SSC corrections, the second row show the Gaussian response matrices without SSC corrections, the third row show the non-Gaussian response matrices with SSC corrections, and the bottom row shows the Gaussian response matrices with SSC corrections, with the left hand column showing the results obtained with the full response matrix and the right hand column showing the results from the approximate form of the response matrix.}
\end{figure*}

\begin{figure*}
\centering
    \begin{minipage}{0.48\textwidth}\includegraphics[width=\textwidth]{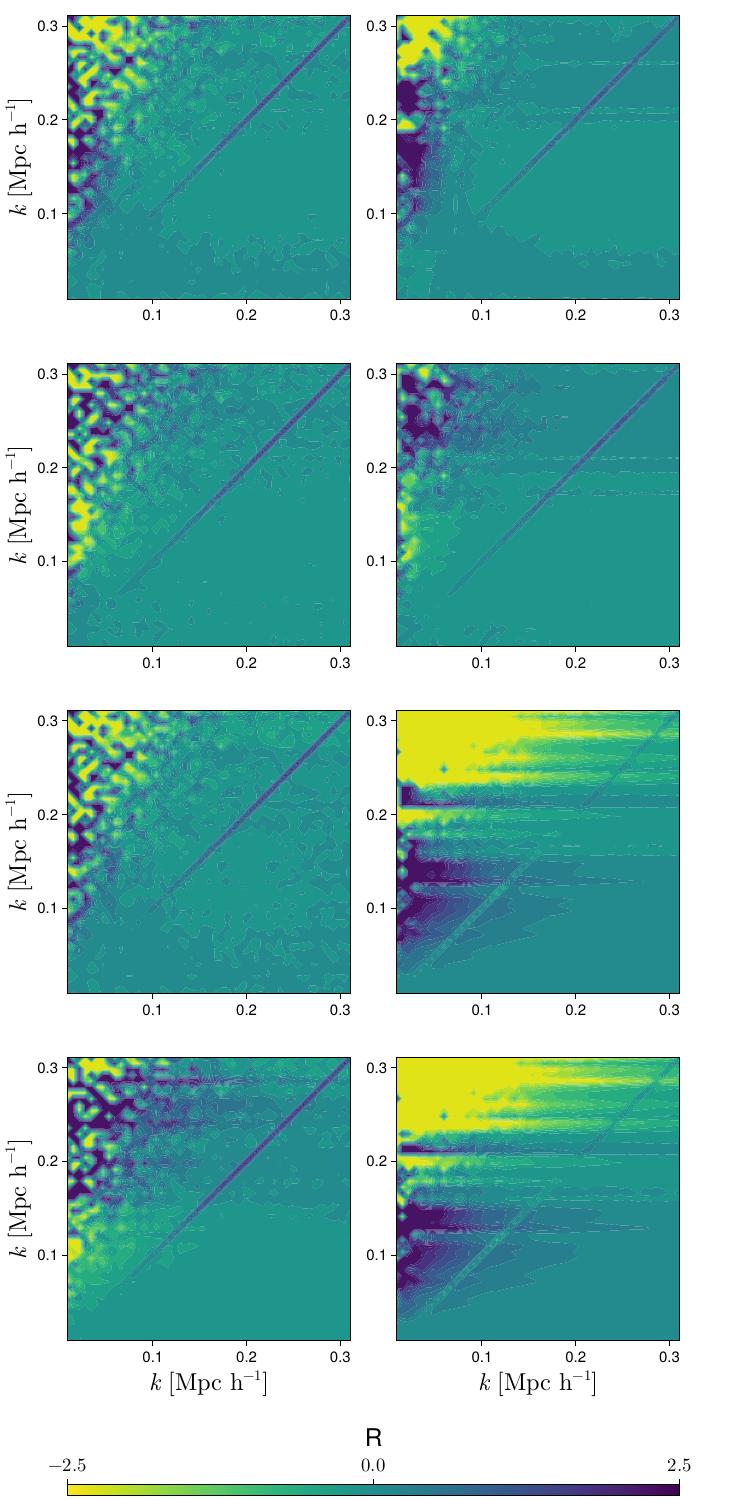}\end{minipage}
    \begin{minipage}{0.48\textwidth}\includegraphics[width=\textwidth]{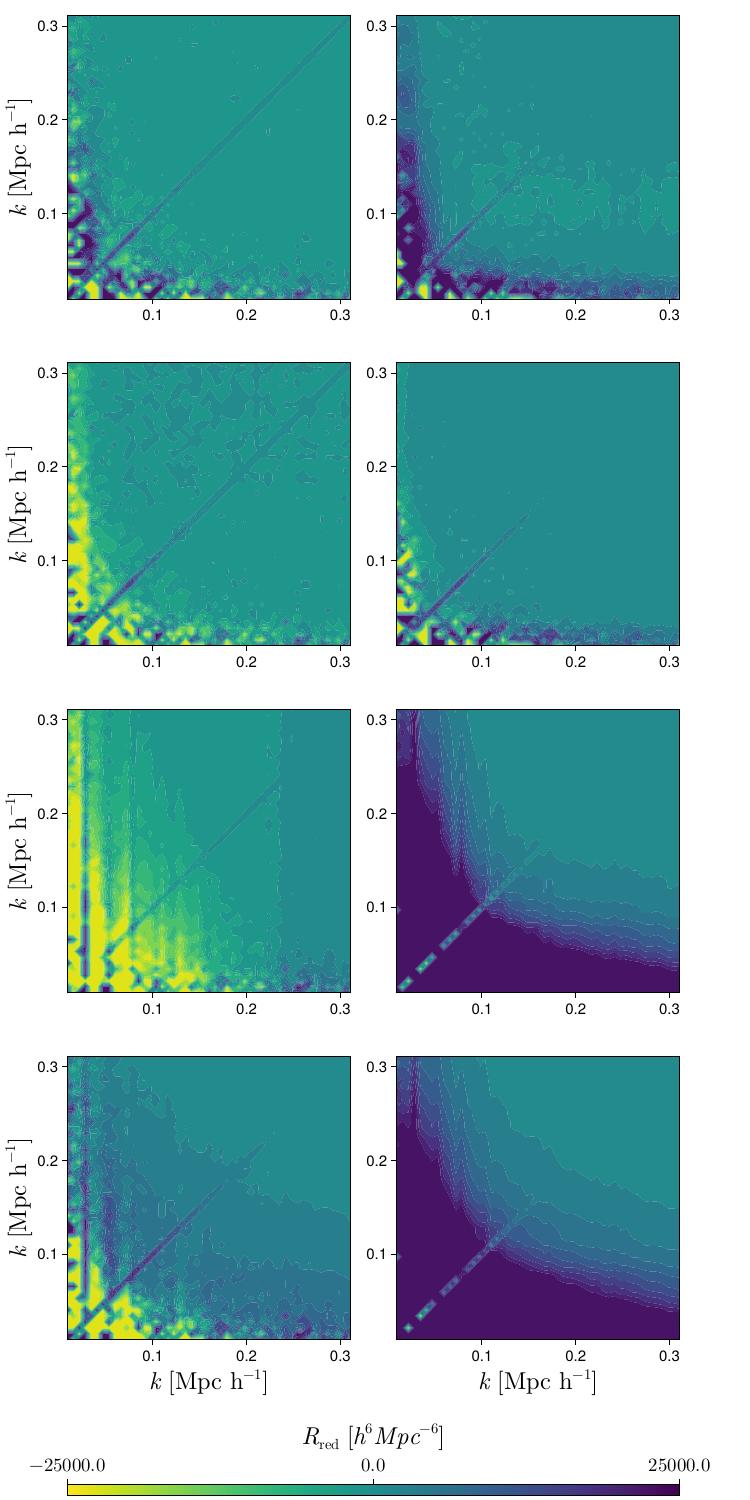}\end{minipage}
    \caption{The response matrices for $n_{s}$ (left hand pair of columns) and the corresponding reduced response matrices (right hand pair of columns).  In each pair of columns, the top row shows the non-Gaussian response matrices without SSC corrections, the second row show the Gaussian response matrices without SSC corrections, the third row show the non-Gaussian response matrices with SSC corrections, and the bottom row shows the Gaussian response matrices with SSC corrections, with the left hand column showing the results obtained with the full response matrix and the right hand column showing the results from the approximate form of the response matrix.}
\end{figure*}

\begin{figure*}
\centering
    \begin{minipage}{0.48\textwidth}\includegraphics[width=\textwidth]{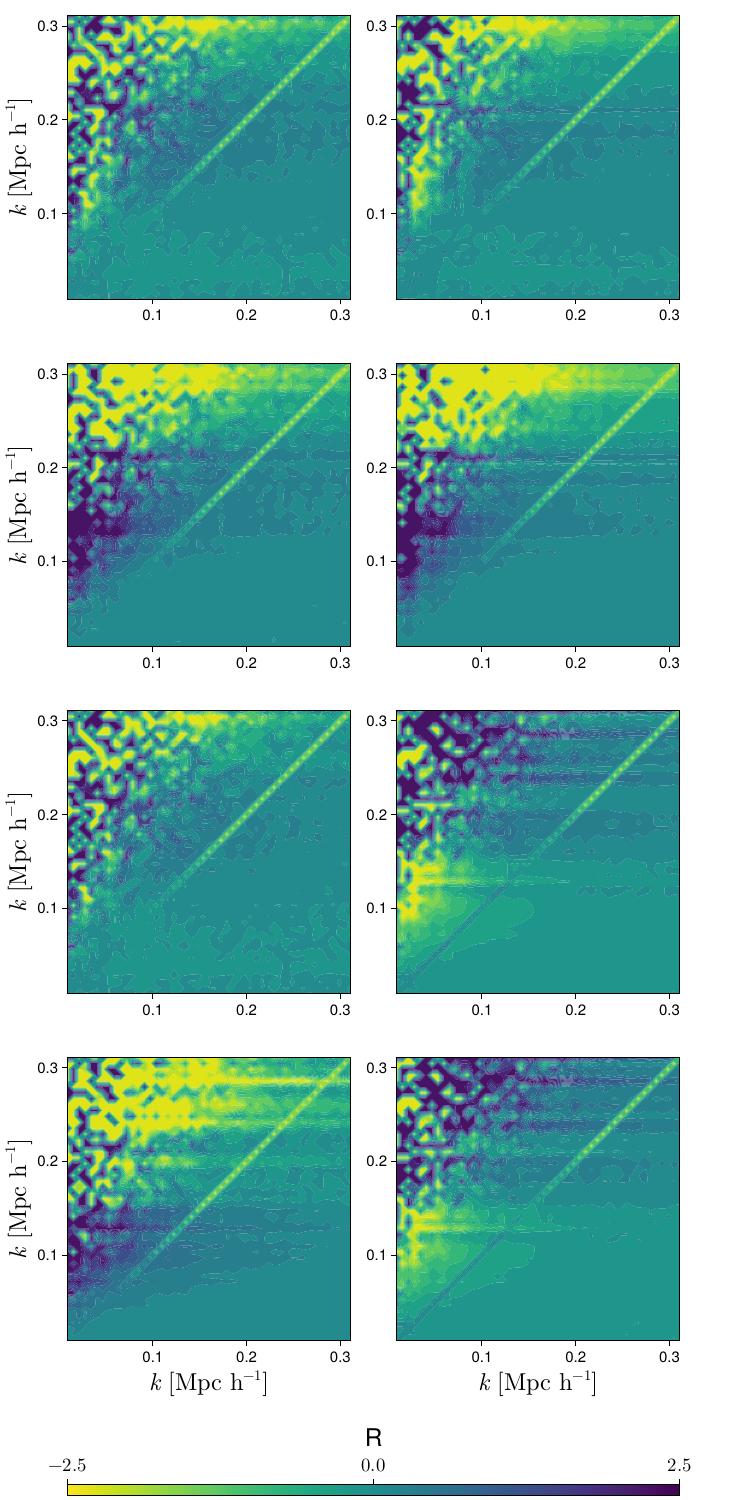}\end{minipage}
    \begin{minipage}{0.48\textwidth}\includegraphics[width=\textwidth]{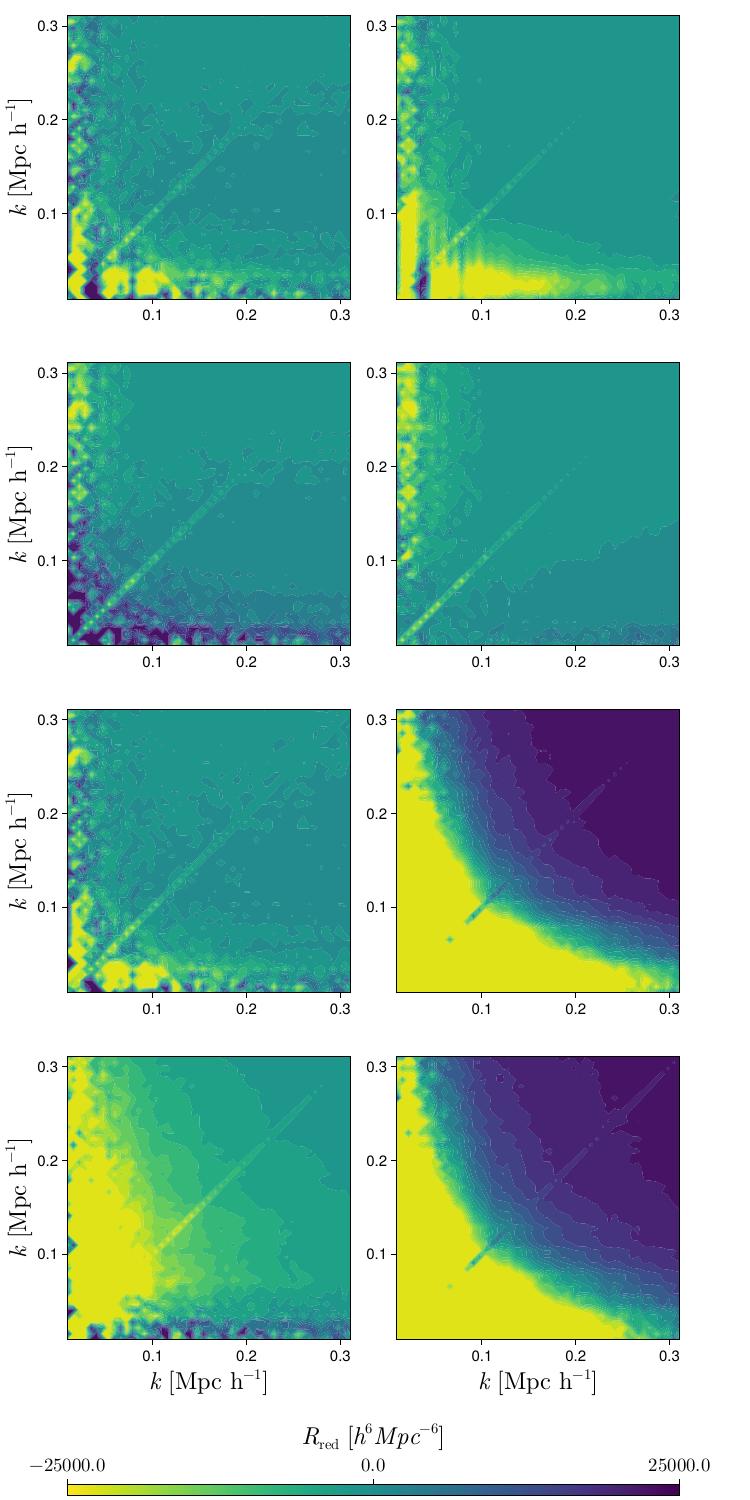}\end{minipage}
    \caption{The response matrices for $\sigma_{8}$ (left hand pair of columns) and the corresponding reduced response matrices (right hand pair of columns).  In each pair of columns, the top row shows the non-Gaussian response matrices without SSC corrections, the second row show the Gaussian response matrices without SSC corrections, the third row show the non-Gaussian response matrices with SSC corrections, and the bottom row shows the Gaussian response matrices with SSC corrections, with the left hand column showing the results obtained with the full response matrix and the right hand column showing the results from the approximate form of the response matrix.}
\end{figure*}

\begin{figure*}
\centering
    \begin{minipage}{0.48\textwidth}\includegraphics[width=\textwidth]{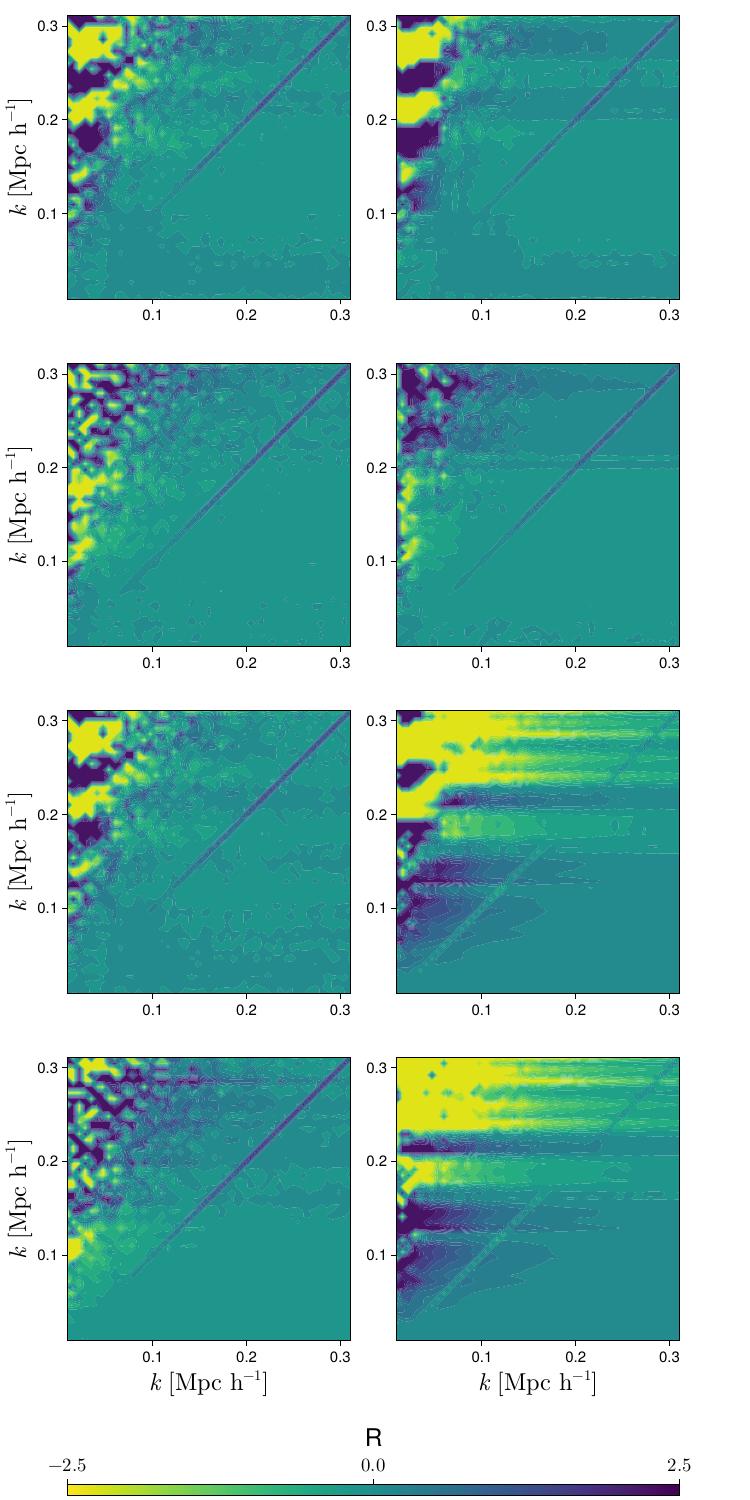}\end{minipage}
    \begin{minipage}{0.48\textwidth}\includegraphics[width=\textwidth]{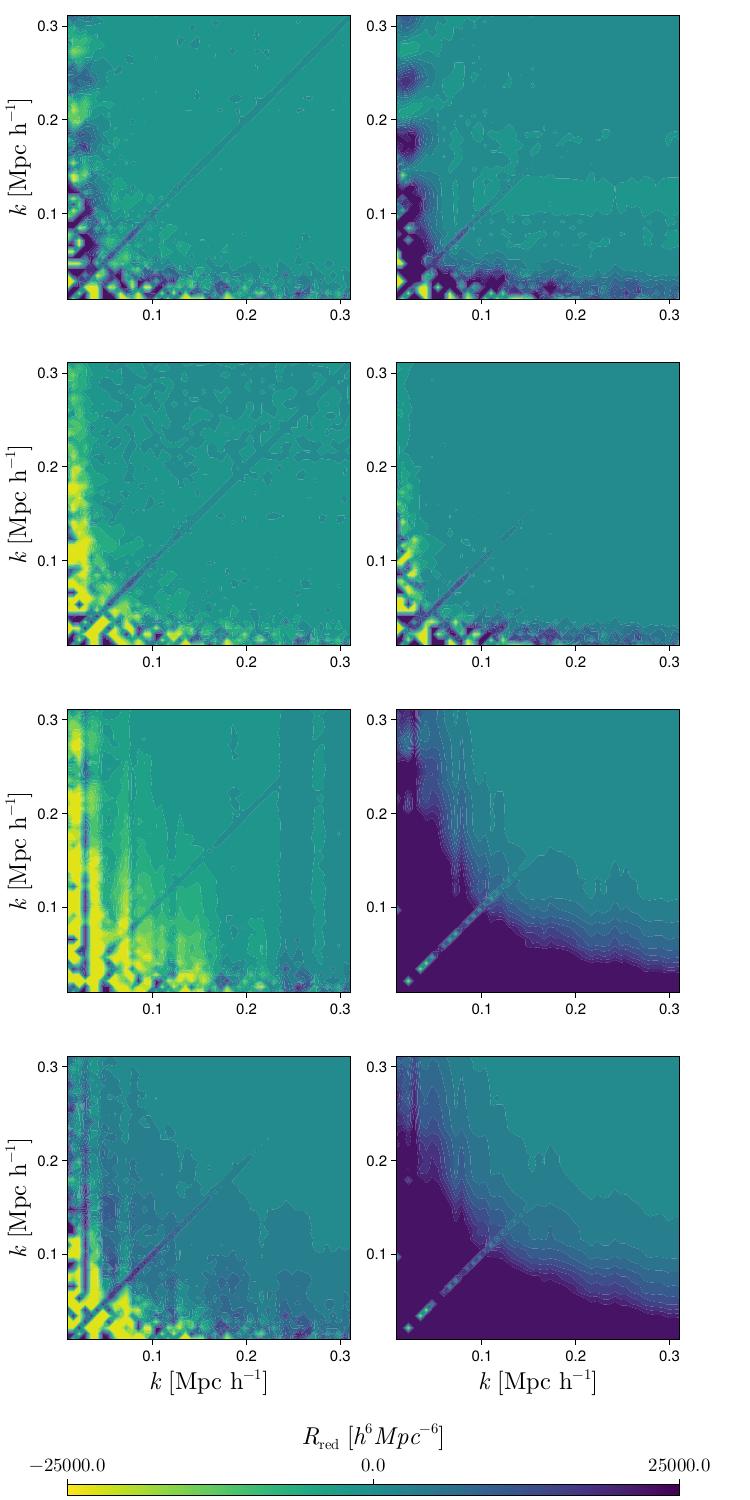}\end{minipage}
    \caption{The response matrices for $h$ (left hand pair of columns) and the corresponding reduced response matrices (right hand pair of columns).  In each pair of columns, the top row shows the non-Gaussian response matrices without SSC corrections, the second row show the Gaussian response matrices without SSC corrections, the third row show the non-Gaussian response matrices with SSC corrections, and the bottom row shows the Gaussian response matrices with SSC corrections, with the left hand column showing the results obtained with the full response matrix and the right hand column showing the results from the approximate form of the response matrix.}
\end{figure*}

\begin{figure*}
\centering
    \begin{minipage}{0.48\textwidth}\includegraphics[width=\textwidth]{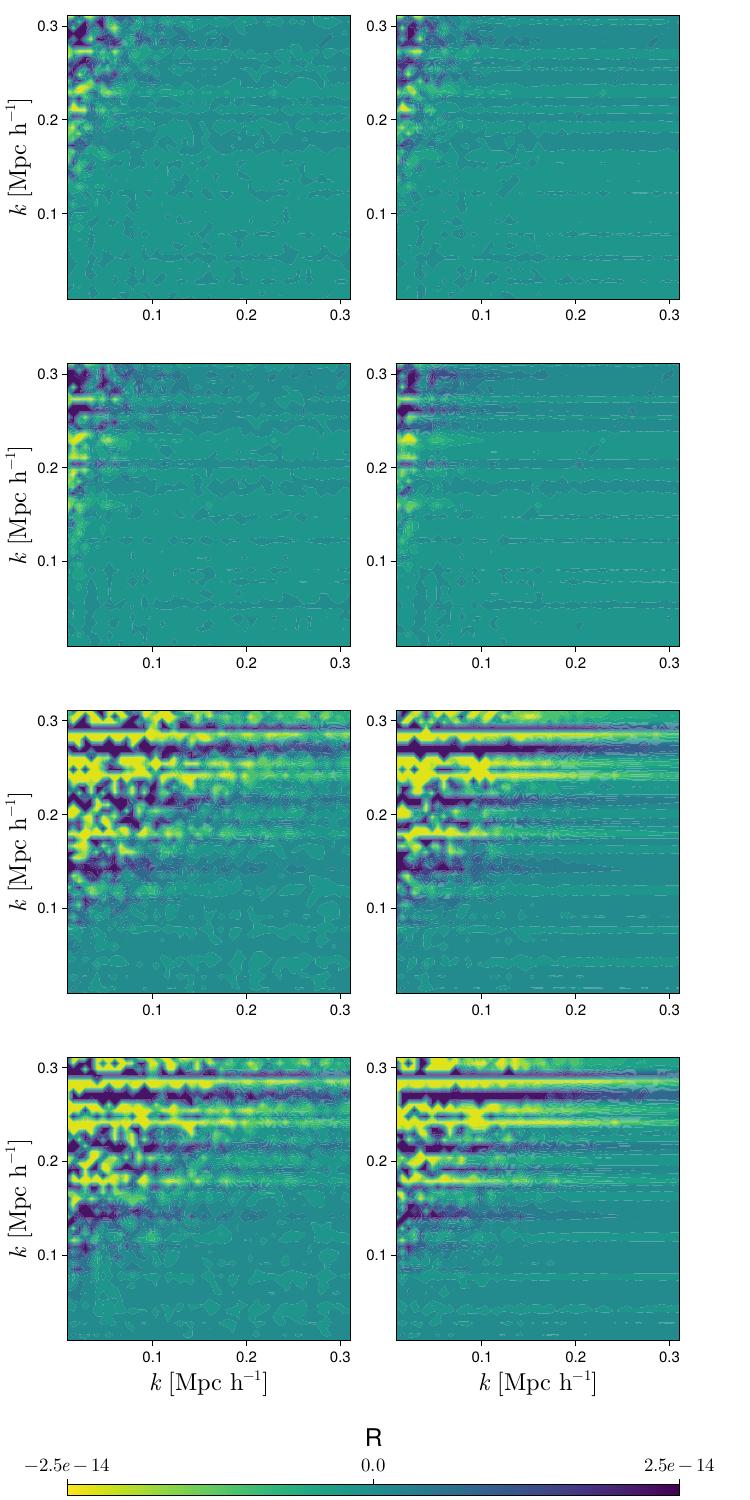}\end{minipage}
    \begin{minipage}{0.48\textwidth}\includegraphics[width=\textwidth]{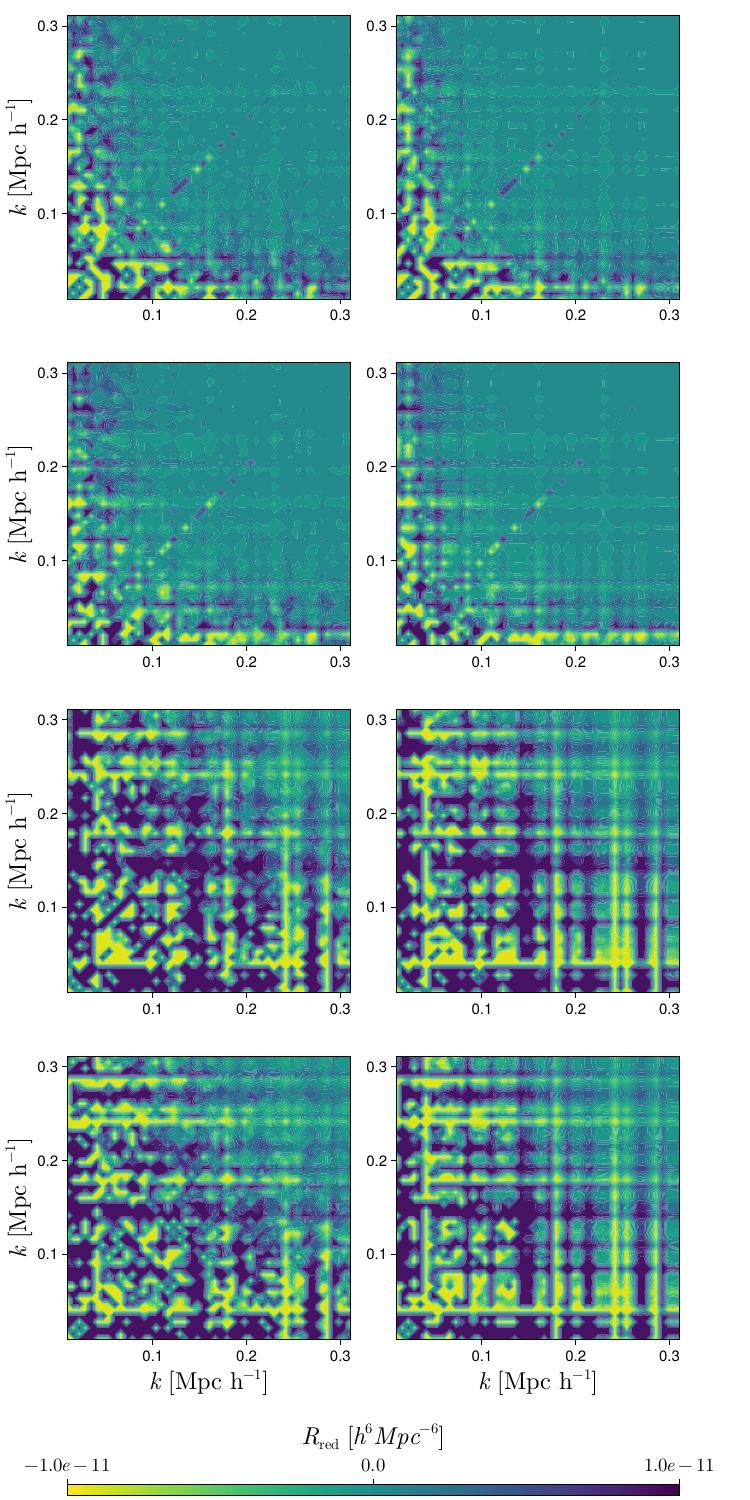}\end{minipage}
    \caption{The response matrices for $w_{0}$ (left hand pair of columns) and the corresponding reduced response matrices (right hand pair of columns).  In each pair of columns, the top row shows the non-Gaussian response matrices without SSC corrections, the second row show the Gaussian response matrices without SSC corrections, the third row show the non-Gaussian response matrices with SSC corrections, and the bottom row shows the Gaussian response matrices with SSC corrections, with the left hand column showing the results obtained with the full response matrix and the right hand column showing the results from the approximate form of the response matrix.}
\end{figure*}

\section*{Data Availability}
The Quijote simulation data used in this paper are publicly available and can be found at \url{https://quijote-simulations.readthedocs.io/} \cite{Villaescusa-Navarroetal2020}.  The covariance matrices calculated for this paper are available upon request.

\end{document}